\documentclass[aip,jcp]{revtex4-2}
\usepackage{graphicx}
\usepackage{amsmath,amssymb}
\usepackage{mathtools}
\usepackage{dsfont}
\DeclareMathAlphabet{\mathbfrak}{U}{euf}{b}{n}
\newcommand{\etal}{\emph{et al.}}
\newcommand\TP[2]{#1\otimes#2}
\newcommand{\unit}{\ensuremath{\mathds 1}}

\newcommand{\bohr}{$\rm\,a_{0}$}
\newcommand{\sign}{\ensuremath{\operatorname{sign}\,}}
\newcommand{\ord}{\ensuremath{\operatorname{ord}\,}}

\newcommand{\di}{\ensuremath{\operatorname{dim}\,}}
\newcommand{\Tr}{\ensuremath{\operatorname{Tr}}}
\newcommand\erw[1]{\langle#1\rangle}
\newcommand\brac[3]{\langle#1|#2|#3\rangle}
\newcommand\overl[2]{\langle#1|#2\rangle}
\newcommand\ket[1]{|#1\rangle}
\newcommand\bra[1]{\langle#1|}
\newcommand\proj[2]{|#1\rangle\langle#2|}
\newcommand{\mcA}{\ensuremath{\mathcal A}}
\newcommand{\mfA}{\ensuremath{\mathfrak A}}
\newcommand{\mfT}{\ensuremath{\mathfrak T}}

\newcommand{\mba}{\ensuremath{\mathbf a}}
\newcommand{\mbb}{\ensuremath{\mathbf b}}
\newcommand{\mbc}{\ensuremath{\mathbf c}}
\newcommand{\mbr}{\ensuremath{\mathbf r}}
\newcommand{\mbt}{\ensuremath{\mathbf t}}
\newcommand{\mbv}{\ensuremath{\mathbf v}}
\newcommand{\mbw}{\ensuremath{\mathbf w}}
\newcommand{\mbA}{\ensuremath{\mathbf A}}
\newcommand{\mbP}{\ensuremath{\mathbf P}}

\newcommand{\bC}{\ensuremath{\mathbb C}}
\newcommand{\bF}{\ensuremath{\mathbb F}}
\newcommand{\bN}{\ensuremath{\mathbb N}}
\newcommand{\bR}{\ensuremath{\mathbb R}}

\newcommand{\sfbA}{\mbox{\boldmath$\mathsf A$}}
\newcommand{\sfbP}{\mbox{\boldmath$\mathsf P$}}
\newcommand{\mfh}{\ensuremath{\mathfrak h}}
\newcommand{\mfH}{\ensuremath{\mathfrak H}}
\newcommand{\mfU}{\ensuremath{\mathfrak U}}
\newcommand{\mfV}{\ensuremath{\mathfrak V}}
\newcommand{\mfW}{\ensuremath{\mathfrak W}}

\newcommand{\Brho}{\ensuremath{\boldsymbol{\rho}}}
\newcommand{\Bsigma}{\ensuremath{\boldsymbol{\sigma}}}
\newcommand{\Bpi}{\ensuremath{\boldsymbol{\pi}}}
\newcommand{\BPhi}{\ensuremath{\boldsymbol{\Phi}}}

\newcommand{\nullv}{\ensuremath{\mathbf 0}} 
\begin{document}

\title{Chemical Bonding in Many Electron Molecules}
\author{Alexander F. Sax}
\email{alexander.sax@uni-graz.at}
\affiliation{Department of Chemistry, University of Graz,  Graz, Austria}

\begin{abstract}
Chemical bonding is the stabilization of a composite molecular system caused by different interactions in and between the subsystems, among the strong kinds of bonding is covalent  bonding especially important. Characteristic for covalent bonding are small atom groups with short distances between the involved atoms, indicating that covalent bonding is essentially a local effect, according to Lewis, this is caused by shared electron pairs. However, the energetic stabilization is an approximately additive  one-electron effect, as was shown by Ruedenberg and coworkers.
In systems composed of many-electron subsystems, the fermionic character of the electrons determines the structure of the electron distribution in a subsystem, and it is decisive for the local interactions between the subsystems. Especially important is the Pauli exclusion principle (PEP), which directs the relative positions of identical electrons. Spin and charge rearrangements are of utmost importance for chemical bonding. Quantum chemical methods like CASSCF (complete active space SCF), also called FORS (fully optimized reaction space), are made to cover all such processes. The standard building blocks of CASSCF wave functions are delocalized molecular orbitals, which cannot display local effects. OVB (orthogonal valence bond) is a method to analyze CASSCF wave functions and to reveal local processes that are responsible for both the energetic aspects of bonding and the spatial structure of the stabilized system. This is shown by analyzing dissociation of ethene, disilene, and silaethene, and the corresponding reverse reactions. Aspects of diabaticity of the reactions and entanglement of subsystems are discussed.
\end{abstract}


\maketitle 
\section{Introduction}
The term bonding is used to describe all stabilization processes in composite systems; the result of bonding is a bonded system or simply a bond. Frequently it is said that a bond exists in a bonded system, as if by the bonding process something, which is called a bond, were added to  or suddenly appeared in the system.\cite{CoulsonCB} Instead of bonding one speaks then of ``making'' or ``creating a bond''. The destabilization of a system can be called debonding, in chemistry parlance one speaks of breaking a bond.\cite{Sax2015} A system must be defined by explicitly stating which objects belong to it and which belong to the environment; the definition of a system must also include a listing of all interactions in the system and, if any, between the system and the environment. If a subset of a system has all properties of a system, it is called a subsystem. The interactions in the system are then the interactions in and between the subsystems.

Molecular chemistry deals molecular systems whose subsystems are atoms or stable atom groups, stable means that the atom group does not spontaneously disintegrate into smaller atom groups or atoms. Stable atom groups, also termed molecular species, can be radicals or molecules, both charged and uncharged; molecules are molecular species that do not react violently with other species of its own kind.\cite{Preuss1963} The term fragments describes any kind of subsystem in a molecular system, it may be reactants and products in real chemical reactions, but also  atom group that can be regarded as reactants in  hypothetical reactions. In chemical reactions, the structure of a molecular systems may change considerably; small fragments may combine to larger ones, or large systems may dissociate into small fragments. Chemical reactions are fundamental to an understanding of bonding and debonding.

To study a physical system means to study how it changes in time; no field of science is interested in systems that never change. The description of a system is done with the help of physical quantities the numerical values of which can be measured; the definition of the system decides which physical quantities are necessary to exhaustively describe a system and its change. In classical physics, it is presupposed that for every physical quantity the numerical value is exactly known at every time, the set of numerical values defines the state of the system at a certain point in time. The values of the physical quantities are thus functions of the time, and the time development of a system is described by the time development of all relevant physical quantities. Knowing the state of a system at a certain time allows to predict what will happen with the system in the future.\cite{Ballentine2009} In quantum theory, system definition is different and will be discussed in the next section.

Two system properties are especially important for the description of bonding, the total energy and the spatial structure of a system. As a scalar quantity, energy allows to set up a scale for comparing systems;  the difference in the total energy between  bonded and  non-bonded systems is the most important measure of the strength of bonding. Every change of the spatial structure is an indicator for a change of a system. Especially important is the change of the distance between fragments, howsoever the distance is defined.  Bonding of initially non-bonded or non-interacting fragments leads always to a reduction of the distances between fragments, the magnitude of the distance in the stable system correlates in general with the strength of bonding, the larger the distance is the weaker is the bonding. A system stabilized by weak molecular interactions can be a complex of stable neutral molecules with many atom pairs having the same distances, as for example in a complex of  two polyaromatic molecules in parallel arrangement. Here, the shortest distance between atoms in different molecules is approximately the sum of the van der Waals radii of the involved atoms. Ignoring the Coulomb interaction between charged systems, the most important strong molecular interaction is covalent bonding. A characteristics of it is the small number of interacting atoms, mostly two to three atoms, seldom more than four atoms.\cite{Levine2005} As a result of covalent bonding, one frequently finds in the product an atom pair AB with a very short distance between them, each atom belonging to a different fragment. The distance between A and B is called the (covalent) bond length of the atom group AB, it is approximately the sum of the covalent radii of the atoms, which are roughly half of the van der Waals radii. In chemistry parlance, formation of a product is explained, by the creation of a covalent AB bond between atoms A and B; when speaking of a bond there is often no clear distinction between the atom group and the result of stabilization. Sometimes one gets the impression that the creation of the AB atom group is the cause of system stabilization and that there is no stabilization without such atom groups.\cite{Hydrophobic}

Lewis proposed that a covalent bond between two atoms consists of a pair of electrons shared between the atoms to which every atom contributes one unpaired electron. If only one electron pair is shared by two atoms, one speaks of a single bond, in case of two or three shared pairs one speaks of double or triple bonds. A fragment with an unpaired electron located at an atom is a radical, if two or three electrons occupy degenerate or nearly degenerate orbitals, one speaks of diradicals or triradicals, if the electrons are located at the same atom they are mono-centered.\cite{Salem1972,Borden1982} These species are, in general, very reactive, diradicals are able to form double bonds, triradicals can form triple bonds.
It is chemical knowledge that the strength of a double bond between same atoms is less than twice the strength of a single bond, and that the two bonds of a double bond are not equivalent. It is a genuine interest of theoretical chemistry to find out, what influences the strength of bonds and how the strength of a bond and the molecular geometry are connected.

In this paper, CAS(4,4) (complete active space MCSCF with four active molecular orbitals and four active electrons) wave functions are used to describe the formation and dissociation of the double bond in ethene, disilene, and silaethen, and then these wave functions will be analysed using the OVB method (orthogonal valence bond). With this method, charge and spin rearrangements at the heavy atoms carbon and silicon, respectively, can be revealed.

\section{Systems and quantum states}
The following presentation of the basic physical concepts is  influenced by Cohen-Tannoudji\cite{Cohen1977}, Levy-Leblond\cite{Levy1990}, Gottfried\cite{Gottfried2003}, Susskind\cite{Susskind2014} and Greiner\cite{Greiner2001}. Dirac's bra-ket notation will be used throughout,  the same symbol $\ket{X}$ will be used for a physical state X and the mathematical object describing the state.

The relationship between system state and measurement in quantum physics is very different from classical physics.
A quantum system can never be characterized by a set of exact numerical values of all physical quantities at a certain time; the best one can know for each physical quantity is the probability that a certain numerical value will be measured. Nevertheless, also with these probabilities one can make predictions about the future of a quantum state. The number of possible numerical values that can be found in a measurement can be countable or uncountable. Whenever a physical quantity $\mcA$ is measured, irrespective of the initial state $\ket{X}$ in which the system is, the outcome of a measurement is one of the possible numerical values of $\mcA$, called an \emph{eigenvalue} $a_i$, and the system will be transformed into the corresponding \emph{eigenstate} $\ket{a_i}$ of $\mcA$. Only bound states of molecular systems are considered in this paper, therefore all  eigenvalue spectra are discrete and can be indexed with integers.  In general, it cannot be predicted which eigenvalue will be measured and into which eigenstate the system will be transformed. One can only measure or calculate the so called \emph{transition  probability} $P(a_i\! \leftarrow\!\! X)$ for the transformation of the initial state $\ket{X}$ into eigenstate $\ket{a_i}$. To avoid any connotation with a temporal evolution of the initial state, one speaks also of a projection of state $\ket{X}$ onto eigenstate $\ket{a_i}$. If after the first measurement the same quantity is measured without changing the measuring apparatus, the same value $a_i$ will be found and the system remains in the eigenstate $\ket{a_i}$, and this holds true for every repeated measurement. One can say that in the first measurement the system is \emph{prepared} in the eigenstate, and that in every repeated measurement this state is \emph{confirmed}. In this case, one has absolute control of the system's state. A state that is specified or controlled as precisely as possible is called a \emph{pure state}\cite{Ballentine2009}. Every eigenstate of a quantity $\mcA$ is a pure state.

A set of eigenstates of a physical quantity $\mcA$ has the following properties: 1) Eigenstates are pairwise \emph{orthogonal}. This means, no system that is initially in an eigenstate of $\mcA$ can be transformed, by measuring $\mcA$, into another of its eigenstates; the system remains unchanged in this state. The transition probability for a transformation between different eigenstates is zero, $P(a_i\! \leftarrow\!\! a_j)=0$. 2) Eigenstates are \emph{normalized}. Since repeated measurement of $\mcA$  in one of its eigenstate does not change the state the transition probability must be one, $P(a_i\! \leftarrow\!\! a_i) =1$. 3) The set of eigenstates must be \emph{complete}. That means, the sum of all  transition probability $P(a_i\! \leftarrow\!\! X)$  must be one,
\begin{equation}\label{equ:completeness}
\displaystyle \sum_i P(a_i \!\leftarrow \!\!X) =1.
\end{equation}
If all three requirements are fulfilled, one says the set of eigenstates of a physical quantity constitutes a \emph{complete set of orthonormal states} (CSOS). The projection of a pure state $\ket{X}$ on the CSOS allows an \emph{exhaustive analysis} of state X. The kets of a CSOS that are used for an analysis are frequently called intermediate states.

But not only eigenstates of a physical quantity can be used for an analysis of an arbitrary pure state, but any set of mathematical objects having the properties of a CSOS. This follows from the fact, that a physical quantity is defined by both the complete set of orthonormal eigenstates  and the corresponding eigenvalues. If different numerical values are attributed to the eigenstates, another physical quantity is defined. But for the analysis of a state only the elements of the CSOS but not the eigenvalues are needed.

Let us assume that calculation of the transition probability for the projection of a pure initial state $\ket{X}$ onto a pure final state $\ket{Y}$ includes  an intermediate measurement of a physical quantity $\mcA$. Going from $\ket{X}$ to $\ket{Y}$ via eigenstate $\ket{a_i}$ of  $\mcA$ are two independent events, so the probability of the composite event is the product of the transition probabilities $P(a_i \!\leftarrow \!\!X)$ and $P(Y \!\leftarrow \!\!a_i)$.  Since the eigenstates $\ket{a_i}$ constitute a CSOS one could assume, that, when this is done for all possible eigenstates, the sum of all these products is the transition probability
\begin{equation}\label{equ:classprob}
P(Y \!\leftarrow \!\!X)  = \sum_i P(Y \!\leftarrow \!\!a_i)P(a_i \!\leftarrow \!\!X) ,
\end{equation}
however, this is not found. Instead one finds in many experiments for varying final states $\ket{Y}$ transition probabilities $P(Y \!\leftarrow \!\!X)$ that differ significantly from what equation (\ref{equ:classprob}) predicts. Indeed, one finds an oscillatory behaviour of the transition probabilities reminiscent of the interference pattern of classical waves, which can never be obtained when the total transition probability is calculated according to equation (\ref{equ:classprob}). To reproduce the experimental findings, the projection of state $\ket{X}$ onto state $\ket{Y}$ must be described by a complex valued function $\overl{Y}{X}$, called quantum amplitude or probability amplitude, with which one can calculate the transition probability as
\begin{equation}\label{equ:QMprob}
P(Y \!\leftarrow \!\!X) = |\overl{Y}{X}|^2
\end{equation}
But then, also the probabilities $P(a_i \!\leftarrow \!\!X)$  and $P(Y \!\leftarrow \!\!a_i)$ must be calculated with quantum amplitudes, and the correct calculation of $P(Y \!\leftarrow \!\!X)$ goes as follows: first calculate $\overl{Y}{X}$ as a sum of products of quantum amplitudes
\begin{equation}\label{equ:quant}
\overl{Y}{X}=\sum_i \overl{Y}{a_i}\overl{a_i}{X}
\end{equation}
and then square the magnitude of the quantum amplitude as in equation \ref{equ:QMprob}.

For the mathematical description of a system one needs a state space, which has the properties of an abstract vector space endowed with a scalar product (see Appendix). A state space is frequently  a Hilbert space. Elements of a state space are called kets, every state of the state space can be represented as a linear combination or coherent \emph{superposition} of other kets.
\begin{equation}\label{equ:lincomb}
\ket{X} = c_1\ket{\phi}_1+ c_2\ket{\phi}_2+\dots
\end{equation}
the coefficients of linear combination $c_1, c_2, \dots$ are elements of the scalar field $\bF$ of the vector space.

Every state space has a complete orthonormal basis (ONB) $\{\ket{b}_i\}_{i\in I}$, which is a CSOS and represents the eigenstates of some physical quantity. In finite dimensional vector spaces completeness of the ONB is trivially fulfilled.
With such a basis, every state $\ket{X}$ can be represented as a superposition of the basis kets
\begin{equation}\label{equ:repX}
\ket{X}=\sum_i c_i\ket{b}_i, c_i \in \bF
\end{equation}
Using this representation of state X, the transition probability $P(Y \!\leftarrow \!\!X)$ consists of two contributions:
\begin{equation}\label{equ:superpos1}
\begin{split}
P(Y \!\leftarrow \!\!X) = |\overl{Y}{X}|^2= \sum_i c_i\overl{Y}{b_i} \sum_j c^*_i\overl{b_j}{Y} =\\
 \sum_i |c_i|^2 |\overl{Y}{b_i}|^2 + \sum_{i<j}\left(c_ic^*_j \overl{Y}{b_i} \overl{b_j}{Y}+ c_jc^*_i \overl{Y}{b_j} \overl{b_i}{Y}\right).
 \end{split}
\end{equation}
The first contribution is a sum of only positive terms, which cannot describe interference effects; this do the terms in the double sum.

Pure states are represented by elements of the state space, and therefore every superposition of pure states is a pure state. A pure state $\ket{X}$ can also be represented the projection operator $\proj{X}{X}$ onto the subspace spanned by the ket $\ket{X}$. This projection operator is the density operator $\Brho$ of the pure state. If one plugs in the representation of $\ket{X}$ \ref{equ:repX}, one gets
\begin{equation}\label{equ:superpos2}
\proj{X}{X} = \sum_i \sum_j c_i c^*_j \proj{b_i}{b_j} = \sum_i  |c_i|^2 \proj{b_i}{b_i} +
\sum_{i<j}\left(c_ic^*_j \proj{b_i}{b_j }+ c_jc^*_i \proj{b_j}{b_i}\right)
\end{equation}
a sum of projection operators onto the basis states multiplied by positive real numbers plus a double sum of transition operators multiplied by complex numbers. Every state that cannot be represented by an expression (\ref{equ:superpos2}) is not a pure state but a mixed state and cannot be element of the state space.

Mixed states are represented by a density operator that is a statistical mixture of projection operators of pure states
\begin{equation}
\Brho = \sum_{k} p_k \proj{k}{k}, \qquad 0 \le p_k\le 1, \qquad \sum_k p_k = 1
\end{equation}
The number of projectors in the sum is arbitrary, it may be smaller or much larger than the dimension of the state space. The $p_k$ are probabilities. If the sum consists of a single non-zero term, the probability must be one and then the density operator represents a pure state.

One should never forget that ``[a]lthough pure states abound in text books and research papers, systems in the real world are rarely in pure states.''\cite{Gottfried2003} If a system is in a mixed state, we have incomplete knowledge about the system, we only know  that it can be in either of the pure states with a corresponding probability.
The following situation demonstrates this. The singlet ground state and the first excited triplet state of the hydrogen molecule have the same dissociation limit. The dissociated system is fourfold degenerate (three triplet states, one singlet state) and it is not know, which of the four states describes the system. It could be either of the four states with 25 percent probability.\cite{McWeeny2000} Note, that in equation (\ref{equ:superpos2}) the first sum on the right side is such a statistical mixture, without the second sum, which represents  interference, $\Brho$ would be the density operator of a mixed state. But in chemistry there are also systems in mixed states where no molecule must be atomized. Salem and Rowland\cite{Salem1972} described the reaction of a diradical with another reactant:  ``In the great majority of systems,[...] the odd electron interaction is sufficient at least for two distinct states to be recognized by the electron-nuclear hyperfine probe of ESR spectroscopy. At the same time it is expected that it is not so large that gross chemical perturbations (such as the approach of a reactant) will discriminate entirely between the two states. Rather, the chemical perturbation will create a "mixture" of the two states, with typical bifunctional behavior.''

In another important remark Gottfried\cite{Gottfried2003} states: ``There is a common and serious misconception that mixtures only arise when pure states are "mixed" by the environment, such as a temperature bath, or by some apparatus, such as an accelerator. Not so: \emph{If a composite system is in a pure state, its subsystems are in general in mixed states.} This is the context in which mixtures are often important in discussions of the interpretation of quantum mechanics, and also in many other contexts.'' The discussion of bonding in molecular systems is such a context. Mixed states are essential for the discussion of entanglement in composite systems, a discussion that has not yet gained  importance in molecular quantum theory.

\section{1-Electron states}
Orbitals are one-electron state functions. \emph{Atomic orbitals} (AOs) $\{\ket{\chi}_i\}_{i\in I_a}$ describe electrons in an atom, \emph{molecular orbital}s (MOs) $\{\ket{\psi}_i\}_{i\in I_m}$ describe electrons in a molecule, and \emph{fragment MOs} (FMOs) $\{{\ket{\varphi^{(k)}}}_i\}_{i\in I_f}$ describe electrons in  fragment $k$, $k=1,2,\dots$. All orbitals are assumed to be normalized. $I_m$, $I_f$ and $I_a$ are the index sets of the respective families of orbitals.

AOs of an element atom are elements of an infinite dimensional Hilbert space $\mfh_a$, the denumerable set of eigenfunctions of the hydrogen-like atom is a  Hilbert basis. Although such a basis is an ONB, it is of minor importance for quantum chemical calculations; more important are finite sets of, in general, non-orthogonal basis functions, called  AO basis sets, the dimension of $\mfh_a$ is then the number of basis functions in the AO basis set, $\di\mfh_a  = M_a$. Whenever finite AO bases are used, all wave functions are projected onto finite dimensional Hilbert subspaces. This is standard in computational chemistry.

The MOs of a molecule with $N_{\rm at}$ atoms can be represented as linear combinations of the AO basis functions of all atoms;  the MO Hilbert space, $\mfh_m$, is the direct sum of all AO Hilbert spaces, $\mfh_m = \bigoplus_{\rm iat=1}^{N_{\rm at}} \mfh^{\rm iat}_a$;
the dimension of $\mfh_m$, that is the number of basis functions for the molecule, is accordingly the sum $M_{\rm mol} = \sum_{\rm iat=1}^{N_{\rm at}} M_{\rm iat}$. Analogously, the dimension of the Hilbert space of FMOs for the $k$-th fragment, $\mfh^{(k)}_f$, is the number of basis functions, $\di\mfh^{(k)}_{\rm fr} = M^{(k)}_{\rm fr}$. If the fragments have no atoms in common, the Hilbert spaces $\mfh^{(k)}_f$ are linearly independent subspaces in the  direct sum  $\mfh_m = \bigoplus_{k} \mfh^{(k)}_f$, and every MO can be represented as linear combination of FMOs of all fragments,  the dimension of $\mfh_m$ is the sum of the dimensions of the fragments,  $M_{\rm mol}=\sum_{k}M^{(k)}_{\rm fr}$.

If the MOs $\ket{\psi}_i$ and the FMOs $\ket{\varphi^{(k)}_i}$ are eigenfunctions of Hermitian operators, they constitute ONBs in the MO and FMO Hilbert spaces, but FMOs from different fragments are not orthogonal. Therefore, Hilbert spaces $\mfh^{(k)}_f$ and $\mfh^{(p)}_f$ are linear independent but not necessarily orthogonal subspaces in $\mfh_m$;  the union of the FMO bases $\bigcup_{k} \{\ket{\varphi^{(k)}_i}\}_{i\in I_k}$ is, in general, a non-orthogonal basis of $\mfh_m$.

MOs are building blocks of many-electron state functions of a molecule; if the MOs are eigenfunctions of Fock operators they contain the information about all intra-molecular interactions. Many-electron state functions that are constructed with the FMOs of non-interacting fragments $\bigcup_{k} \{\ket{\varphi^{(k)}_i}\}_{i\in I_k}$ contain only the intra-fragment interactions but no inter-fragment interactions. If these FMOs are orthogonalized, they do no longer describe the intra-fragment interactions as exactly as the original FMOs, the deviation from the original FMOs depends on the chosen orthogonalization method. The localized fragment MOs (LFO) $\{\ket{\widetilde{\varphi}}^{(k)}_i\}_{i\in I_k}$ that are used in the OVB method are obtained with an orthogonal transformation from CASSCF MOs (Procrustes transformation\cite{Sax2012}); therefore they contain the information about all interactions in the molecule, both intra- and inter-fragment interactions. But, by construction, they are as similar to the FMOs as possible, and therefore they should represent the intra-fragment interactions as much as possible. The LFOs are elements of the Hilbert space  $\widetilde{\mfh}^{(k)}_f$, the direct sum of these Hilbert spaces is again the MO Hilbert space, $\mfh_m = \bigoplus_{k} \widetilde{\mfh}^{(k)}_f$. The union of the LFOs $\bigcup_{k}\{\ket{\widetilde{\varphi}}^{(k)}_i\}_{i\in I_k}$ is an orthogonal basis of $\mfh_m$.

A spin 1/2 quantum particle can only be in either of two orthogonal spin states $\ket{\sigma}$, traditionally designated as $\ket{\alpha}$ and $\ket{\beta}$; the spin state space $\mfh^{\rm spin}$ is two dimensional. To consider also the spin of an electron in orbital $\ket{\psi}\in \mfh$, one needs the \emph{spin orbital} $\ket{\widehat{\psi}} = \ket{\psi} \otimes \ket{\sigma}$ , which is the outer product of  orbital $\ket{\psi}$ and spin state $\ket{\sigma}$; spin orbitals are elements of the Hilbert space $\widehat{\mfh}= \mfh\otimes \mfh^{\rm spin}$, its dimension is twice the dimension of $\mfh$.
Two spin orbitals are orthogonal if the spin states are different. If the spin states are equal, spin orbitals are orthogonal only if the orbitals are orthogonal,
\begin{equation}
\overl{\widehat{\psi}_i}{\widehat{\psi}_j} = \overl{\psi_i}{\psi_j}\overl{\sigma_i}{\sigma_j}=\delta_{ij}^{\rm orb} \delta_{ij}^{\rm spin}.
\end{equation}
All MOs and FMOs considered in this paper are considered to be orthogonal.

\section{Many electron states}
Several CSOS can be used for an analysis of $n_e$-electron states. The most famous one is a CSOS of Slater determinants, which are normalized, totally antisymmetric products of $n_e$ spin orbitals, $\frac{1}{{\sqrt{n_e}}!}\widehat{\phi}_1\wedge\widehat{\phi}_2\wedge\dots \wedge\widehat{\phi}_{n_e}$, and therefore elements of the tensor space $\widehat{\mfA}^{\otimes {n_e}}$ defined over the spin orbital vector space $\widehat{\mfh}$ with dimension $2\di\mfh$. In the following, the space of $n_e$-electron states will be designated by $\widehat{\mfH}^{n_e}$, its  dimension is $\binom{2\di\mfh}{{n_e}}$. In general, only  $n_{\rm active} < \di\mfh$ active orbitals will be used in actual calculation, correspondingly smaller is the dimension of $\widehat{\mfH}^{n_e}$.  Slater determinants are eigenkets of the operator of the total spin projection, $S_z$, but they are, in general, not eigenkets of the total spin operator. However, proper linear combinations of Slater determinants  are eigenkets of it, they are called \emph{configuration state functions} (CSF) and designated by $\ket{\Phi({n_e},S,M_S)}$ with $S$ the total spin quantum number and $M_S$ the spin magnetic quantum number. CSFs $\ket{\Phi({n_e},S,M_S)}$ are elements of a subspace of $\widehat{\mfH}^{n_e}$ having the dimension
\begin{equation}\label{CASnm}
\displaystyle \frac{2S+1}{n_{\rm active}+1}\binom{n_{\rm active}+1}{\frac{{n_e}}{2}-S}\binom{n_{\rm active}+1}{\frac{{n_e}}{2}+S+1}.
\end{equation}
The CSFs $\ket{\Phi}$ constitute a CSOS in this subspace.

Alternatively,  CSFs can defined as the antisymmetrized product of an outer product of ${n_e}$ orbitals, some can occur twice, and ${n_e}$-spin eigenfunctions $\ket{\Theta({n_e},S,M_S)}$.
\begin{equation}\label{equ:CSF}
\ket{\Phi({n_e},S,M_S)} = \mcA \ket{\phi_1 \phi_2 \dots \phi_{n_e}} \ket{\Theta({n_e},S,M_S)}
\end{equation}
One has to consider spin degeneracy, that is, there may be several linearly independent spin eigenfunctions to a given $S$ and $M_S$.\cite{Pauncz1979}

Multiplying an arbitrary molecular state $\ket{\Psi}$ from left with the unit operator, made with a CSOS of CSFs $\{\ket{\Phi}_i\}_{i\in I}$, gives
\begin{equation}\label{equ:analysis}
\Bigl(\sum_{i\in I}  \proj{\Phi_i}{\Phi_i}\Bigr) \ket{\Psi} = \sum_{i\in I}  \ket{\Phi_i} \overl{\Phi_i}{\Psi}.
\end{equation}
This is an analysis of state $\ket{\Psi}$ with the CSOS of CSFs as intermediate states. If the quantum amplitudes are variationally optimized, one speaks of full CI (FCI), the quantum amplitudes are called CI coefficients; CASSCF is the FCI method where both the CI coefficients and the orbitals are simultaneously optimized.  CSFs as intermediate states are better suited for an interpretation of the transition probabilities $|\overl{\Phi_i}{\Psi}|^2$ than Slater determinants.

\section{Many electron states of composite systems}
To understand bonding in composite systems it is necessary to know how charge and spin distributions change during the stabilization process.
A spin distribution is characterized by the coupling of the fragment spin states to a resultant spin state of the molecule. The description of charge distributions needs CSFs with varying numbers of electrons in the fragments. The construction of molecular CSFs from fragment CSFs is described below.
\subsection{Non-orthogonal FMOs}
The advantage of the definition of CSFs in equation (\ref{equ:CSF}) is that it can be easily adapted to composite molecular systems.
If a molecule is composed of two fragments A and B with $n_e^A$ and $n_e^B$ active electrons, respectively, $n_e = n_e^A+n_e^B$, and if the fragment CSFs, $\ket{\Phi^A(n_e^A,S_A,M_A)}$ and $\ket{\Phi^B(n_e^B,S_B,M_B)}$, are known, one gets composite CSFs by linear combinations of  products of the fragment CSFs.
Coupling the spins of the two fragments gives  resultant spins with spin quantum numbers $S$ that can have   values $|S_A-S_B| \le S  \le S_A+S_B$ and $M=M_A+M_B$.
\begin{equation}
\ket{X_{AB}({n_e},S,M_S)} = \sum_{M_A,M_B} \overl{S_A S_B M_A M_B}{SM}\; \ket{\Phi_A(n_e^A,S_A,M_A)} \otimes \ket{\Phi_B(n_e^B,S_B,M_B)}
\end{equation}
$\overl{S_A S_B M_A M_B}{SM}$ are the Clebsch-Gordan coefficients.

To be totally antisymmetric, the inter-fragment antisymmetrizer $\mcA_R$ (\ref{equ:mcAR})  must be applied to $\ket{X_{AB}({n_e},S,M_S)}$
\begin{equation}\label{equ:FMOCSF}
\ket{\Phi_{AB}({n_e},S,M_S)} = \mcA_R \ket{X_{AB}({n_e},S,M_S)}
\end{equation}
and the CSF must be normalized. If the FMO Hilbertspaces $\mfh^{(k)}_f$, $k=1,2,\dots$ are not orthogonal the CSFs $\ket{\Phi_{AB}({n_e},S,M_S)}$ are also not orthogonal, they are a basis of $\mfH_m$ but no ONB and therefore no CSOS.

When a molecule dissociates, the fragments can be neutral radical species with one, two or even more singly occupied FMOs,  or cation/anion pairs derived from the neutral species.  The active FMOs are  partially occupied by the active electrons. It is much less demanding to describe  the dissociation or creation of single bonds than to describe  bonding between fragments with more than one active FMO, because here rearrangements of electrons and spins both in and between the fragments have to be considered. Understanding the properties of diradicals\cite{Salem1972,Borden1982} is essential for the study of dissociation and recombination reactions, and valence bond (VB) theory is the theoretical method that stresses the local character of bonding between radical fragments by using geminals to describe the coupling of unpaired electrons located in active orbitals. In conventional VB, these orbitals are AOs or atom centered hybride orbitals. If FMOs are used, and if the active electrons occupy atom centered FMOs, whereas all other FMOs are delocalized MOs, one has a mixture of VB and MO method, termed FMO-VB.

\subsection{Orthogonal FMOs}
Given the CSOS of CASSCF CSFs, if all MCSCF MOs are replaced by  LFOs $\{ \ket{\widetilde{\varphi}}^{(k)}_i  \}_{i\in I_f}$ (see equation (\ref{equ:CSF})) the set of CSFs constitutes again a CSOS.

\begin{equation}\label{equ:OCSF}
\ket{\widetilde{\Phi}({n_e},S,M_S)} = \mcA \ket{\widetilde{\varphi}^{(1)}_1\widetilde{\varphi}^{(1)}_2 \dots\widetilde{\varphi}^{(1)}_1\widetilde{\varphi}^{(1)}_2} \ket{\Theta({n_e},S,M_S)}.
\end{equation}
However, $\ket{\widetilde{\Phi}({n_e},S,M_S)}$ can also be interpreted as the result of coupling fragment CSFs made with LFOs as shown in (\ref{equ:FMOCSF}). Computationally, one creates the configurations for a state with spin quantum number $S$ by distributing ${n_e}$ electrons among $n_{\rm active}$ LFOs,  sets up the CI matrix of the Hamiltonian in the CSF basis and diagonalizes it. From the CI coefficients  one gets the transition probabilities, the diagonal elements of the CI matrix are the expectation values of the Hamiltonian calculated with the CSFs. This describes the OVB analysis of an MCSCF wave function.

\section{Covalent chemical bonding}
There are two important issues of chemical bonding, one is the origin of energetic stabilization, and the second is the origin of molecular structure.

\subsection{Energetic aspects}
The first issue is, in my opinion, solved. Ruedenberg and coworkers\cite{Ruedenberg2007,Ruedenberg2009,Schmidt2014} showed with high-level quantum chemical calculations that energetic stabilization of H$_2^+$ is a 1-electron effect, thus electron pairs are not essential for the stabilization of molecular species.  It was also shown by Ruedenberg that interference of AOs increases the probability to find an electron between the atoms, this shift away from the parent atom can be called delocalization. This is the origin of the famous accumulation of charge in the midbond region attributed to the shared electron pair in the Lewis theory. Delocalization enlarges the region of high position probability and by this the kinetic energy decreases. The pulling of the naked  proton on the electron of the hydrogen atom prohibits the electron to spread into the space as it would do in the unperturbed atom. By this pulling, the charge distribution, represented by the 1s AO, contracts but, at the same time, there is a polarization of the spherical charge distribution towards the pulling proton described by a p-type AO. Although the contraction of the 1s AO favors the Coulomb attraction and disfavors the atomic kinetic energy, both  effect enhance  delocalization and the decrease of the molecular kinetic energy and, in contrast to traditional explanation, an increase of the repulsive Coulomb interaction, but nevertheless a decrease of the total energy.  It is surprising that the important role of kinetic energy for chemical bonding was and is still ignored, even by physicists, like Cohen-Tannoudji\cite{Cohen1977}, Levy-Leblond\cite{Levy1990}, Gottfried\cite{Gottfried2003}, or Thirring\cite{Thirring1979}, who showed and stressed the importance of kinetic energy for the stability of the hydrogen atom by using the Heisenberg inequalities.  But as soon as they discuss chemical bonding, only the role of  the potential energy is mentioned, at best, one can find rough estimates of potential and kinetic energy using the virial theorem. However, a profound discussion of chemical bonding needs high-level quantum chemical calculations; based on such calculations  it was possible to show that small deformations of the electron distributions in molecules have large impact on the magnitude of kinetic and potential energy, and only with deformed quasi-AOs can the antagonistic interplay of kinetic and potential energy be correctly described. This demonstrates that subtle changes in the shape of AOs involved in the description of bonding may have large impact on the energetics and thus on the interpretation of the origin of stabilization.
The Hilbert space of the delocalized MOs $\mfh_m$ is  the direct sum of the  AO spaces for both atoms, $\mfh_m = \mfh_a^A \oplus \mfh_a^B$. The larger the dimension of the AO spaces is, the better can the  MOs as superpositions of these AOs describe contraction, polarization and delocalization. From these MOs it is possible to create new AO spaces of basis functions that represent contracted and hybridized orbitals, so called quasi-AOs, which constitute a modified AO space $\mfh^q_a$. Calculations with MOs that are linear combinations of quasi-AOs allow to  analyse the influence of contraction and polarization on kinetic, potential and total energy.

For the H$_2$ molecule and the two-atomic molecules B$_2$, C$_2$, N$_2$, O$_2$, and F$_2$, it was shown by Ruedenberg \etal{} that bonding can be explained in the same way as bonding in H$_2^+$.\cite{Schmidt2014,Schmidt2014a} For all molecules it was found that contraction of the AOs enhances delocalization and thus the decrease of the kinetic energy. Dilithium was not studied but would be interesting because it is the higher homologue of dihydrogen stabilized by the singlet coupled  2s valence electrons. Unexpectedly, the dilithium cation is about 25 percent more stable than dilithium, which also questions the validity of the claim that electron pairs cause covalent bonding.

\subsection{Molecular structure and the fermionic character of electrons}
In every atom with more than two electrons the Pauli exclusion principle (PEP) is valid; the tendency of identical fermions, that is those having the same spin projection, to avoid coming spatially close is of utmost importance, in atoms it is responsible for the shell structure of the electron cloud. If a shell is filled, every additional electron must occupy states that have their maximum of the radial density at larger distances from the nucleus, because whatever the spin projection of the additional electron is, there is already at least one of the same kind that prohibits the new electron to come too close. But also within a shell the PEP keeps electrons with identical spin as far apart as possible. This behaviour is especially important for the spatial arrangement of valence electrons. Element atoms in the second period of the periodic table show very similar maxima $r_{\rm max}$ of the radial probability distribution for the 2s and the 2p subshells, therefore one can safely say that 2s and 2p valence electrons are in the same electron shell at the distance $r_{\rm max}$, or, simply spoken, on the same sphere with radius $r_{\rm max}$; two and three electrons will prefer to stay on a great circle of the sphere. Lennard-Jones\cite{Lennard1954} showed that the probability to find two identical electrons on a great circle will be maximal for diametrically opposed positions, and three identical electrons will prefer a trigonal arrangement;  four identical electrons will prefer a tetrahedral arrangement on the sphere. This can be shown, with moderate effort, with wave functions, that comprise, for up to three spins  only complex exponentials, which depend on one angle, only when four spins on a sphere are considered, spherical harmonics are needed that are functions of two angles. However, in all these cases, only spherical harmonics are necessary to get the angular behavior of hybrid orbitals, the radial wave function is irrelevant. In a filled valence shell, e.g., in nobel gas atoms,  four identical $\alpha$ electrons will be tetrahedrally arranged as will be the four identical $\beta$ electrons. Coulomb repulsion maximizes the distances between $\alpha$ and $\beta$ electrons so that the eight electrons occupy the corners of a cube. Averaging over all orientations of the cube yields the spherical electron distribution. This geometric arrangement of the electrons in a filled valence shell is reminiscent of the cubical atom proposed by Lewis\cite{Lewis1916} in 1916, which was regarded as outdated. At that time, spin was not yet known, but after the role of spin was recognized and the spatial distribution of identical electrons was described by Lennard-Jones, Linnett (1961) postulated the structure of two interpenetrating tetrahedra.\cite{Linnett1961} With this model, it is possible to give a concise explanation of the angular structure of several hydrides. To understand the structure of the water molecule, one starts with the O$^{2-}$, which is isoelectronic with neon. If a proton approaches the oxygen dianion it attracts electrons, but because of the PEP it will never be two $\alpha$ or two $\beta$ electrons but always an $\alpha$-$\beta$ pair. By this, one vertex of the $\alpha$ and one vertex of the $\beta$ tetrahedron are brought into coincidence, forming a Lewis electron pair between hydrogen and oxygen, whereas six vertices occupy the vertices of a regular hexagon with alternating $\alpha$ and $\beta$ electrons. If a second proton approaches the OH$^-$ atom group, two further vertices with different electrons are brought into coincidence, forming a second Lewis pair between hydrogen and oxygen, but by this, also the remaining two vertices are forced into coincidence and the four electrons occupy as lone pairs these vertices.\cite{Popelier2001} The HOH bond angle is the result of the repulsion between the four electron pairs, which have very different spatial extensions: lone pair electrons occupy a much smaller, compact domain, whereas the domain of the bonding electron pairs is stretched between the atoms with each electron of the bonding pair close to ``its'' atom. This was shown by Savin and coworkers\cite{Scemama2007} in their study of the maximum probability domains in several molecules. The two $\alpha$ electrons and the two $\beta$ electrons that occupy the lone pair domains have a much higher ``Pauli repulsion'' than the electrons in the bonding domains. Of course there is repulsion due to the charge, but because electrons are fermions and for identical electrons the PEP holds, it is the PEP that dominates the spatial arrangement of the electrons. Lennard-Jones said already 1954 in all clarity about the Pauli principle ''\dots a property which holds for all electronic systems, whether they are atoms, molecules or solids: \emph{Electrons of like spins tend to avoid each other}. This effect is most powerful, much more powerful than that of electrostatic forces. It does more to determine the shapes and properties of molecules than any other single factor. It is the exclusion principle which plays the dominant role in chemistry.''\cite{Lennard1954} Daudel \emph{et al.} simply stated: ``This shows that the `Pauli repulsion' between electrons possessing the same spin is very significant. In fact, this \emph{repulsion is the main origin of bond angles}.\cite{Daudel1984} The relevance of the PEP for the VSEPR model (valence shell electron pair repulsion) was recently appreciated by Gillespie.\cite{Gillespie2008}
In their study, Savin \etal{} found the interpenetrating tetrahedra of the cubical atom as a second structure of very high probability.

With the help of Linnett's spin tetrahedra one can  see, that formation of lone pairs is forced, it is the result of external perturbations. In the FH molecule, as in OH$^-$, the six valence electrons not involved in bonding, prefer a hexagonal arrangement where every electron has maximum distance to its neighbors, but not three lone pairs. In H$_2$O, NH$_3$, or CH$_4$, it is the lower symmetry of the perturbation potential of the hydrogen atoms surrounding the heavy central atoms that forces the valence electrons to occupy pairwise distinct spatial domains.  This external perturbation destroys  the spherical symmetry of the central atom so that its atomic states, which are as eigenstates of the angular momentum operator orthogonal to each other, can superpose and hybridize. This is nothing but the polarization described by Ruedenberg \etal. Hybrid orbitals are the result of the external perturbation but no intrinsic property of an atom, therefore they are not the cause of the molecular structure but its result. Which and how many AOs hybridize depends on the symmetry and the strength of the external perturbation potential. In chemistry, explanation of molecular geometry starts however with selecting an atom, frequently termed the ``heavy atom'', to which other atoms, for example hydrogen atoms, are bonded. Then it is claimed that a certain type of hybrid orbitals of the heavy atom can best explain the molecular geometry, as if offering hybrid orbitals were a decision of the heavy atom. This reversion of cause and effect is physically wrong, but acceptable for model-based stereochemical reasoning. Energetic stabilization due to the formation of ``covalent X-H bonds'' can be explained in the spirit of Ruedenberg with quasi-AOs (hybrid orbitals) of heavy atom and the atom bonded to it.

Bonding in molecules with heavy atoms from periods higher than period 2 needs some more consideration. Hybridization is claimed to occur between AOs of comparable energy\cite{Housecroft2005}, AOs of element atoms in periods 3 and higher should accordingly much better hybridize than AOs of element atoms in period 2, because the orbital energy differences of 2s and 2p AOs for C, N, and O are 18 eV, 21 eV, and 24 eV, respectively, whereas for the 3s and 3p AOs of Si, P, and S they are only 13 eV, 15 eV, and 16 eV, respectively. The differences for atoms of period 4 are similar to those of period 3. So it is not the energetic similarity that enables hybridization; it is the similarity of the positions r$_x$ of the maxima of the radial densities of the s and p subshells.\cite{WAGoddard1978} Since the thickness of the shells is not known, I replace, faute de mieux, the electron density in the shell by the surface density $\sigma$ defined as the reciprocal value of the surface area of a sphere with radius r$_x$. The factor $4\pi$ is omitted.  The results for elements carbon, nitrogen, oxygen, silicon, phosphorus, and sulfur are given in Table \ref{tbl:atoms}. Summarizing the results one can say: In period 2, the p maxima  are 5 and 8 percent smaller than the s maxima, and therefore are the p surface densities between 9 and 22 percent larger than the s surface densities. In period 3 are the p maxima about 20 percent larger than the s maxima, and the p surface densities are about 30 percent smaller than the s surface densities. And, all surface densities for period 3 are only about 30 percent of the corresponding densities for period 2. All densities are normalized to one electron. Considering the similarity of the position of the maxima in period 2 it seems to be justified to speak of one shell containing 4, 5 or 6 valence electrons, respectively, whereas s and p subshells in period 3 are spatially much more separated so that the s subshell is always occupied by two electrons and the p subshell by 2, 3 or 4 electrons, respectively. Only perturbations of high symmetry and strength force hybridization of s and p AOs, as the silicon atom demonstrates. A perturbation of low $C_{2v}$ symmetry by two hydrogen atoms is obviously not able to hybridize 3s and 3p AOs in silylene, the 3s AO remains doubly occupied, only the two electrons in p AOs are involved in bonding. The spatial orthogonality of the p AOs explains the HSiH bond angle of 92 deg\cite{NIST_SiH2} in the ground state of silylene. The same reasoning explains the bond angle in germylene, stannylene and plumbene.\cite{Aldridge2001} On the other hand, a tetrahedral perturbation caused by four hydrogen atoms forces hybridization and SiH$_4$, GeH$_4$, SnH$_4$ and PbH$_4$ are tetrahedral. Obviously, already the perturbation by three hydrogen atoms forces hybridization and causes a pyramidal structure of the XH3 species, but not of CH$_3$.

\begin{table}
\caption{\label{tbl:atoms}Orbital energies $\varepsilon$ in eV, maxima of radial densities, r$_x$, in \bohr, r$_x^2$ in \bohr$^2$, surface density $\sigma={\rm r}_x^{-2}$ in \bohr$^{-2}$}
\begin{tabular}{l|r|r|r|r|r|r|r|r}
 & $\varepsilon$(2s) & $\varepsilon$(2p) & r$_x$(2s) & r$_x$(2p) &  r$^2_x$(2s) &  r$^2_x$(2p) & $\sigma$(2s) &   $\sigma$(2p)\\
Carbon($^3$P) & -19.15  & -1.47  & 1.233 & 1.183 & 1.520 & 1.399 & 0.658 & 0.715\\
Nitrogen($^4$S) &-25.64 & -4.73 & 1.033 &  0.954 &  1.067 & 0.910 & 0.937 & 1.099  \\
Oxygen($^3$P) &-33.71 & -9.78& 0.886 &  0.802 &  0.785 & 0.643 & 1.274& 1.555\\[2ex]
 & 3s & 3p & r$_x$(3s) & r$_x$(3p) &  r$^2_x$(3s) &  r$^2_x$(3p) & $\sigma$(3s) &   $\sigma$(3p)\\
Silicon($^3$P) & -14.64  & -1.73  & 1.793 & 2.186 & 3.215 & 4.779 & 0.311 & 0.209\\
Phosphorus($^4$S) &-18.89 & -4.19 & 1.583 &  1.869 & 2.506 & 3.493 & 0.399 & 0.286 \\
Sulfur($^3$P) &-23.86 &-7.61& 1.420 &  1.648 &  2.016 & 2.716 & 0.496 & 0.368
\end{tabular}
\end{table}

It must be stressed, that the explanation of molecular geometry using the Linnett tetrahedra has the quality of a Gedankenexperiment, because starting from a highly charged anion with nobel gas electron structure to which protons are added is rather unphysical, just consider the C$^{4-}$ ion. But it points to the fact that if many identical electrons are in the system the fermionic character of electrons is crucial for the geometry and energy of a molecule, charge is of minor importance when spin dominates. And spin dominates whenever many fermions are confined to small spatial domains, as in the valence shell of period 2 atoms, in contrast to the same number of electrons in the much more extended valence shell of atoms from higher periods. Non-identical electrons can come close together, which increases the Coulomb repulsion; if, by whatever mechanism, the number of identical electrons increases, that is, when spin-flips change low-spin states  into high-spin states, more identical electrons will avoid each other and the Coulomb repulsion is automatically reduced. Spin reorganization is an important process in chemical reactions, it changes the local spin arrangements and is relevant for the changes in the geometries of the reactants. And this, in turn, helps to optimize the energetic stabilization as described by Ruedenberg.

MCSCF methods with correctly chosen active MOs and the correct number of active electrons, known as CASSCF or FORS methods,  allow to calculate all local spin and charge rearrangements that are essential for a certain reaction, but that are hidden because of the use of delocalized MOs. Methods like conventional VB, which uses localized AOs as building blocks, can make local processes visible, but the non-orthogonality of the AOs can also hide important aspects of the electronic structure.\cite{Sax2015,Angeli2013}
But it must be clearly said: A method like OVB is not a tool for the calculation of molecular geometries, energies, or other properties, it is method to analyze CASSCF wave functions.

\section{Basics of OVB}
The technical details of OVB are described in two papers\cite{Sax2012,Sax2015}, the method was used to study symmetry aspects of chemical reactions\cite{Sax2017}, a paper on the addition  of carbene-like fragments to molecules with double bonds is in preparation\cite{Strasser2022}, as is a study on the C$_2$ molecule.\cite{Sax2022}

The bonded system consists of a  molecule with a double bond in the singlet ground state, it has four active electrons and four active MOs, the bonding and the antibonding $\sigma$ MOs and the bonding and antibonding $\pi$ MOs. The dissociated system consists of two fragments A and B, which are prototype diradicals with two active FMOs, the $s$ and the $p$ lonepair FMOs and two active electrons. To describe a smooth transition from a doubly bonded molecule to the fragments, or the recombination of the fragments, CAS(4,4) state functions are necessary where the four active electrons are distributed among the four active MOs. The number of CSFs describing the system as calculated with the expression in (\ref{CASnm}), that is the dimension of the FCI problem, is 20, this is the maximum number found when the system has $C_1$ or $C_s$ symmetry, if the symmetry is higher than $C_s$, the number of CSFs is smaller. In detail, in $D_{2h}$ (planar ethene or disilene) it is eight, in $C_{2h}$ (trans-bent disilene) it is 12, in $C_{2v}$ (planar silaethene) it is 12.

\begin{figure}
\caption{ \label{fig:CAS44}The scheme of the CSFs. Insert: The scheme of active FMOs. }
\includegraphics[width=\textwidth]{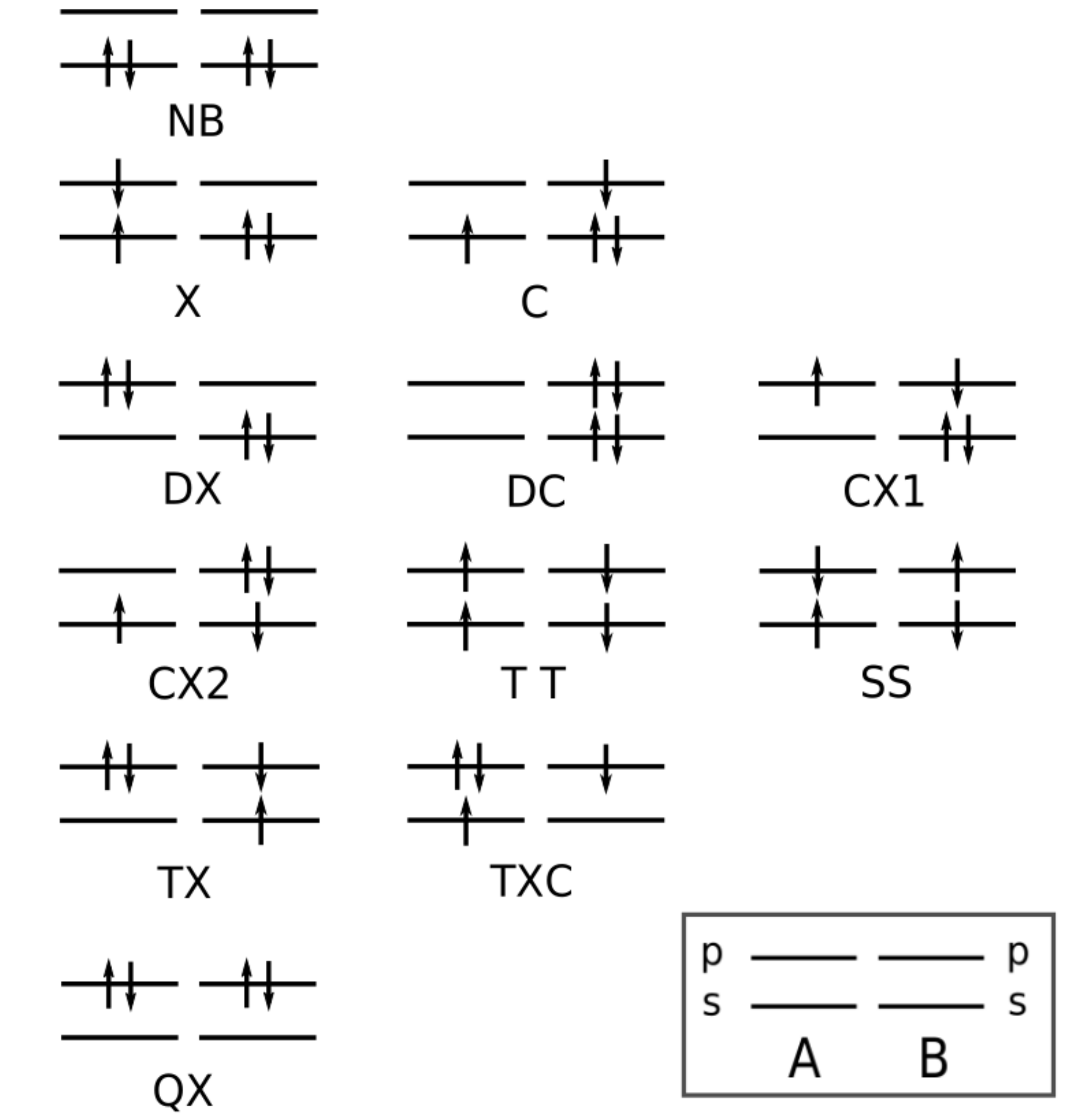}
\end{figure}

The occupation of the four active FMOs is schematically represented in Figure \ref{fig:CAS44}, each scheme corresponds to a composite CSF, see equation (\ref{equ:FMOCSF}).
for example, TT represents the singlet coupling of two fragment or local triplets
\begin{equation}
\ket{\Phi_{AB}(4,0,0)} = \mcA_R \sum^1_{\mathclap{\substack{M_A, M_B =-1\\
          M_A+M_B=0}}} \overl{1 1 M_A M_B}{00}\; \ket{\Phi_A(2,1,M_A)} \;\ket{\Phi_B(2,1,M_B)}
\end{equation}
Note, all CSFs but NB, TT, SS and QX have symmetric counterparts, which are not shown in Figure \ref{fig:CAS44}.
Furthermore,  the same scheme can be used to represent the CSFs made with delocalized MOs, just replace, e.g., the FMOs by the $\sigma/\sigma^*$ and the $\pi/\pi^*$ pairs.

For the understanding of the coupling of carbenes fragments, one must know that a neutral carbene can be in three singlet states with orbital occupations $\rm s^2, p^2$ and $s^1p^1$,  and in one triplet state, having also $s^1p^1$ orbital occupation. Carbene cations and  anions can be in two doublet states, and doubly charged carbene cations and anions can be in a single singlet state.

Inspection of the occupation schemes for the molecular system shows that the active electrons can be divided among the fragments in three ways: in the first case, each fragment houses two electrons, the fragments are neutral; in the second case, one fragment is a cation, the other an anion; in the third case,  one fragment is a doubly charged cation, the other is a doubly charged anion. To show the occupation of the fragments, the notation $\ket{\Phi^{n_A,n_B}_{AB}}$ is used.

An important feature of all kinds of orthogonal VB is, that CSFs made with orthogonal orbitals retain their chemical characteristics, that is, they describe either a neutral or a charged electron distribution, they do not describe charge shifts or delocalization, as it is possible with non-orthogonal orbitals. A charge shift must be described by linear combination of neutral and ionic CSFs, the square of the CI coefficients will be measures for the shift. In conventional VB, CSFs made with non-orthogonal orbitals are not orthogonal and there is no unique measure of the delocalization described by them. As was frequently shown for the hydrogen molecule, the Heitler-London or covalent CSF and the ionic CSF are orthogonal only for large distances between the hydrogen atoms, the overlap between them increases with decreasing H-H distances and they become identical if the distance is zero. So neither CSF has a definite characteristics, neither is purely neutral or ionic, it depends on the distance between the interacting atoms.\cite{Malrieu2006,Sax2015} The explanation is, that the charge shift is represented by the overlap of the AOs; the mathematical form of the CSFs is always the same and does not reflect the change in the charge distribution, that is ``the physics'', against common belief.\cite{Schmalz2002}  I dubbed CSFs  representing electron distributions that change with the reaction coordinate ``chameleon'' CSFs. The success  of the Heitler-London CSF is based exactly on this behavior, covalent bonding is based on delocalization and overlapping AOs describe this intrinsically, although it is hidden by the mathematical form of the CSF. Thorough discussion of the difference between conventional VB and OVB can be found in several papers of Malrieu and coworkers,\cite{Malrieu2006,Malrieu2008,Angeli2013,Sax2015} but also in older books on Quantum chemistry, e.g., in the first edition of Elementary Quantum Chemistry by Pilar\cite{Pilar1968} where he writes about the results obtained by McWeeny in the 1950ies\cite{McWeeny1954}: ``In conclusion, the use of OAO's [orthogonal AOs] in the VB method leads to a clearer electrostatic picture of chemical bonding but destroys the chemists's simple concepts of covalent and ionic character.'' And this is not appreciated in some communities.\cite{Gallup2002}

The following CSFs are neutral:  $\ket{\Phi^{2,2}_{AB}}$: NB, X, DX, TT,SS, TX, QX.
NB means No Bond, this CSF represents, e.g., two silylenes in their singlet ground states with both electrons in the s-type lone pair FMO. X represents a single excitation, DX, TX and QX represent double, triple, and quadruple  excitations, respectively. Excitations are always meant with respect to NB, which is regarded to represent a state of lowest energy. Note, that this is not valid for methylene, which has a triplet as ground state. TT was explained above, and SS represents the singlet coupling of two local excited singlet states. TX is a double excitation in one fragment and a single excitation in the other one. QX means double excitations in both fragments.

Mono-ionic CSFs $\ket{\Phi^{1,3}_{AB}}$ are: C, CX1, CX2, TXC.
C is a single transfer of an electron from the doubly occupied s FMO of one fragment to the empty p FMO of the other fragment (single charge transfer). CX1 means a single charge transfer from fragment A to fragment B, and a single excitation of the remaining electron in fragment A.
CX2 is a single charge transfer from A to B and a single excitation in B. TXC means single excitations in one fragment, a double excitation in the other fragment and an additional single charge transfer from s FMO to s FMO.

There is only one bi-ionic CSF $\ket{\Phi^{0,4}_{AB}}$, namely DC describing a double charge transfer from one fragment to the other.

For the CSFs an occupation representation is used in the Figures, e.g., NB$\;\approx\; \ket{2020}$, TT$\;\approx\; \ket{\alpha\alpha\beta\beta}$, CX1$\;\approx\; \ket{0\alpha 2\beta}$. As mentioned, most CSFs have symmetric counterparts, which must be linearly combined to have the correct spatial symmetry. In the legend of the figures, only one component CSF is mentioned.

An OVB analysis of a chemical reaction, as presented in this paper, starts with the calculation of the potential energy curve for the dissociation or the recombination in a  molecular state defined by the number of active electrons, the number of active MOs, and the spin quantum numbers $S$ and $M_S$. For a set of inter-fragment distances $R$, called the reaction coordinate, all  geometry parameters that are allowed to change are optimized. At each inter-fragment distance, the optimized delocalized CASSCF MOs are the basis of the MO Hilbert space $\mfh_m$. Then the coordinates of each fragment are taken from optimized geometry of the molecule and FMOs are calculated with a proper method, e.g., RHF of UHF. The union of the FMOs is the non-orthogonal basis $\bigcup_{k} \{\ket{\varphi^{(k)}_i}\}_{i\in I_k}$. With the Procrustes transformation the CASSAF MOs $\{\ket{\psi}_i\}_{i\in I_m}$ are transformed into orthogonal LFOs  $\{ \ket{\widetilde{\varphi}}^{(k)}_i  \}_{i\in I_f}$ and with the LFOs one gets the  CSFs $\ket{\widetilde{\Phi}({n_e},S,M_S)}$ which can be interpreted as coupled orthogonal fragment CSFs (\ref{equ:FMOCSF}). This CSOS allows the analysis of the molecular state, for this the Hamilton matrix in the CSF basis is calculated and diagonalized.

The computational details are as follows: The 6-31G basis with two d sets at the heavy atoms was used. The X-Y distance, X,Y = C, Si, was changed
in 0.05 \AA{}{} increments.  Converged MOs at  geometry R$_i$ are used as starting MOs for the calculation at geometry R$_{i+1}$. Together with the small increments in the X-Y distance, this allows to distinguish different diabatic states. All calculations were done with a local copy of GAMESS.\cite{Gamess}

\section{Investigated systems}
In this paper, the dissociation reactions of ethene, silaethene, and disilene into carbene-like fragments, $\rm XYH_4 \rightarrow  XH_2+ YH_2$, Y = C, Si are discussed. The three systems  were chosen, because the dissociation and the recombination reactions should occur very differently, according to chemical reasoning. For the recombination reactions, the electronic ground state of the reactants is decisive. Carbene has a triplet ground state with two lone pair FMOs each occupied by a single electron. The two unpaired electrons  are able to couple with other unpaired electrons to yield two bonding electron pairs. This will be studied in the recombination of two carbenes.  Silylene, on the other hand has a singlet ground state, no unpaired electrons and should therefore have no tendency for making new bonds between the silicon atoms. To do this, it is necessary to uncouple the electron pair in the lone pair orbital. This will be studied in the recombination reaction of two silylenes. And the recombination reaction of one carbene and one silylene will show which local spin and charge reorganization are necessary to form silaethene in its singlet ground state. The dissociation of stable molecules should create high spin fragments, which should change to low spin, if this is the ground state of the fragments.

The dissociation reaction will be termed In-Out reaction, the recombination reaction Out-In reaction.
The combination of geometry data, transition probabilities for the CSFs, also called the weights of the CSFs, and expectation values of the energy calculated with the CSFs will help to understand, what happens during these reactions. In the figures, data are shown only for important CSFs, meaning their transition probabilities are larger than a given threshold somewhere along the whole reaction coordinate $R$.

\subsection{Reaction $\rm C_2H_4 \rightarrow  2 CH_2$ in $D_{2h}$ }
Dissociation and recombination take place in planar geometry, the singlet ground state has $A_g$ symmetry; the eight CSFs or linear combinations of CSFs having $A_g$ symmetry are the neutral CSFs  NB, TT, SS, QX, DX+DX, and the ionic linear  combinations DC+DC, CX1+CX1, CX2+CX2. From these CSFs only the following four are important (threshold is 0.1) for the description of the ground state of the molecular system: TT, CX1+CX1, CX2+CX2, and DC+DC.

\begin{figure}[ht]
\caption{Potential energy curve for dissociation and recombination reaction}
\includegraphics[width=0.45\textwidth]{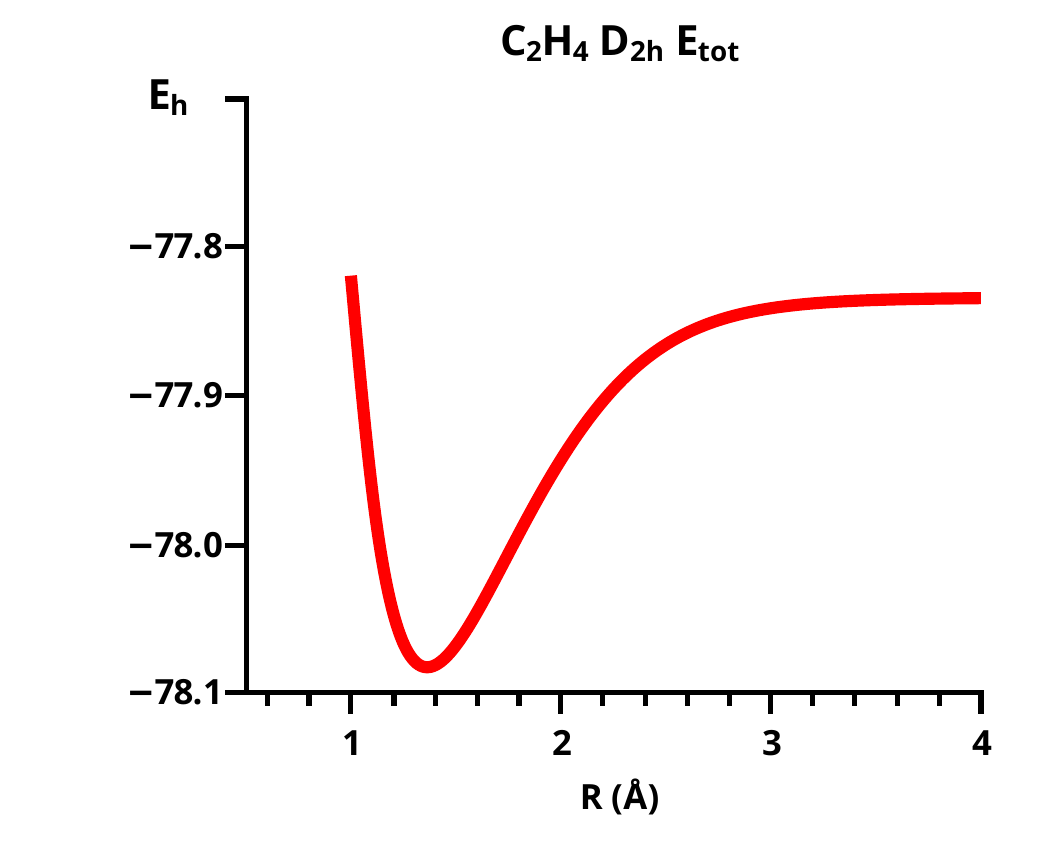}
\label{fig:E_ethylene}
\end{figure}
The total energies for both reactions are identical, (Figure \ref{fig:E_ethylene}) dissociation and recombination proceed completely equally.  The change of the geometry parameters are as expected.
\begin{figure}[ht]
\caption{CH length and HCH angle as function of the C-C distance. The geometry parameters of triplet carbene are inserted.}
\includegraphics[width=0.45\textwidth]{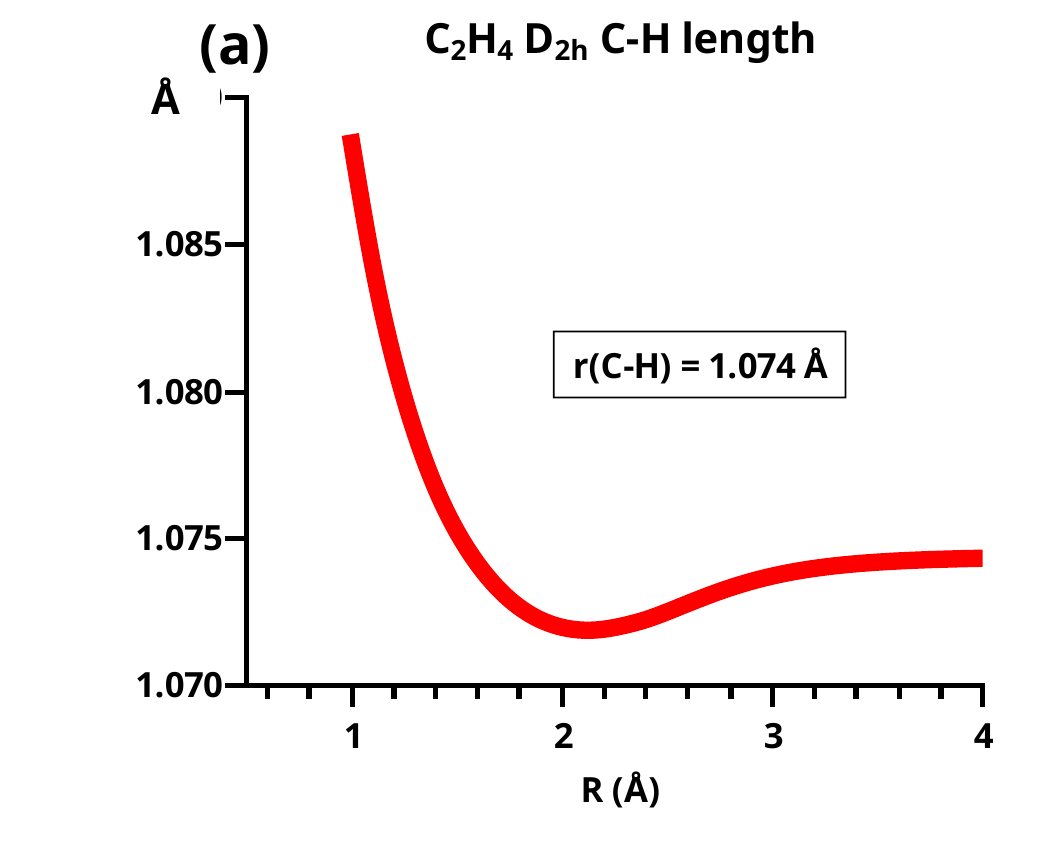}
\includegraphics[width=0.45\textwidth]{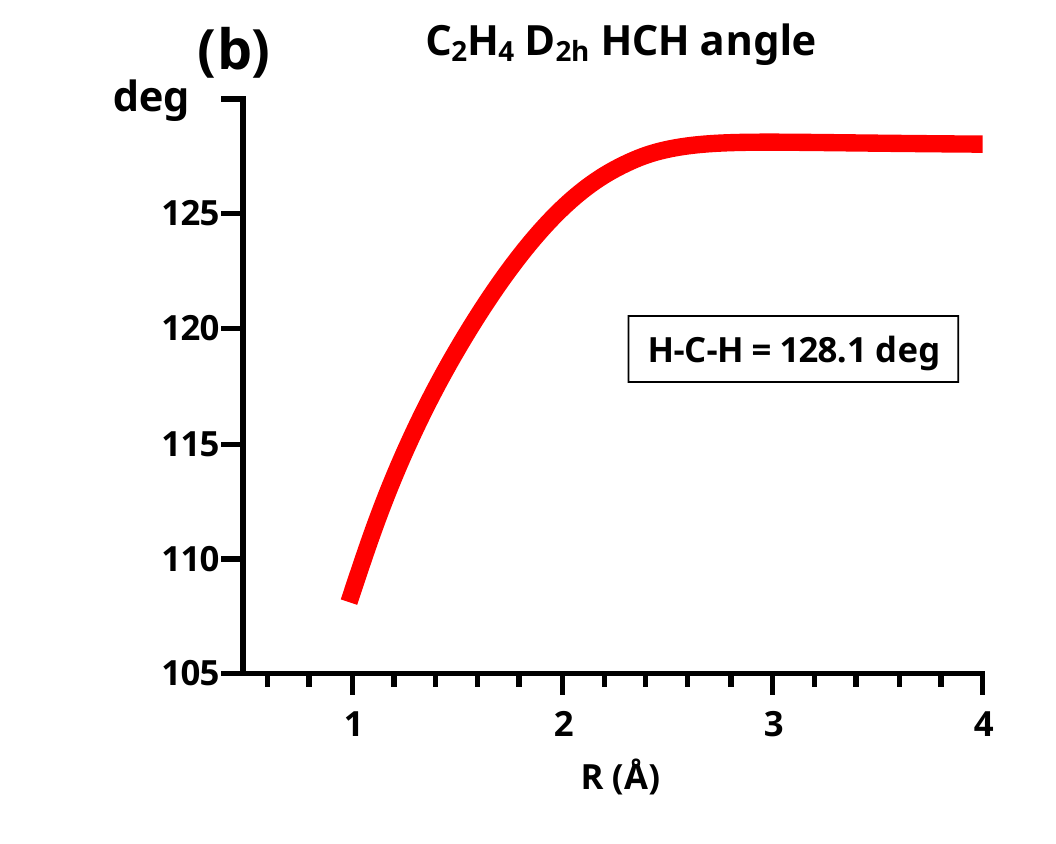}
\label{fig:geom_ethylene}
\end{figure}
Because the reactions take place in $D_{2h}$ symmetry, only the HCH angle and the CH  length are relevant geometry parameters (Figure \ref{fig:geom_ethylene}).
At large C-C distances, both the CH length and the HCH bond angle have the typical values of triplet carbene. The bond angle is constant for C-C distances larger then 2.5\,\AA{}, then it decreases monotonously. The CH distance starts to decrease already at a C-C distance of about 3.0\,\AA{}, it reaches a minimum at about 2.0\,\AA{} and then increases again. The minimum of the total energy lies at a C-C distance of about 1.4\,\AA{}.
\begin{figure}[ht]
\caption{a) Energies and  b) transition probabilities for the CSFs (threshold of 0.1).}
\includegraphics[width=0.45\textwidth]{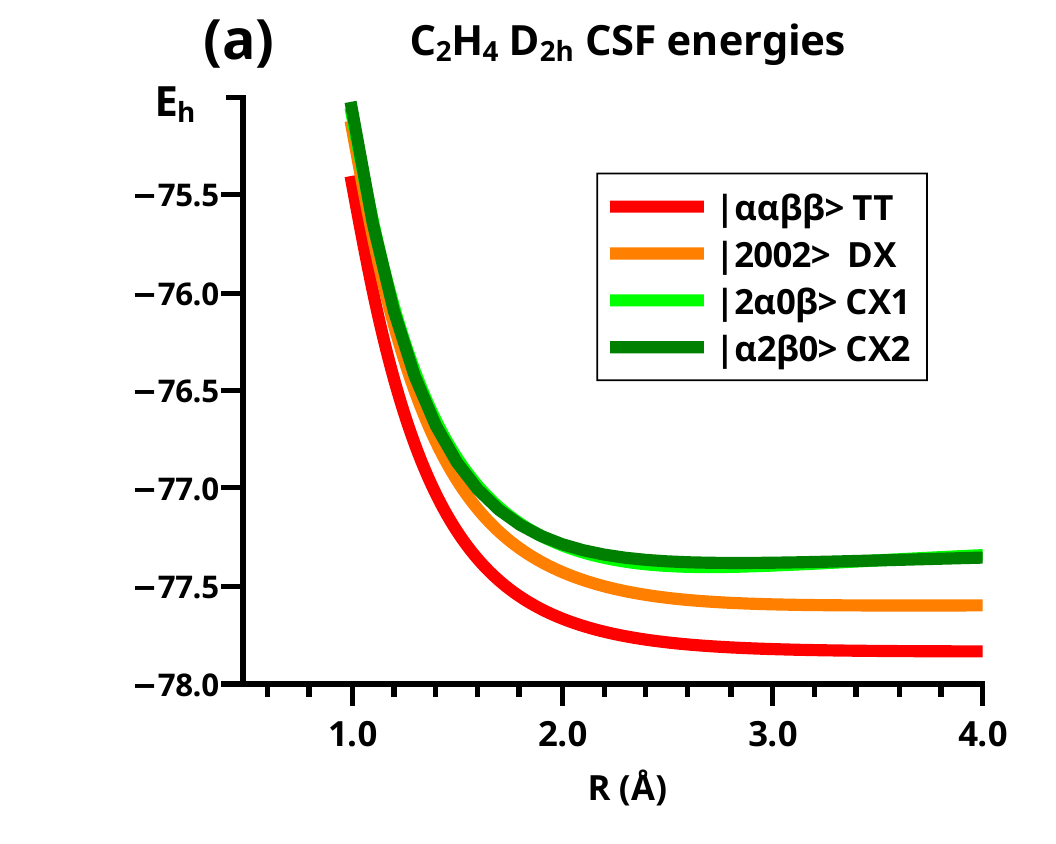}
\includegraphics[width=0.45\textwidth]{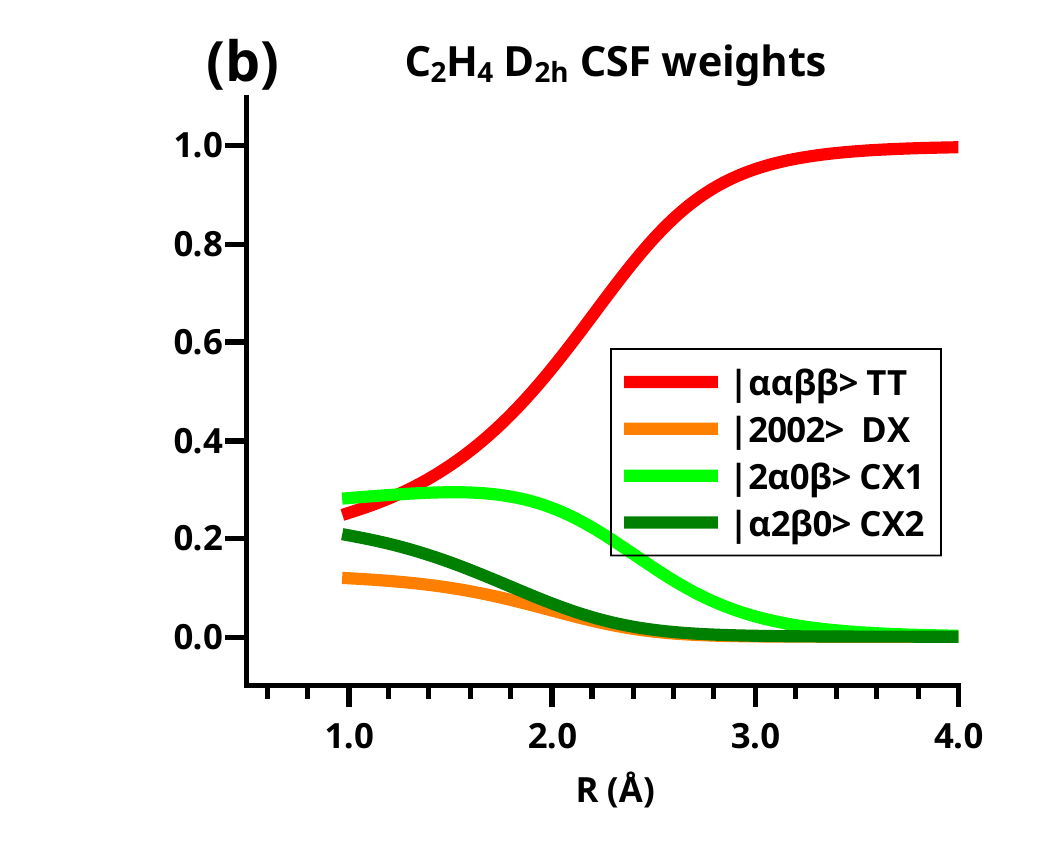}
\label{fig:OVB_ethylene}
\end{figure}
The transition probabilities (Figure \ref{fig:OVB_ethylene} b) demonstrate chemical bonding as it is taught in chemistry: If two reactants have unpaired electrons in a high spin state  new bonds can be formed, in this case the double bond in ethene. The relative decrease of the weights of the neutral TT and the increase of the weights of the ionic CSFs CX1, CX2, and DC reflects the increase of delocalization during covalent bonding. The monotonically increasing energies for decreasing C-C distance is typical for OVB CSFs (Figure \ref{fig:OVB_ethylene} a).

\subsection{Reaction $\rm Si_2H_4 \rightarrow  2 SiH_2$ in $D_{2h}$}
In $D_{2h}$,  the singlet ground state has $A_g$ symmetry, the eight CSFs of this symmetry are the same as mentioned  for the ethene reactions. The four important linear combinations of CSFs are the same as for ethene plus the neutral  NB.  In-Out and Out-In reactions of the $A_g$ ground state  proceed differently, as the potential energy curves show. The minimum of the total energy is at a Si-Si distance of 2.20\,\AA.

\begin{figure}[ht]
\caption{Energy curves for dissociation and recombination reaction.}
\includegraphics[width=0.45\textwidth]{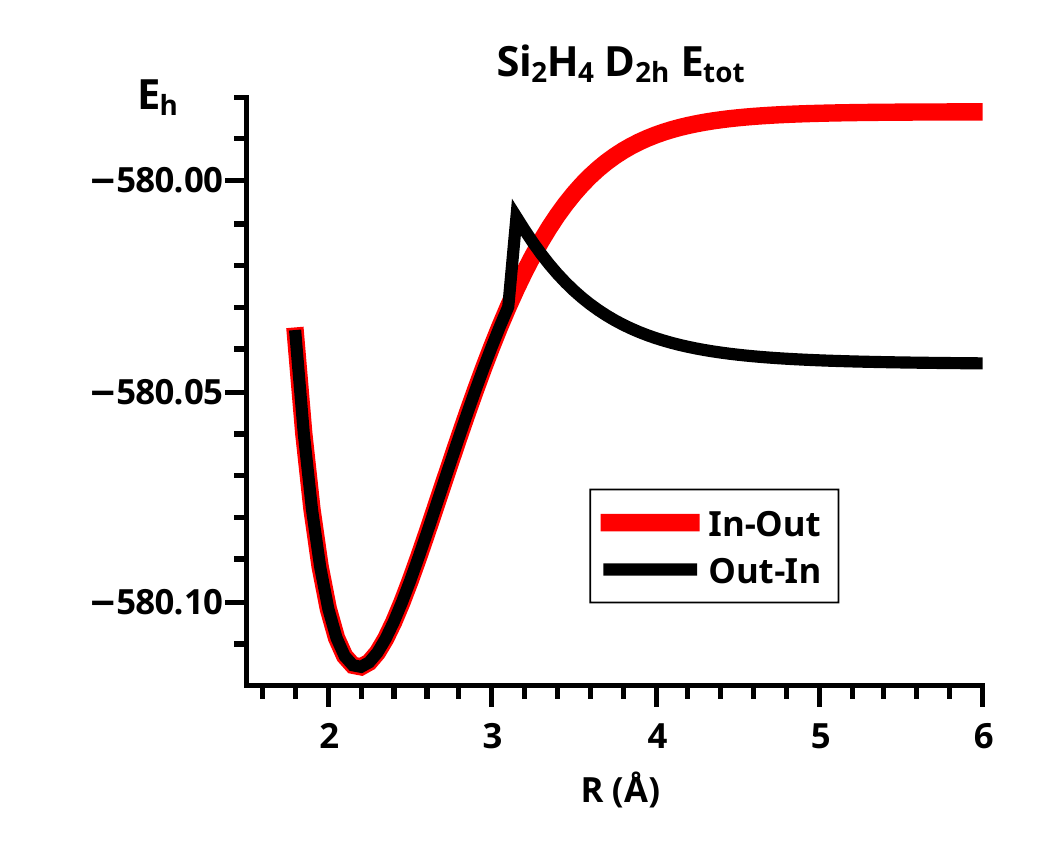}
\label{fig:E_disilen_plan}
\end{figure}

In-Out  occurs similar to ethene dissociation in a completely smooth way, both silylene fragments have at large Si-Si distances the typical triplet geometries, which change only little when going to small Si-Si distances.
The recombination of two silylenes in their singlet ground states is not possible. For large Si-Si distances, the energy of the molecular system lies 157\,kJ/mol lower than the energy of two triplet silylenes, and this is exactly twice the singlet triplet splitting in silylene  calculated with the CAS(4,) wave function. With decreasing Si-Si distance, the total energy increases, the potential energy curve crosses at about 3.3\,\AA{} the curve describing the dissociation, increases further and at about 3.1\,\AA{} the system jumps to the lower lying In-Out state (Figure \ref{fig:E_disilen_plan}).

\begin{figure}[ht]
\caption{SiH length and HSiH angle as function of the Si-Si distance. The geometry parameters of singlet and triplet silylene are inserted.}
\includegraphics[width=0.45\textwidth]{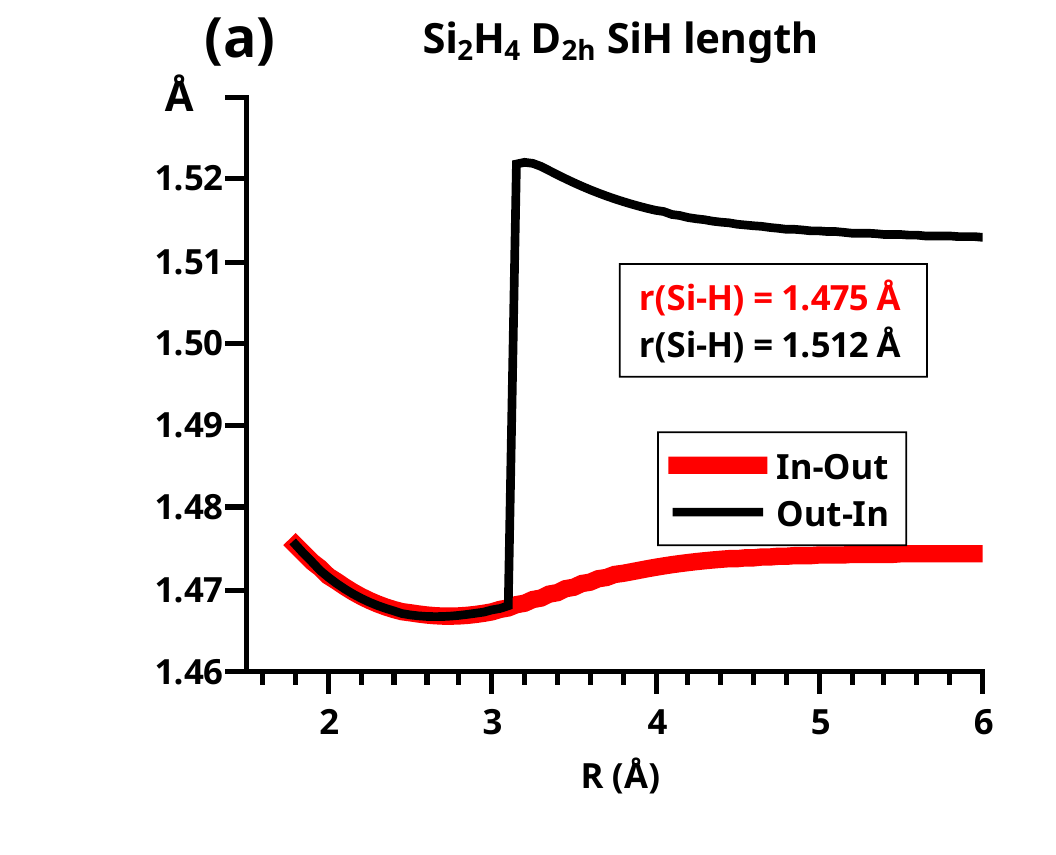}
\includegraphics[width=0.45\textwidth]{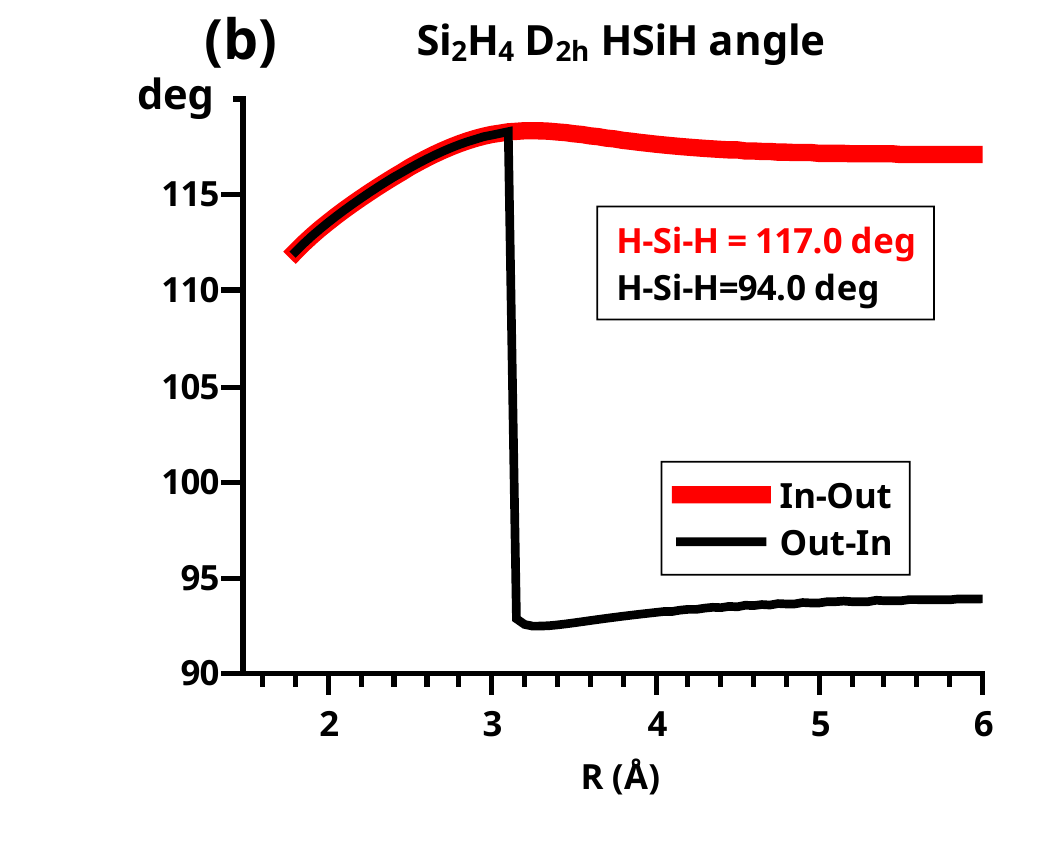}
\label{fig:geom_disilen_plan}
\end{figure}

The  geometry parameters of singlet silylene (Figure \ref{fig:geom_disilen_plan}) change moderately when the total energy increases. At a Si-Si distance of about 3.1\,\AA{} the geometry changes abruptly to the triplet geometry,  the fragments change from low spin to high spin geometry.

\begin{figure}[ht]
\caption{Dissociation reaction. a) Energies and  b) transition probabilities for the CSFs (threshold of 0.1).}
\includegraphics[width=0.45\textwidth]{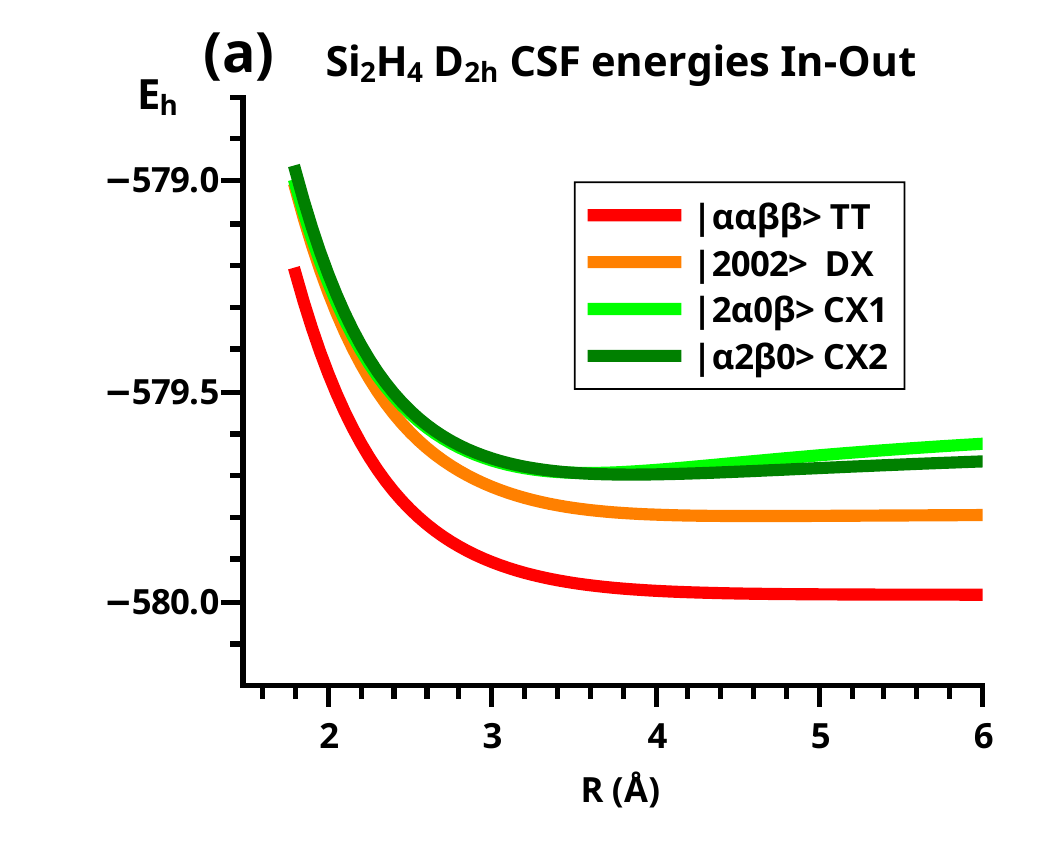}
\includegraphics[width=0.45\textwidth]{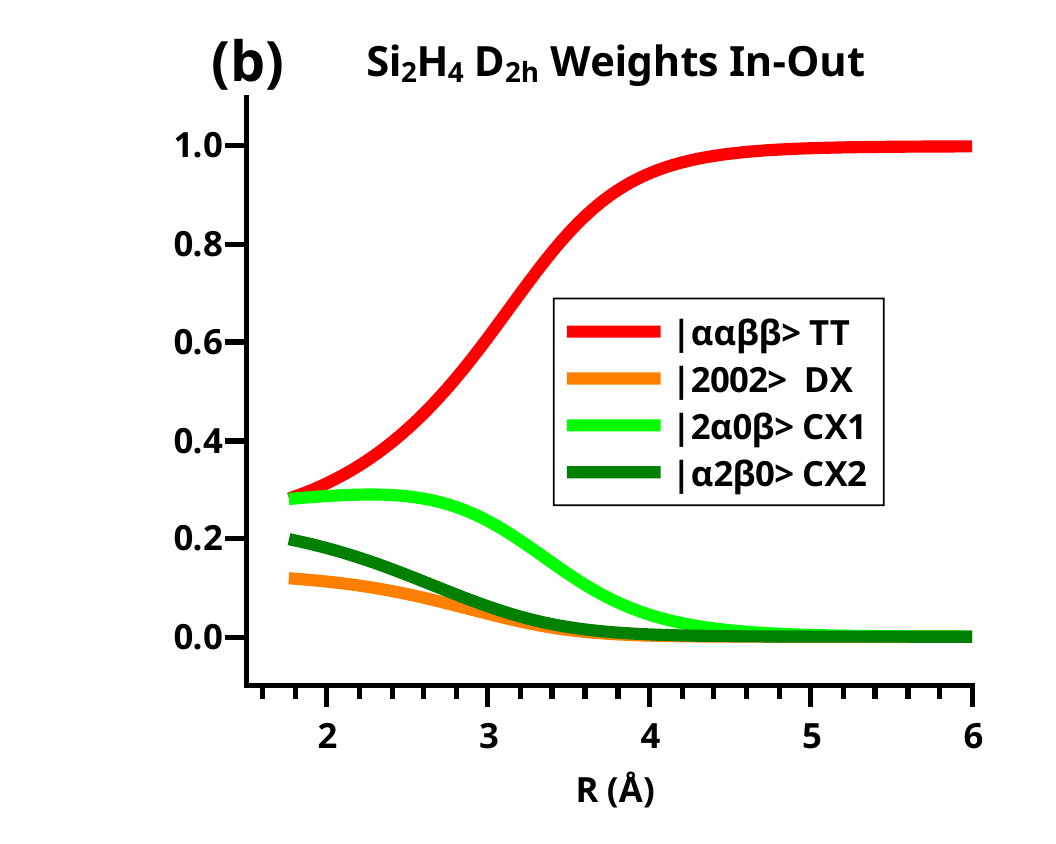}
\label{fig:OVB_disilen_plan_IO}
\end{figure}

This description is corroborated by the weights and energies of the CSFs.
In-Out (Figure \ref{fig:OVB_disilen_plan_IO}) is dominated by TT at large Si-Si distances, at shorter distances, the weights of  CX1, CX2, and DC increase. The CSF energies increase monotonously with decreasing Si-Si distance, the mono-ionic CSFs have a shallow minimum around the equilibrium geometry of disilene.
\begin{figure}[ht]
\caption{Recombination reaction. a) Energies and  b) transition probabilities for the CSFs (threshold of 0.1).}
\includegraphics[width=0.45\textwidth]{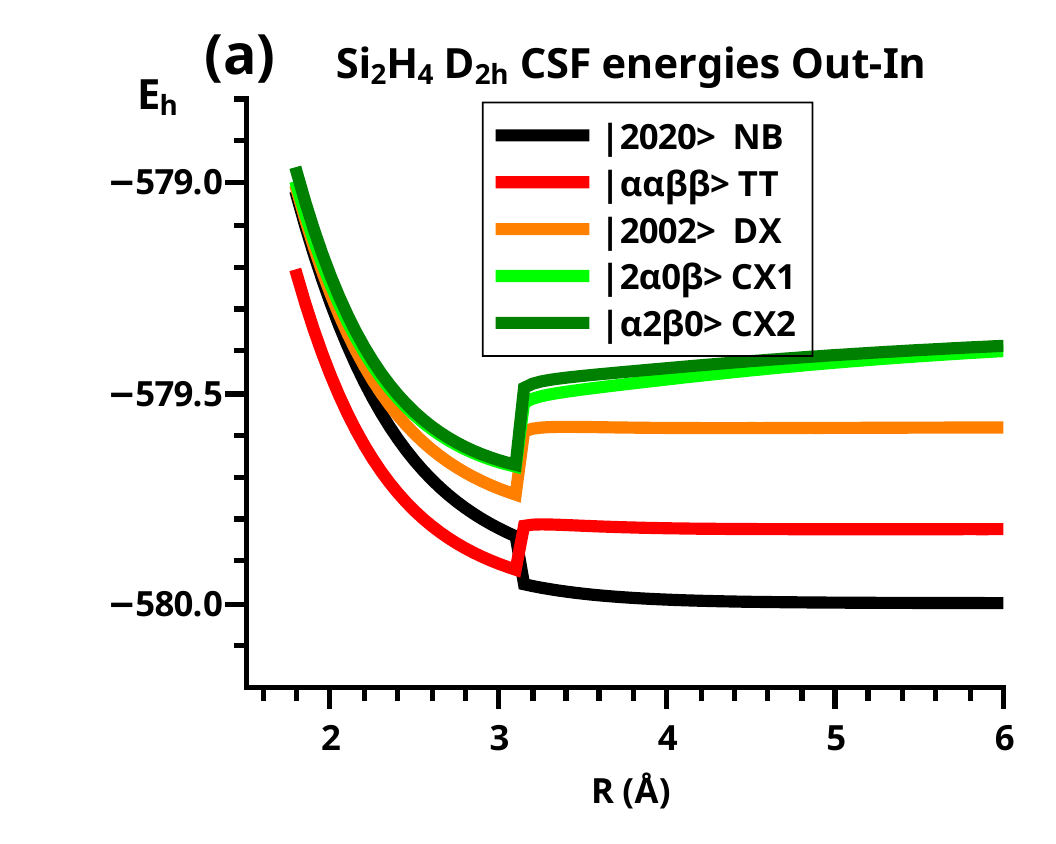}
\includegraphics[width=0.45\textwidth]{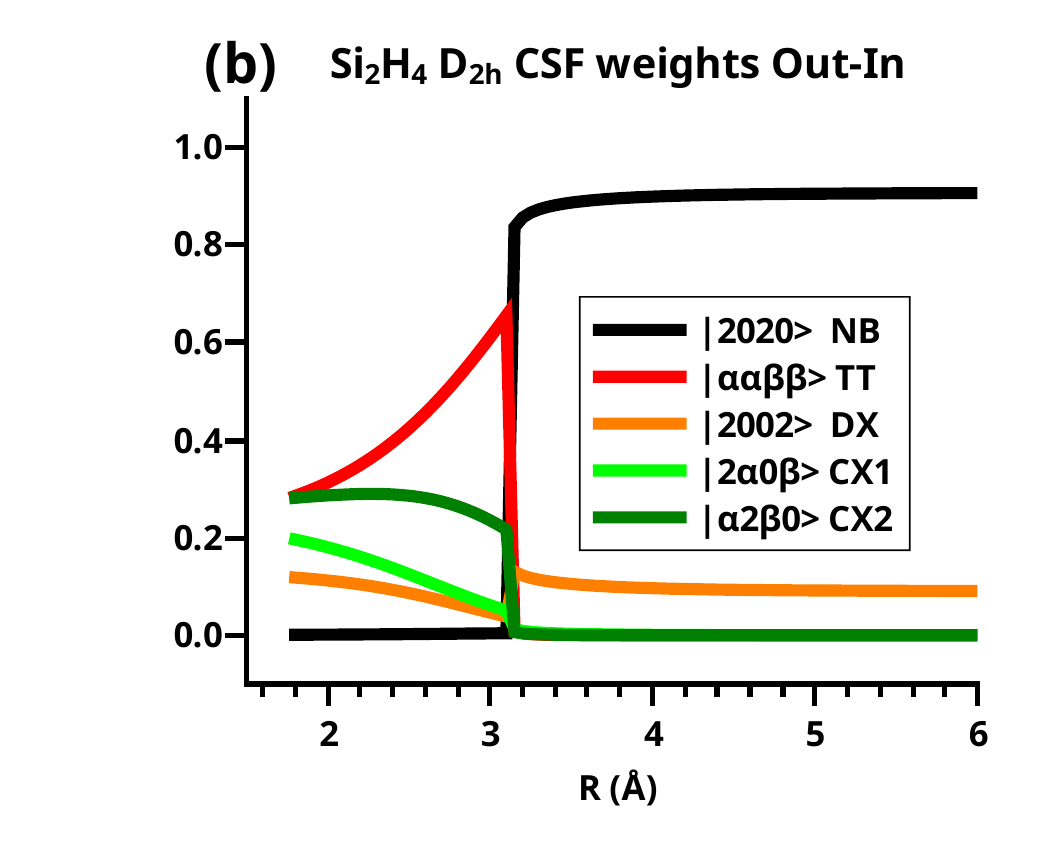}
\label{fig:OVB_disilen_plan_OI}
\end{figure}
Out-In (Figure \ref{fig:OVB_disilen_plan_OI}) is dominated by  NB at large Si-Si distances, DX+DX describes the angular correlation of the lone pair electrons in singlet silylene; at a Si-Si distance of 3.1\,\AA{}, NB is replaced by TT and the ionic CSFs that describe delocalization when two triplets are interacting. But also the CSF energies indicate the spin rearrangement: in a low spin state, the lone pair electrons are non-identical electrons and can come close, the Coulomb repulsion can only be reduced by enlarging the domain where the lone pair electrons reside. That means, the lone pair orbitals must expand. In a high spin state, the PEP prohibits the electrons from coming too close, the Coulomb repulsion is automatically smaller and the domain can contract. When the spins in silylene change from low spin to high spin and the lone pair orbitals contract, this is extremely unfavorable for  the singlet coupled electrons as described by  NB. Accordingly, the energy must increase, as one can see. That TT benefits from the contraction is no surprise, that all other CSFs describe charge distributions that also benefit from the orbital contraction is a result of the analysis.

\subsection{Reaction $\rm Si_2H_4 \rightarrow  2 SiH_2$ in $C_{2h}$}
In $C_{2h}$ symmetry, there are 12 CSFs or linear combinations of fragment CSFs having $A_g$ symmetry, the eight CSFs that describe the reactions in $D_{2h}$ plus the four linear combinations C+C, X+X, TX+TX, TXC+TXC. With respect to threshold 0.1, only seven CSFs are important.

\begin{figure}
\caption{Energy curves for dissociation and recombination reaction.}
\includegraphics[width=0.45\textwidth]{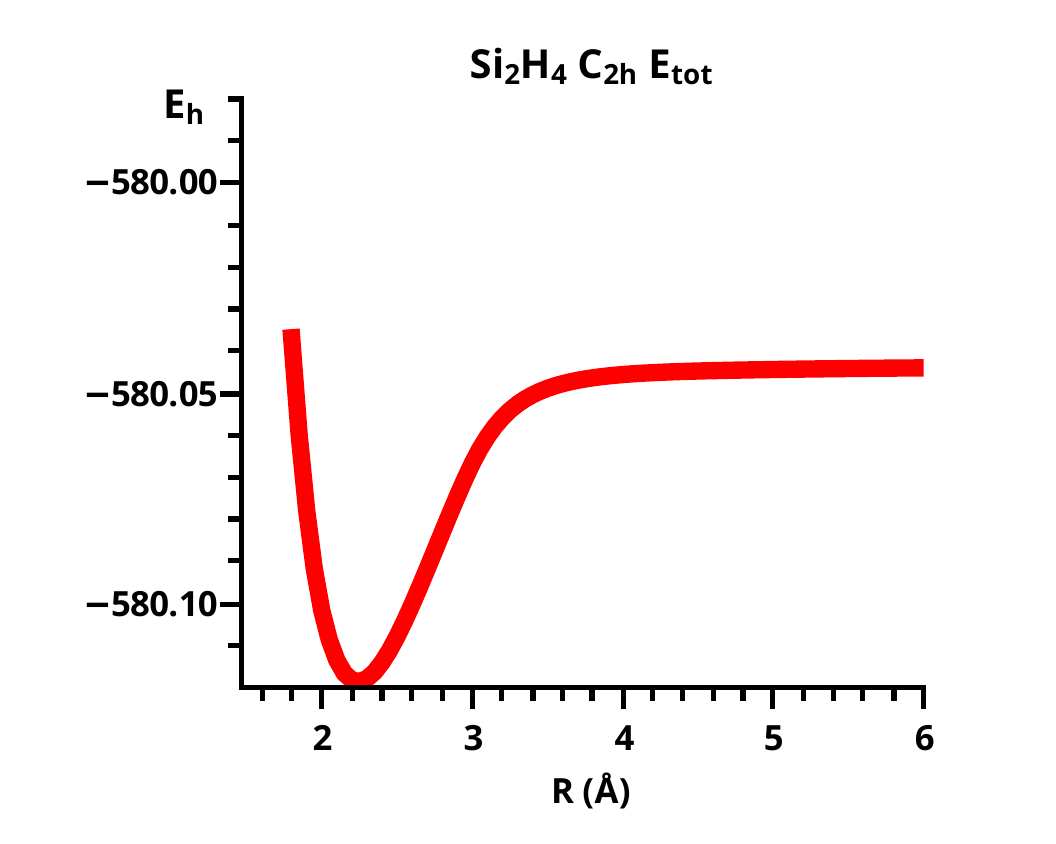}
\label{fig:E_disilene_bent}
\end{figure}

\begin{figure}
\caption{a) SiH length, b) HSiH angle and c) out-of-plane angle as function of the Si-Si distance. }
\includegraphics[width=0.32\textwidth]{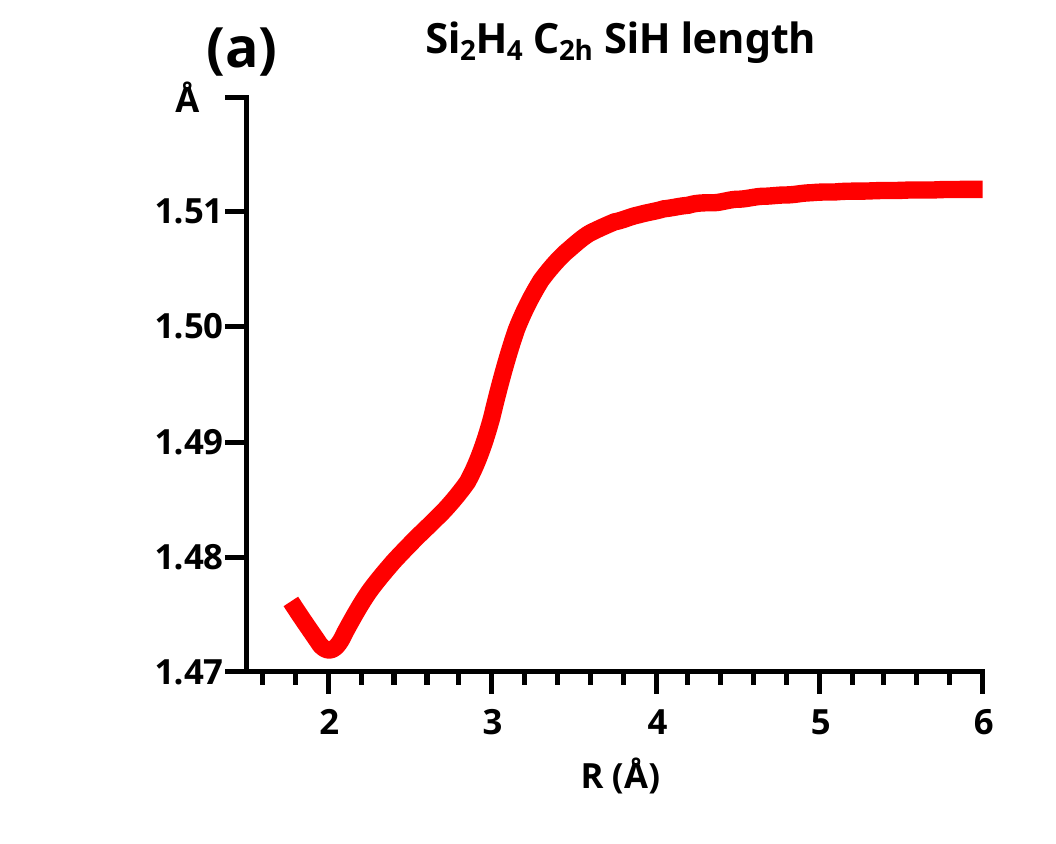}
\includegraphics[width=0.32\textwidth]{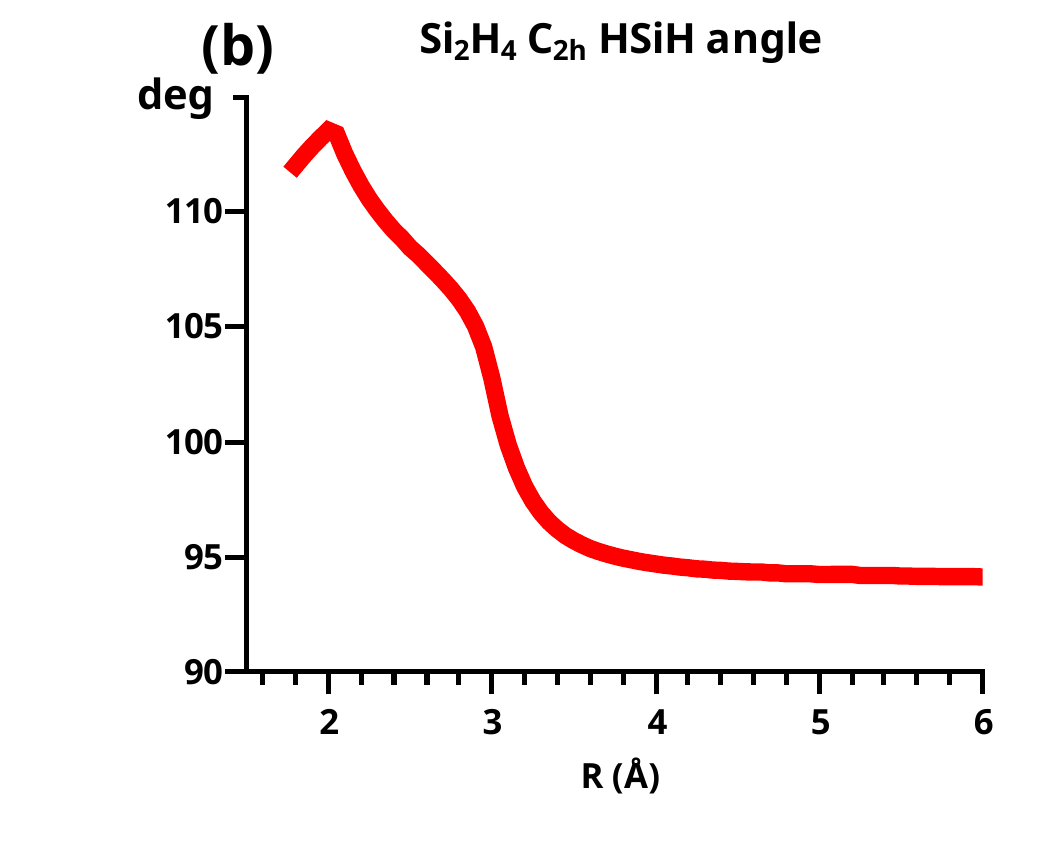}
\includegraphics[width=0.32\textwidth]{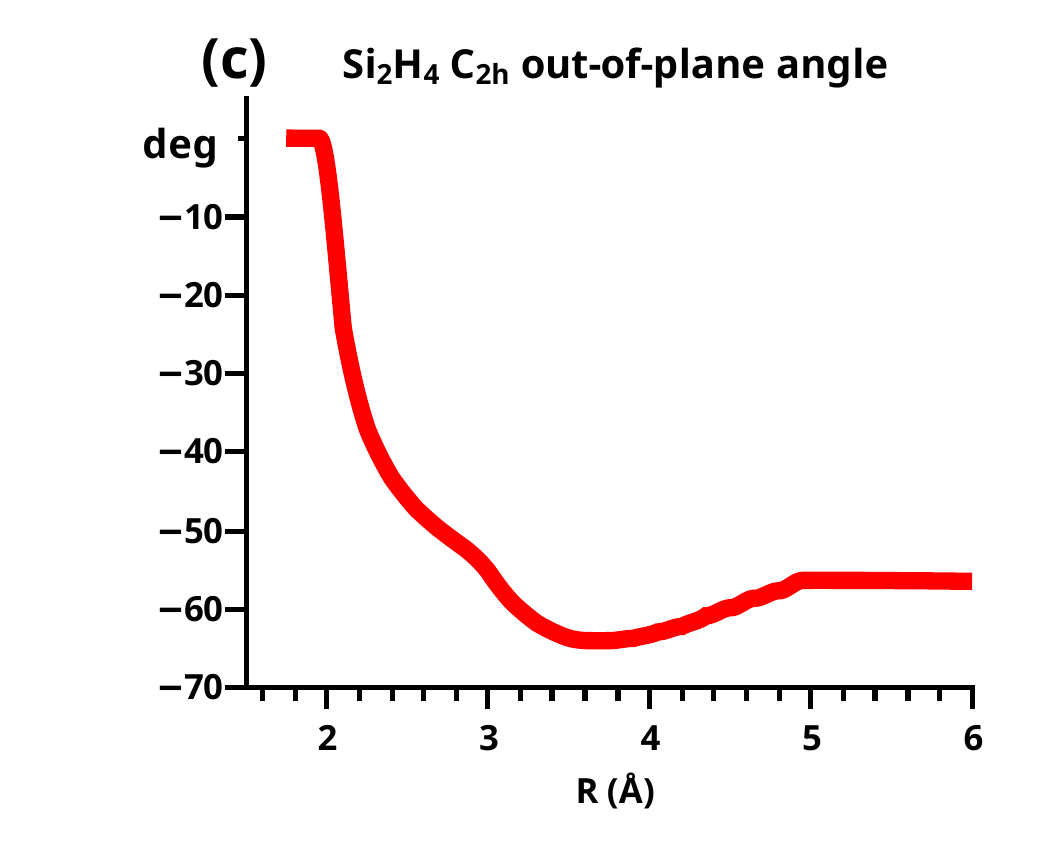}
\label{fig:geom_disilene_bent}
\end{figure}

In-Out and Out-In proceed in an identical way, the potential energy curves (Figure \ref{fig:E_disilene_bent}) are equal and smooth. However, the change of the geometry parameters as a function of the reaction coordinate are described by wobbled curves (Figure \ref{fig:geom_disilene_bent}). At the theoretical level chosen, the equilibrium geometry of disilene is non-planar, the equilibrium Si-Si distance is slightly larger than that of planar disilene, it is about 2.25\,\AA. If the Si-Si distance is shortened, disilene becomes planar. When the Si-Si distance is enlarged, the non-planarity becomes more pronounced and the geometry of the silylene fragments changes to that of singlet silylene, but not as smooth as expected. The CSF weights show that as soon as disilene becomes non-planar the two ionic CSFs C+C and TXC+TXC become immediately important, the weight of the neutral CSF NB increases much slower. Pyramidalization at the silicon atoms means a change from sp$^2$ hybridization to either sp$^3$ hybridization, as in case of the silyl radical,  or no hybridization as in case of the silyl radical anion or the silylene radical. As discussed above, in atoms of periods 3 and  higher the s-subshell in free atoms is always doubly occupied and hybridization of s-AOs and p-AOs needs a perturbation of lower than spherical  symmetry and of sufficient strength. Obviously, the perturbation is strong enough when the  Si-Si distance is shorter than 2.0\,\AA,  then sp$^2$ hybridization and planarization of the silicon atoms are possible; when the Si-Si distance increases, the tendency to doubly occupy the s-subshell dominates and pyramidalization is favorable. The description of the pyramidalization at the silicon atoms needs the ionic CSFs that are  in $D_{2h}$ of $B_{1g}$ symmetry.

\begin{figure}
\caption{a) Energies and  b) transition probabilities for the CSFs (threshold of 0.1).}
\includegraphics[width=0.45\textwidth]{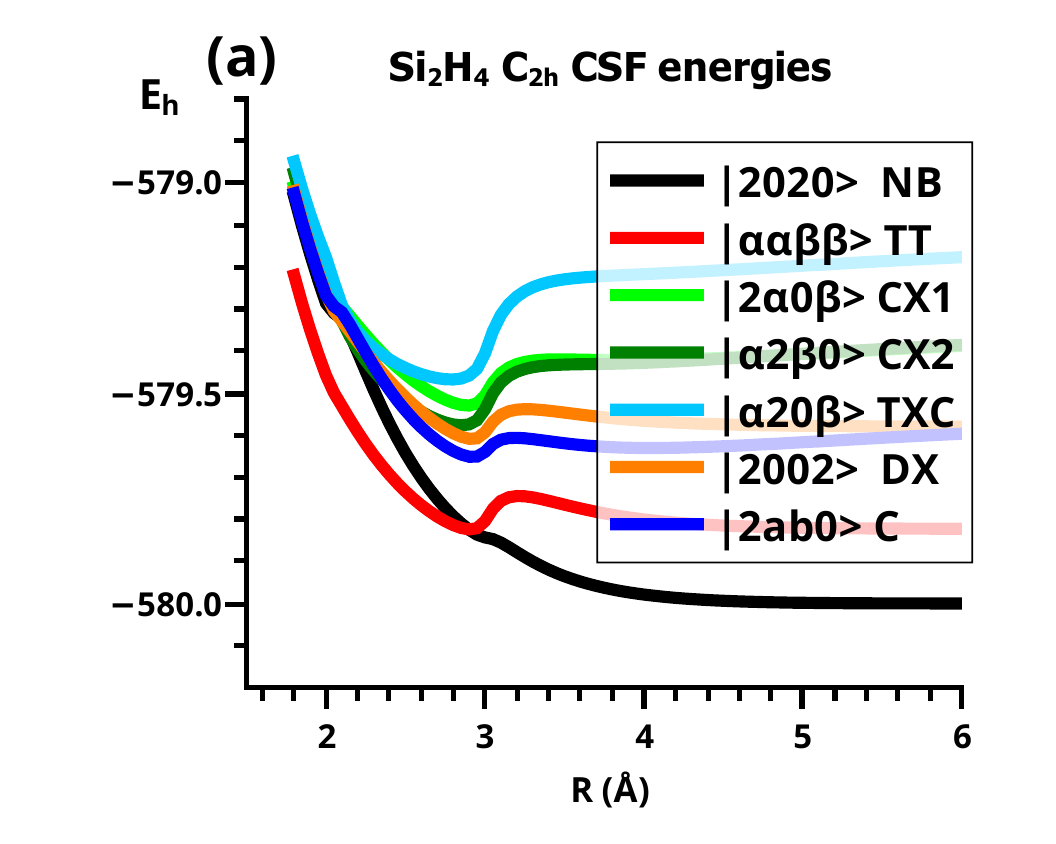}
\includegraphics[width=0.45\textwidth]{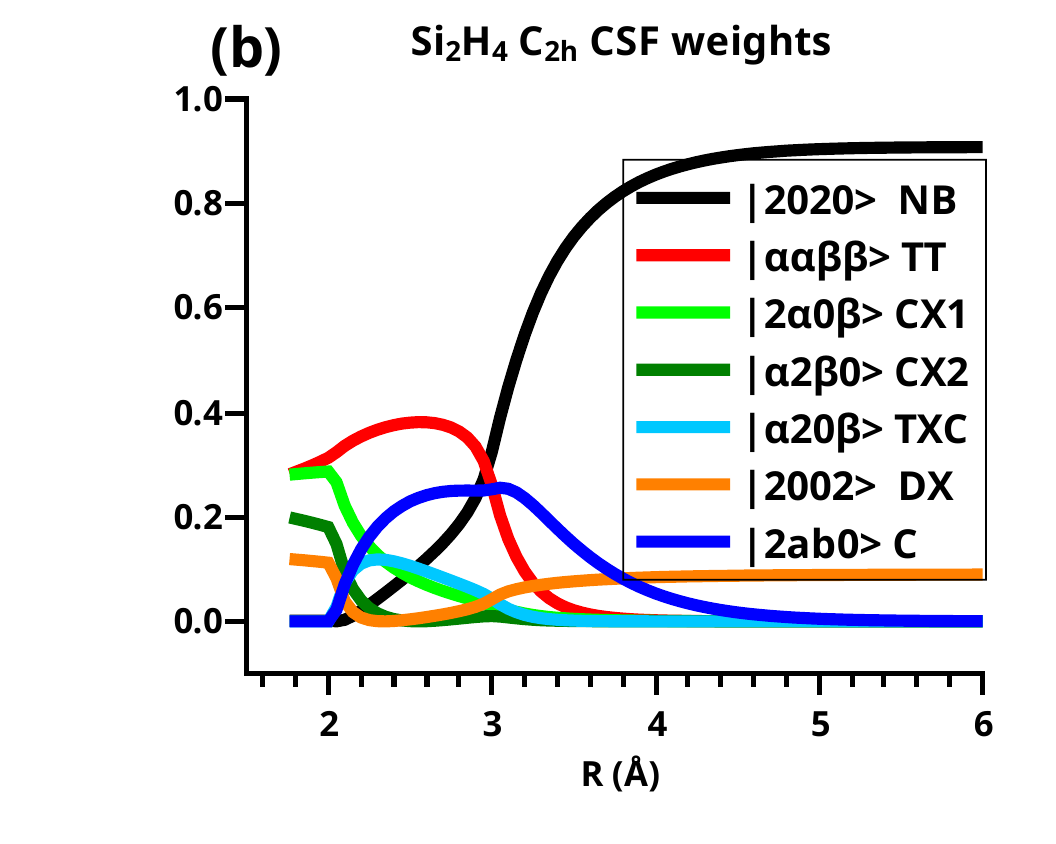}
\label{fig:OVB_disilene_bent}
\end{figure}

Out-In starts from two singlet silylenes in a pronounced trans-bent structure, one can see for large distances that NB dominates and the correlating CSF DX+DX. Already at a Si-Si distance larger than 4.0\,\AA{}, the contribution of C+C increases, whereas the weights of NB and DX+DX are reduced. TT  becomes important only at a Si-Si distance of about 3.5\,\AA{}, at about 3.0\,\AA{}  NB, TT and C+C have equal weights (Figure \ref{fig:OVB_disilene_bent} a). As the shape of the energy curves indicate, (Figure \ref{fig:OVB_disilene_bent} b) at this distance a contraction of the orbitals occurs, and the geometry parameters show that silylene changes from low spin to high spin, but all changes do not occur  as abruptly as in case of the recombination in planar geometry. There, the weight of TT has a maximum value of 0.7, in non-planar geometry the maximum value is only 0.4; the maximum weight of C+C is close to 0.3  and this value is nearly constant for all Si-Si distances between 2.4 and 3.4\,\AA{}.

\begin{figure}
\caption{Energy curves for dissociation and recombination reactions for planar and trans-bent disilene.}\label{fig:disietots}
\includegraphics[width=0.45\textwidth]{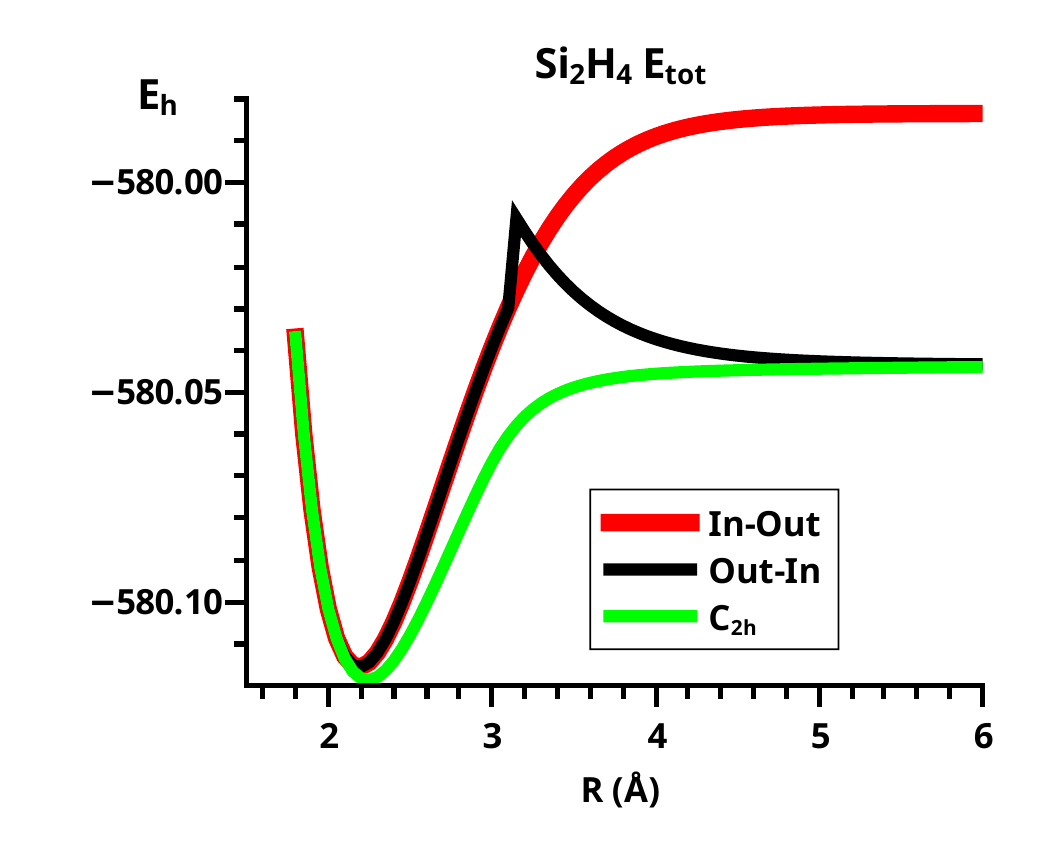}
\end{figure}

Figure \ref{fig:disietots} shows how deviation from planarity enlarges the equilibrium distance and allows a smooth dissociation into ground state silylenes.

\subsection{Reaction $\rm CSiH_4 \rightarrow  CH_2 + SiH_2$ in $C_{2v}$}
\begin{figure}
\caption{Energy curves for dissociation and recombination reaction.}\label{fig:C2vEtots}
\includegraphics[width=0.45\textwidth]{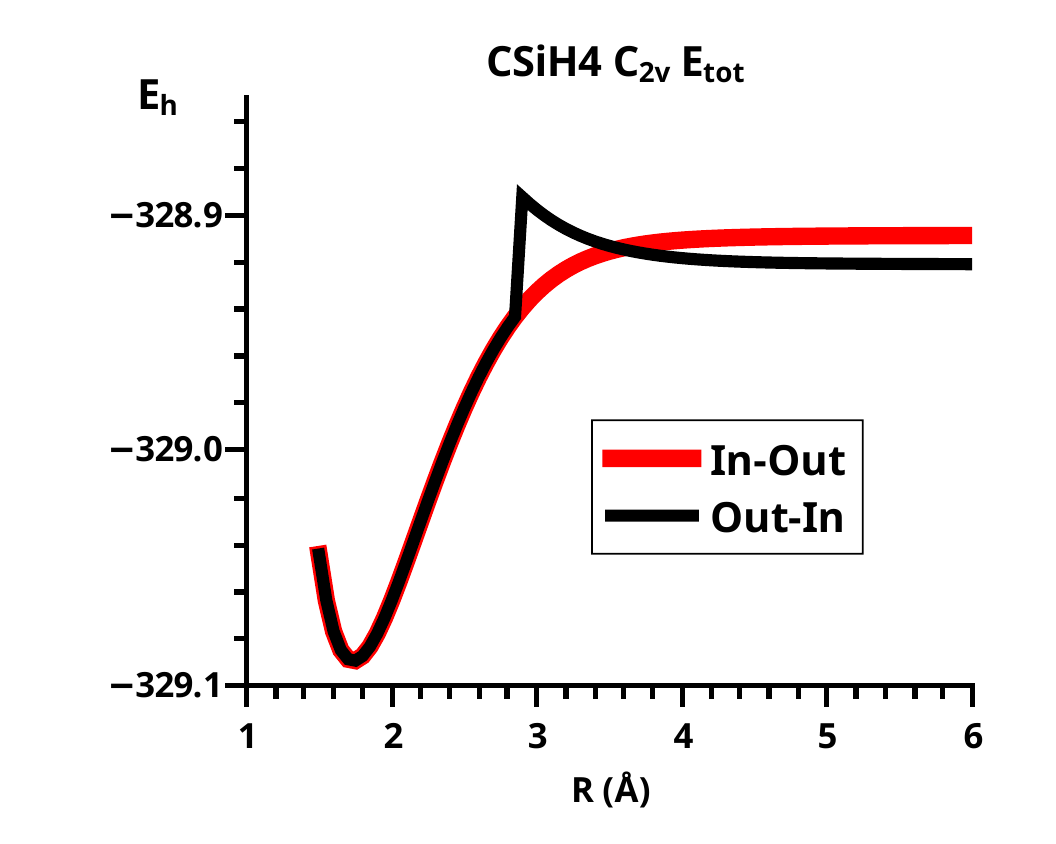}
\end{figure}

Silaethene is of interest, because the fragments silylene and carbene have different ground state multiplicities.
In $C_{2v}$,  the singlet ground state has $A_1$ symmetry, the 12 CSFs having this symmetry are: NB, TT, SS, QX, DX(C), DX(Si), DC(C), DC(Si), CX1(C), CX1(Si), CX2(C), CX2(Si), the atom symbol indicates either the atom with additional electrons or where the excitation occurs. Since there are no linear combinations of CSFs, the squares of the CI coefficients are smaller than in the systems discussed so far, and all CSFs with weights larger than 0.1 are considered as important. In-Out and Out-In  proceed differently, as the potential energy curves show,\ref{fig:C2vEtots}. For In-Out, the following seven CSFs are  important: TT, CX1(Si), CX2(Si), DX(Si), CX1(C), CX2(C), DX(C); for Out-In, also NB is important.

\begin{figure}
\caption{a) SiH length, b) CH length  as function of the C-Si distance.}
\includegraphics[width=0.45\textwidth]{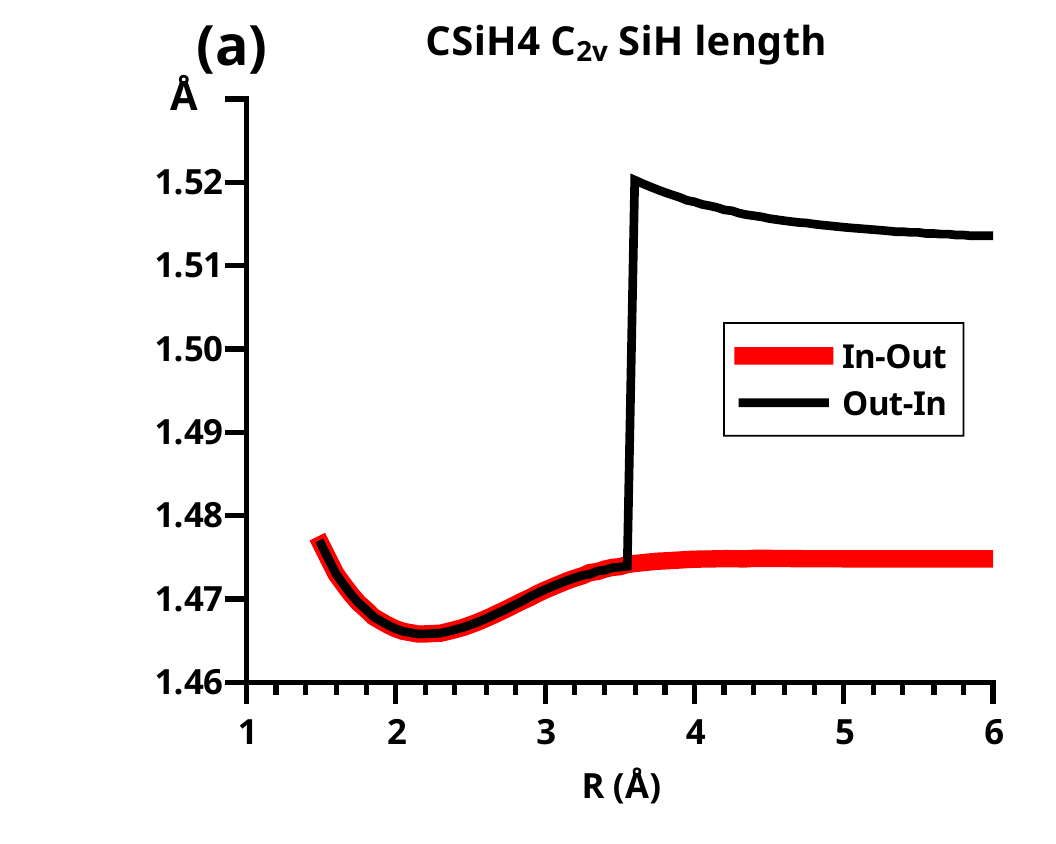}
\includegraphics[width=0.45\textwidth]{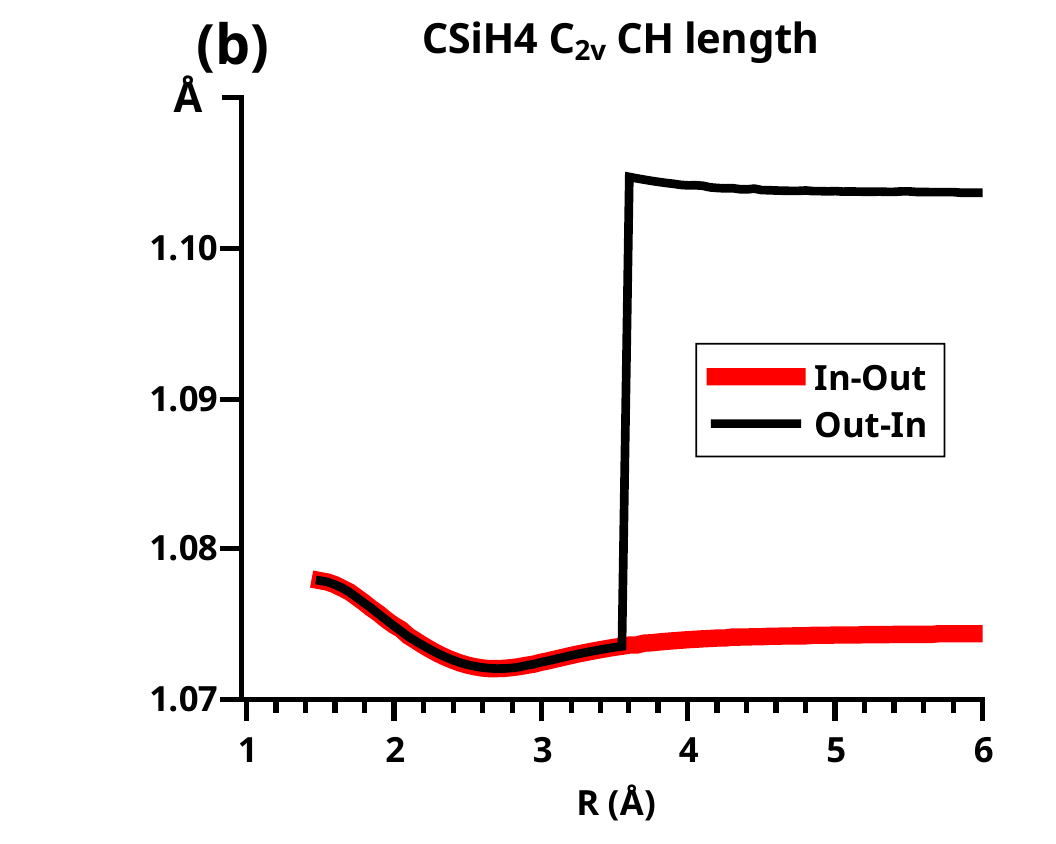}
\label{fig:length_C2v}
\end{figure}

\begin{figure}
\caption{HSiH and HCH bond angles  as function of the C-Si distance.}
\includegraphics[width=0.45\textwidth]{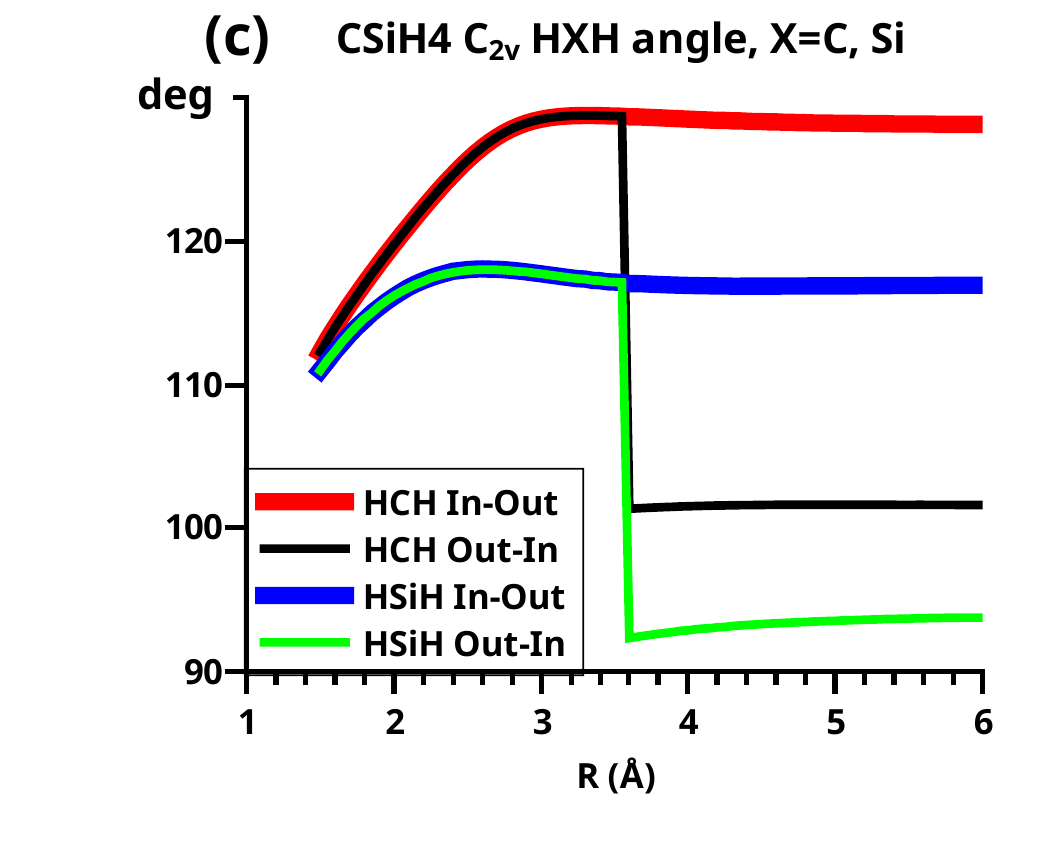}
\label{fig:angles_C2v}
\end{figure}

In-Out leads to fragments in the  triplet state, similar to the dissociations of ethene or planar disilene, the reaction proceeds in a completely smooth way, as the geometry parameters show (Figures \ref{fig:length_C2v} and \ref{fig:angles_C2v}). The ground state of carbene is a triplet state but for silylene it is an excited state; the singlet-triplet splitting in silylene (about 79\,kJ/mol) is roughly twice as large as that in carbene (about 46\,kJ/mol). The ground state the dissociated system will consist of both fragments in their lowest singlet state, in the first excited state both fragments will be in their lowest triplet states, this state will lie  about 33\,kJ/mol above the ground state.  In agreement with this estimate, one finds an energy difference of roughly 35\,kJ/mol between In-Out and Out-In  at a C-Si distance of 6.0\,\AA.
The recombination of the two singlet fragments is not possible without spin rearrangement.  When the C-Si distance decreases the energy increases; at about 3.5\,\AA{} the energy curves of In-Out and Out-In cross, but only at a C-Si distance of  about 2.9\,\AA{} the system jumps to the lower lying state with both fragments in their respective triplet states.

Spin rearrangement is reflected also in the changes of bond lengths and bond angles in both fragments (Figures \ref{fig:length_C2v} and \ref{fig:angles_C2v}).

\begin{figure}
\caption{Dissociation reaction. a) Energies and  b) transition probabilities for the CSFs (threshold of 0.1).}\label{fig:C2vCSFIO}
\includegraphics[width=0.45\textwidth]{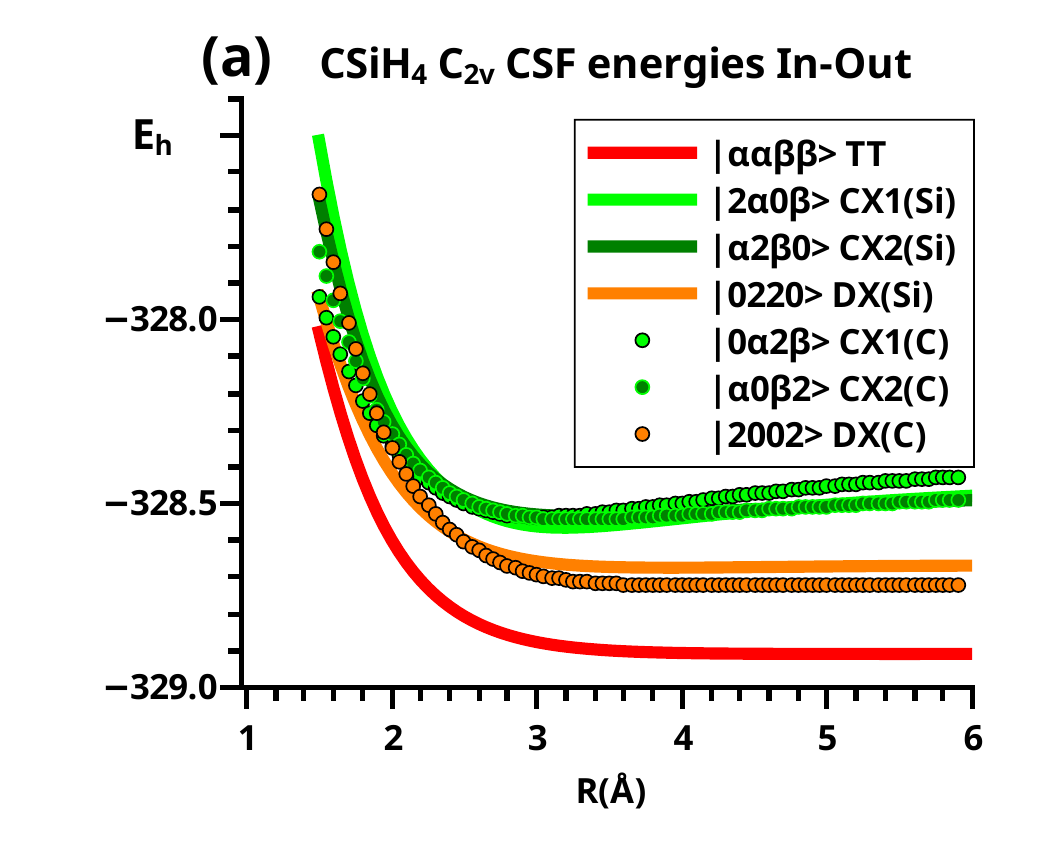}
\includegraphics[width=0.45\textwidth]{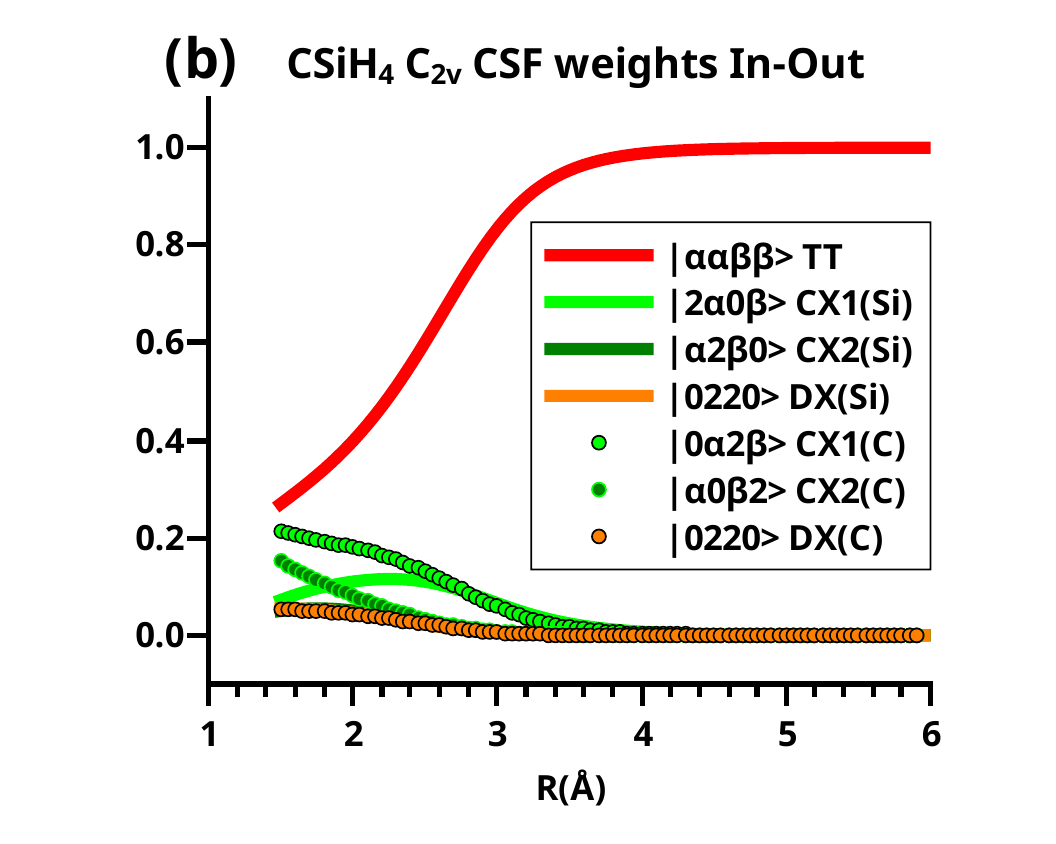}
\end{figure}

\begin{figure}
\caption{Recombination reaction. a) Energies and  b) transition probabilities for the CSFs (threshold of 0.1).}\label{fig:C2vCSFOI}
\includegraphics[width=0.45\textwidth]{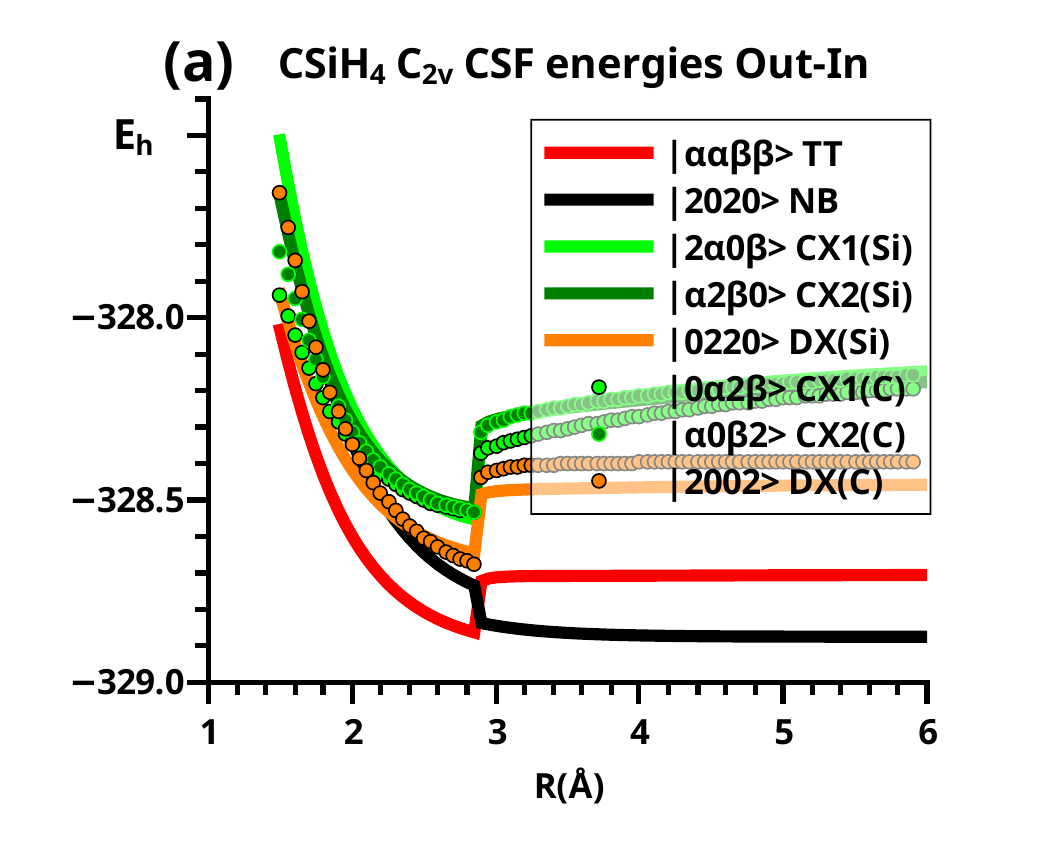}
\includegraphics[width=0.45\textwidth]{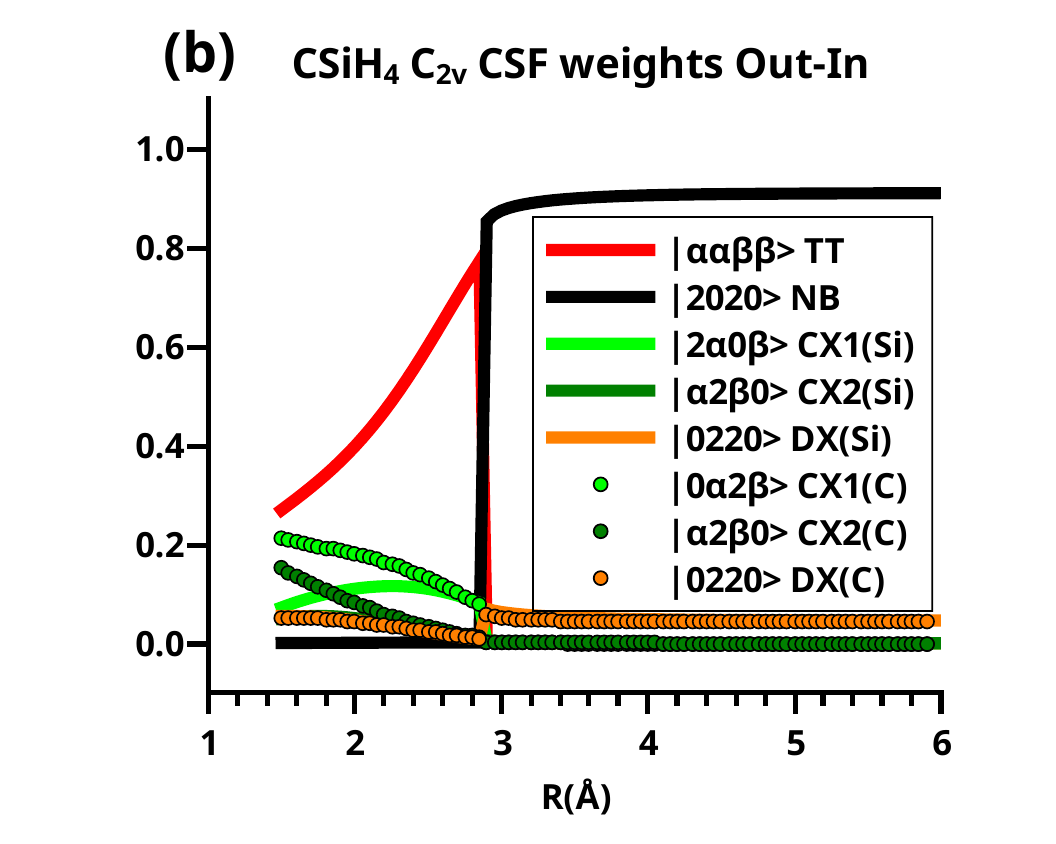}
\end{figure}

Weights and energies of the CSFs show that at large C-Si distances both In-Out and Out-In are dominated by very few CSFs (Figures \ref{fig:C2vCSFIO} b) and \ref{fig:C2vCSFOI} b)); In-Out is dominated solely by TT, Out-In by NB and the DX(Si) and DX(C)  describing angular correlation in the singlet states of the fragments. At short distances dominate TT and the ionic CSFs that describe delocalization and polarization during bonding. Now, the difference between carbon and silicon becomes obvious: the weights of CX1(C) and CX2(C) are considerably higher than CX1(Si) and CX2(Si), this reflects the charge shift towards carbon during bonding, one can also say this is so because carbon is more electronegative than silicon.  DX(Si), DX(C) and CX2(Si) have nearly the same weights, the curves in Figures \ref{fig:C2vCSFIO} and \ref{fig:C2vCSFOI} overlay exactly.
As for planar disilene, Out-In starts with both fragments in singlet states; NB dominates, DX(C) and DX(Si) describe the angular correlation of the lone pair electrons in the singlet fragments. At a C-Si distance of 2.9\,\AA{},  NB is replaced by TT and the ionic CSFs that describe delocalization during bonding of two triplets. But also the CSF energies (Figure \ref{fig:C2vCSFOI} a) indicate the change in the fragment multiplicity at this C-Si distance: the energy of all CSFs but NB decrease after the spin flip, however NB does not contribute at C-Si distances smaller than 2.9\,\AA{}.

\subsection{Reaction $\rm CSiH_4 \rightarrow  CH_2 + SiH_2$ in $C_{s}$}
\begin{figure}
\caption{Energy curves  for dissociation and recombination reaction in $C_s$ symmetry.}\label{fig:CsEtots}
\includegraphics[width=0.45\textwidth]{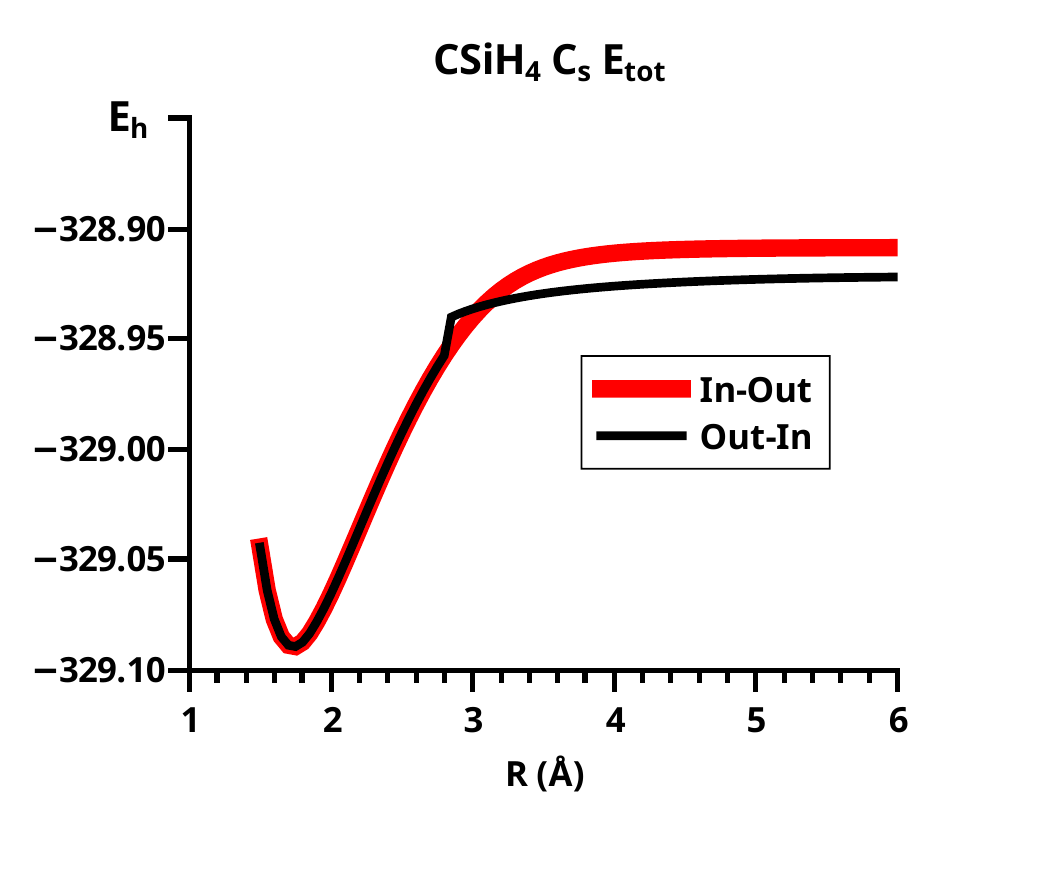}
\end{figure}

Deviation from planarity leads to a trans-bent structure of $C_s$ symmetry. In contrast to disilene,  In-Out and Out-In occur in different ways, the reaction asymptote of In-Out is characterized by the coupled triplets but not by the ground state of coupled singlets. The spin and charge rearrangement during Out-In occur at the same C-Si distance as in $C_{2v}$ symmetry, but there is no increase in energy when the fragments approach (Figure \ref{fig:CsEtots}).

\begin{figure}
\caption{a) SiH length, b) CH length  as function of the C-Si distance.}
\includegraphics[width=0.45\textwidth]{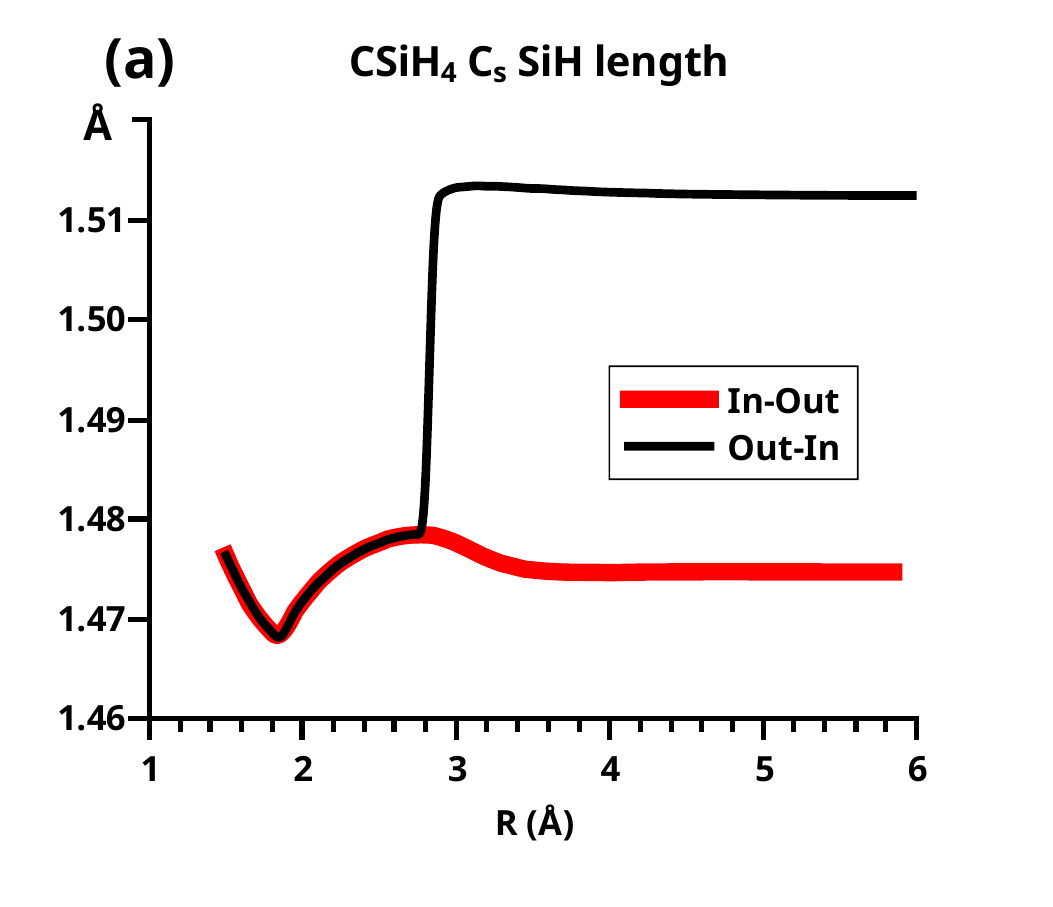}
\includegraphics[width=0.45\textwidth]{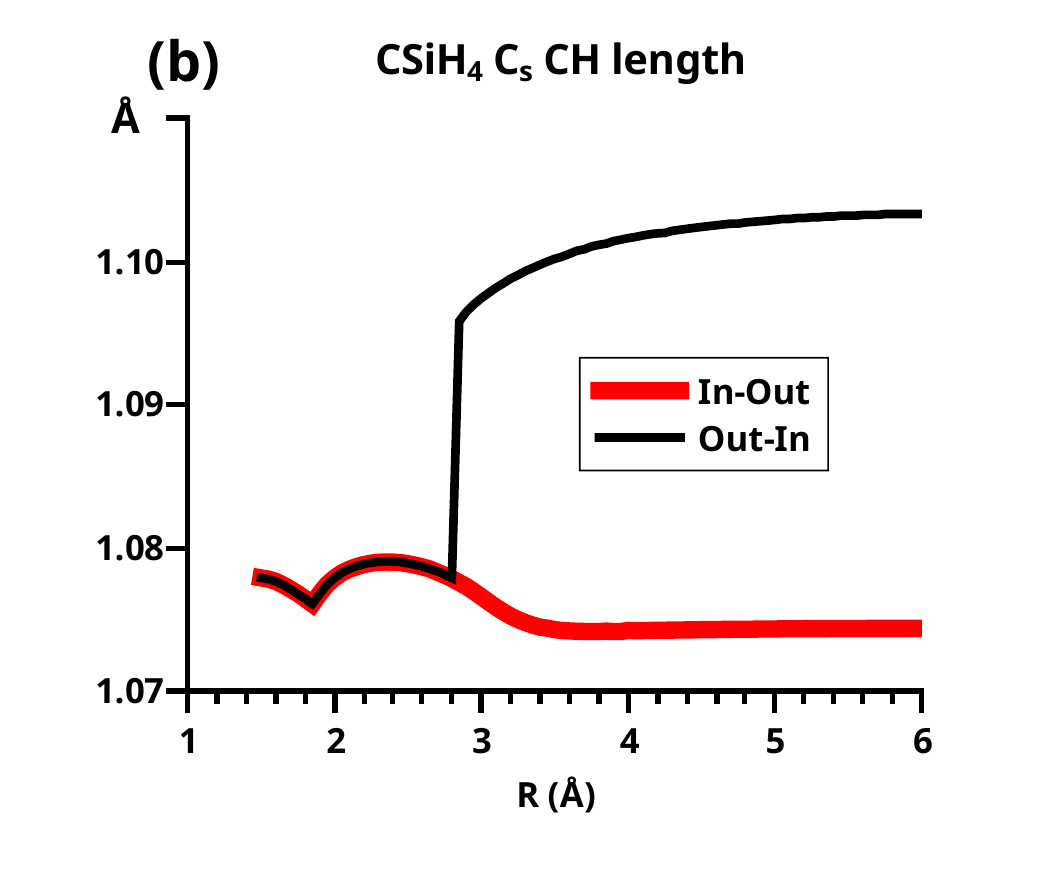}
\label{fig:length_Cs}
\end{figure}
\begin{figure}
\caption{HSiH and HCH bond angles  as function of the C-Si distance.}
\includegraphics[width=0.45\textwidth]{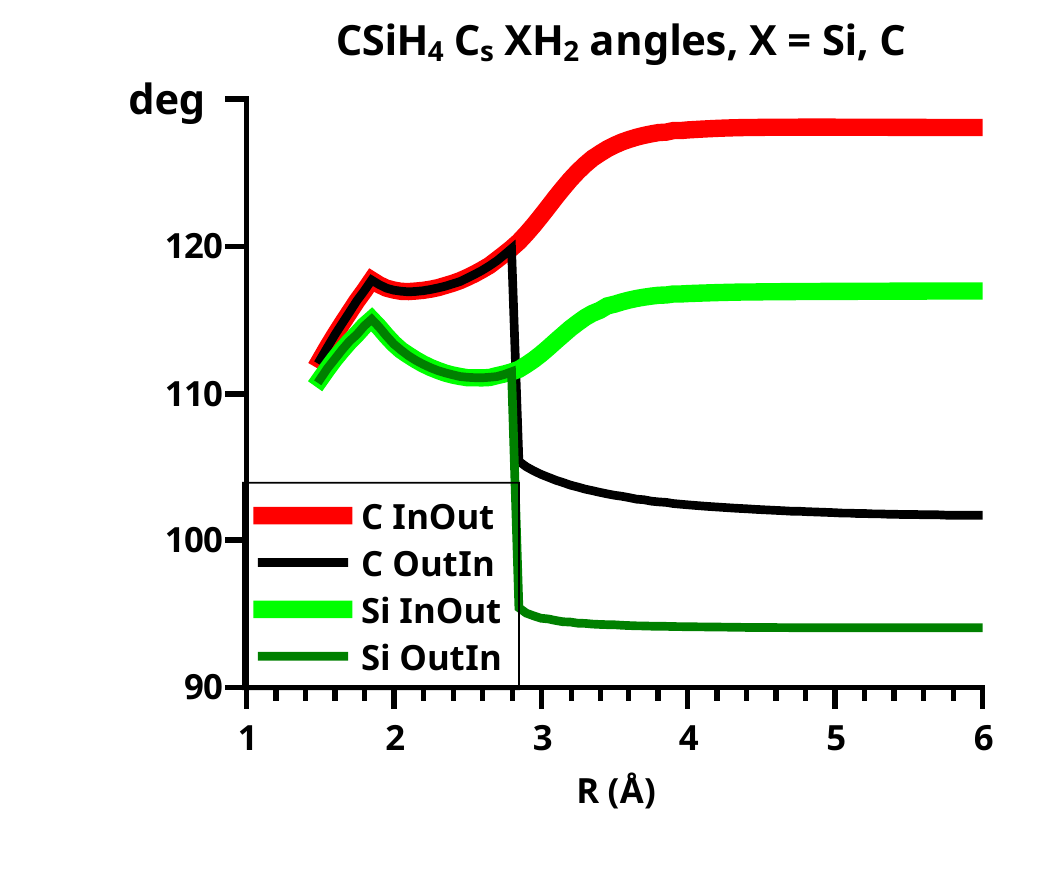}
\label{fig:angles_Cs}
\end{figure}
\begin{figure}
\caption{Out-of-plane angles as function of the C-Si distance. }
\includegraphics[width=0.45\textwidth]{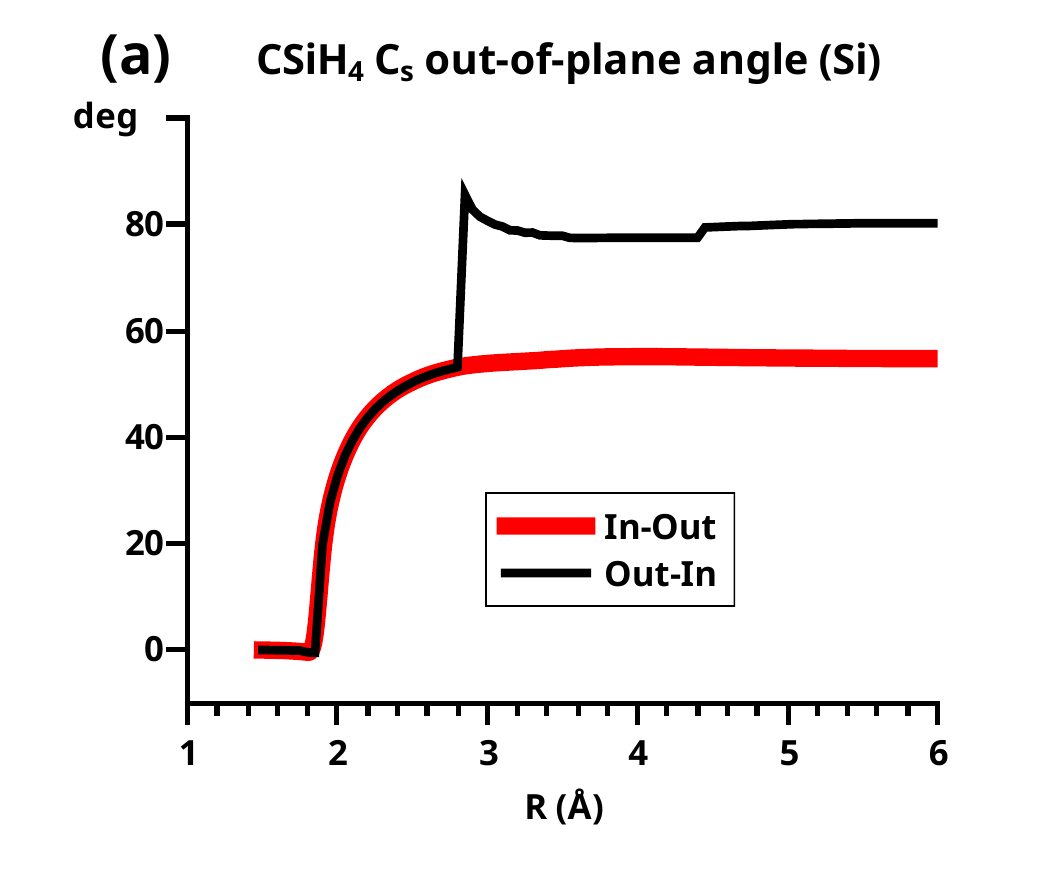}
\includegraphics[width=0.45\textwidth]{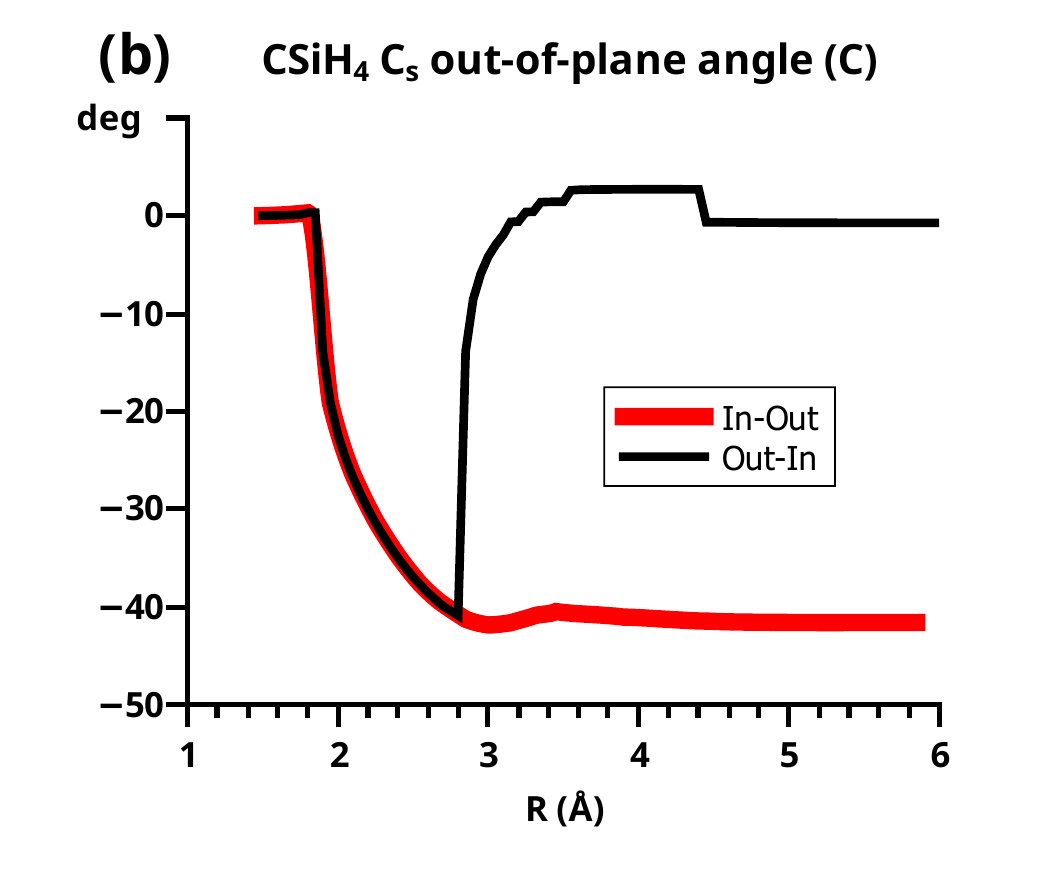}
\label{fig:oop_Cs}
\end{figure}

For C-Si distances smaller than 1.8\,\AA, the molecule is planar, for larger distances the molecule is trans-bent. In the planar structure, the bond HXH bond angles but also the XH distances are comparable to those in ethene and planar disilene at X-X distances (X=C,Si) shorter than the respective  equilibrium distances. Enlarging the C-Si distance in the trans-bent structure,  the HXH angles decrease, go through a minimum and then increase; after 4.0\,\AA{} they have the constant value of the triplet fragments. The XH distances first increase, go through a maximum and then decrease to the triplet values. At the carbon atom, the out-of-plane angle increases to about 40 degrees  and to 60 degrees at the silicon atom, for C-Si distances larger than 3.0\,\AA, these values remain constant. Starting Out-In from a different geometry, all geometry parameters remain unchanged until the distance between the fragments allows interactions that influence also the local geometries. This is  at a  C-Si distance of about 3.0\,\AA{}, where the system jumps to the state describing the dissociation, and all geometry parameters change to the In-Out values. See Figures \ref{fig:length_Cs}, \ref{fig:angles_Cs} and \ref{fig:oop_Cs}.

\begin{figure}
\caption{CSF energies and weights for the dissociation reaction|}\label{fig:CsCSFIO}
\includegraphics[width=0.45\textwidth]{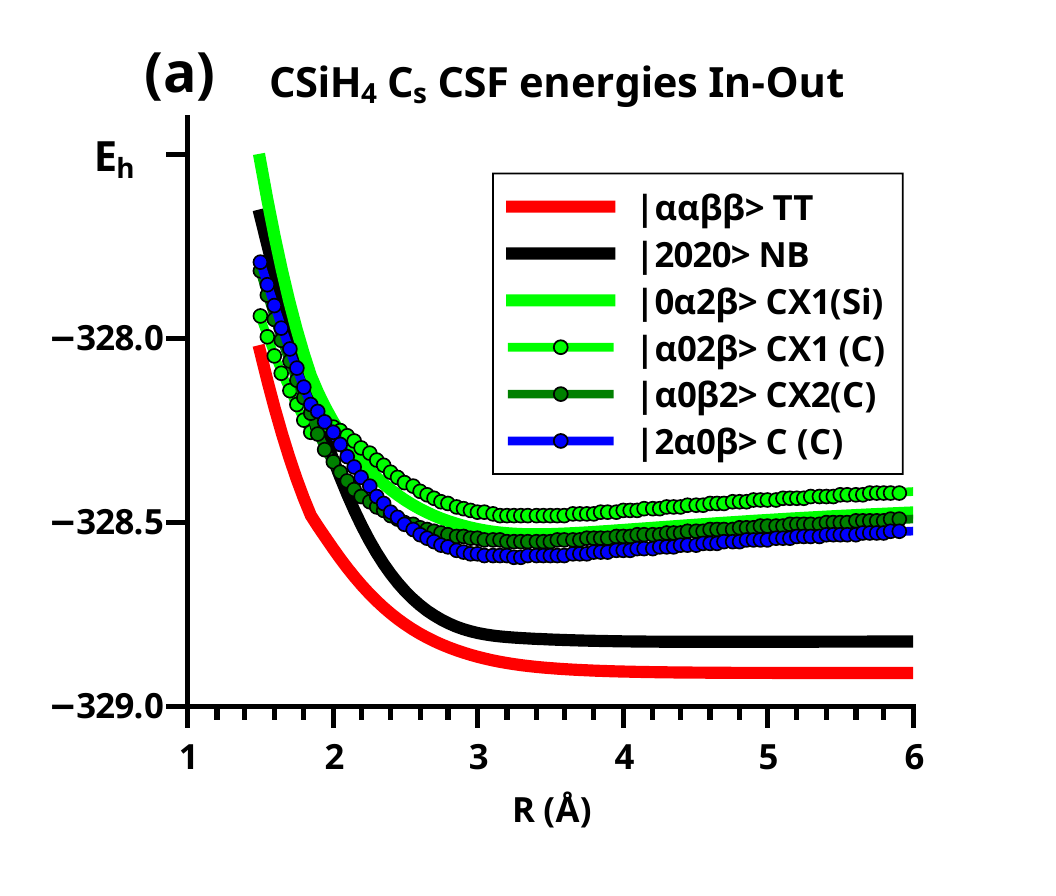}
\includegraphics[width=0.45\textwidth]{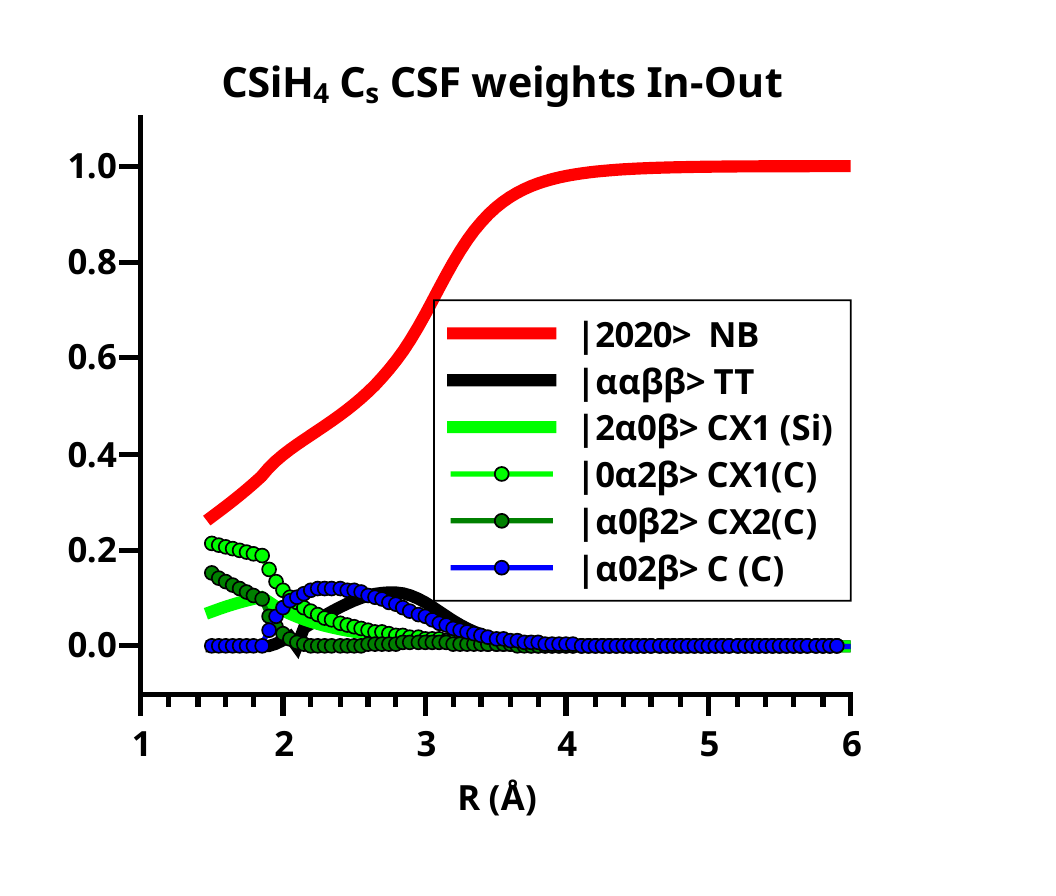}
\label{fig:OVB_Cs_IO}
\end{figure}

\begin{figure}
\caption{CSF energies and weights for the recombination reaction|}\label{fig:CsCSFOI}
\includegraphics[width=0.45\textwidth]{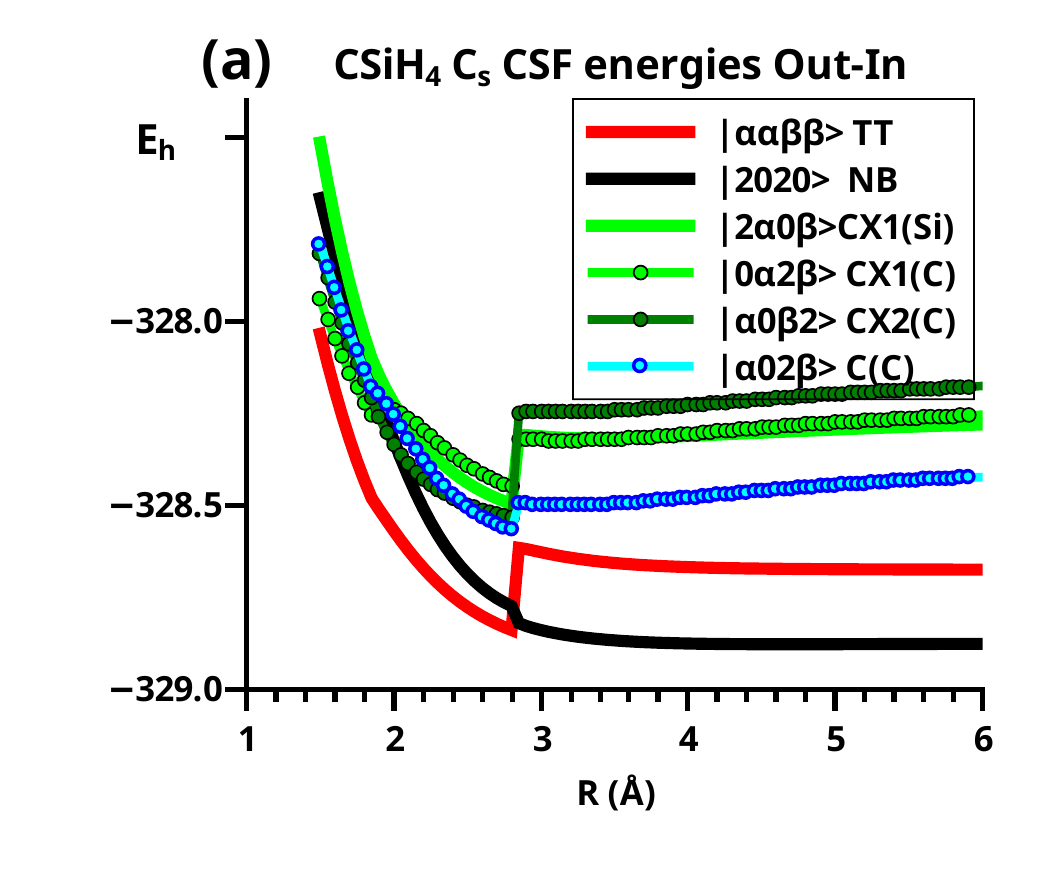}
\includegraphics[width=0.45\textwidth]{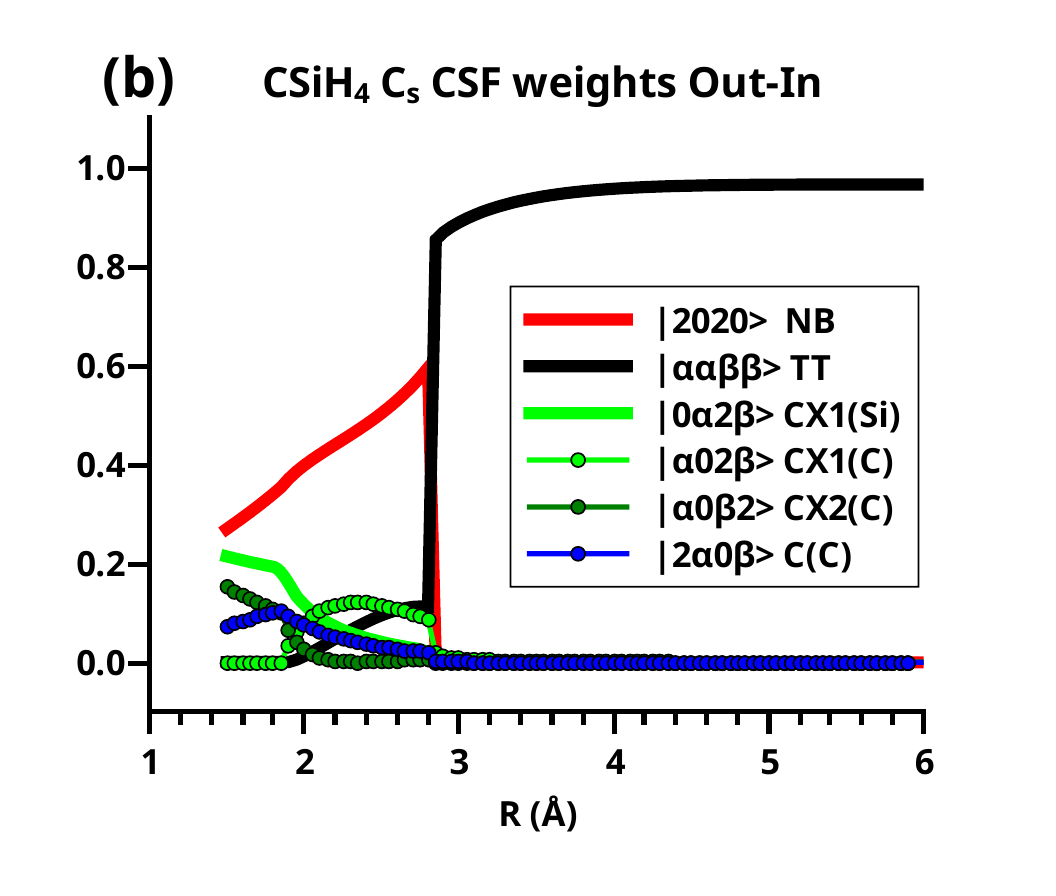}
\label{fig:OVB_Cs_OI}
\end{figure}

The weights of the CSFs corroborate what the geometry parameters suggest, namely that for large C-Si distances the states of In-Out and Out-In are dominated by either TT or NB. See Figures \ref{fig:OVB_Cs_IO} and \ref{fig:OVB_Cs_OI}. Bonding needs high-spin arrangement, which avoids repulsion of the fragments because of repulsion of identical electrons; at a C-Si distance of about 3.0\,\AA{} a change from low-spin to high-spin is energetically favorable. Pyramidalization allows charge rearrangements at carbon and silicon that are not possible in the planar molecule, they are represented by the  ionic CSFs C(C) and C(Si), which describe a shift of charge towards the respective atoms.  To reduce intra-atomic Coulomb repulsion, spin rearrangement is necessary,  X(C)  supports the shift of electrons into regions compatible with tetrahedral spin arrangement, at least at the carbon atom with a relatively small valence shell. The large weight of  NB at C-Si distances smaller than 2.9\,\AA{} is again compatible with a charge distribution at the silicon atom  that can be characterized as consisting of a doubly occupied s-shell and a partially occupied p-shell, or, in other words, silicon does not prefer  hybridization. Only when the C-Si distance smaller than 2.0\,\AA[] the perturbation of the electron distribution by the surrounding atoms is large enough to force hybridization, then the high-spin arrangement causes planarization at both heavy atoms.


\begin{figure}
\caption{Transition probabilities for short C-Si distances.}\label{fig:CswCSFscal}
\includegraphics[width=0.45\textwidth]{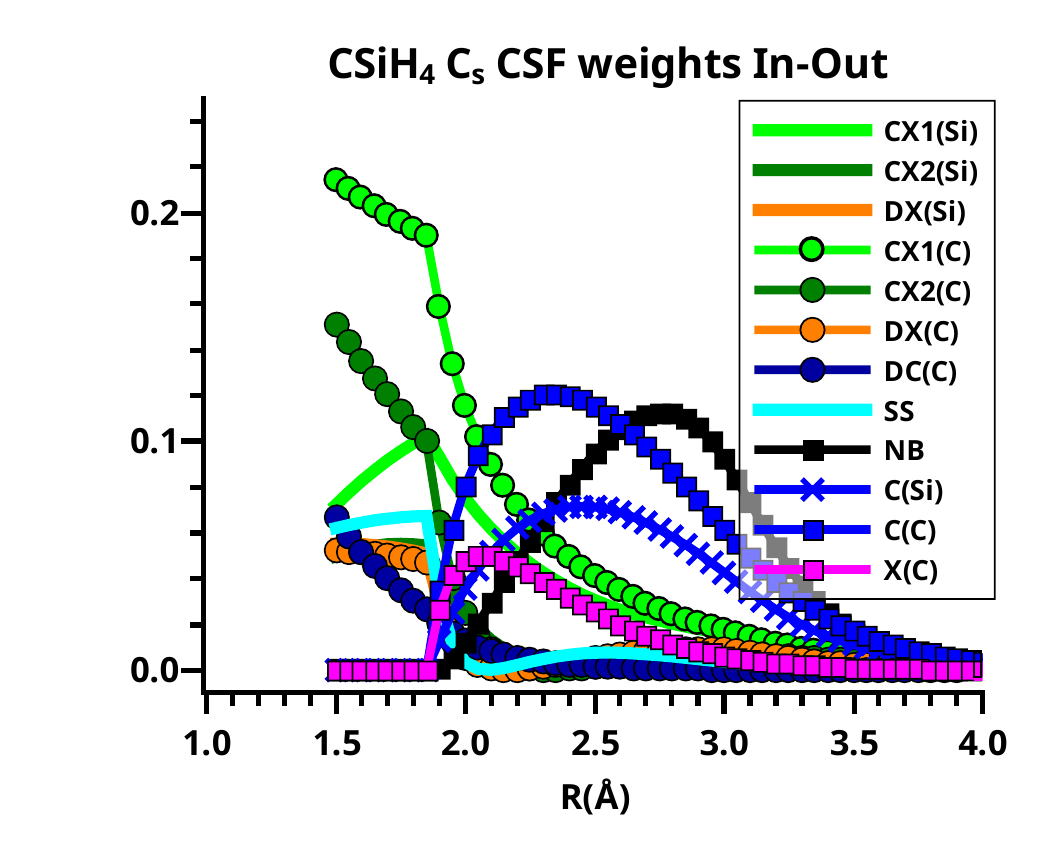}
\end{figure}

The changes in the spin- and charge distributions cannot be described by a single dominant CSF, many CSFs are necessary, as can be seen in Figure \ref{fig:CswCSFscal}, which also shows how the distributions change with the distance between the heavy atoms and with the symmetry of the molecular system.

\begin{figure}
\caption{Energy curves for dissociation and recombination reactions for planar and trans-bent silaethene.}\label{fig:CsEtotsc}
\includegraphics[width=0.45\textwidth]{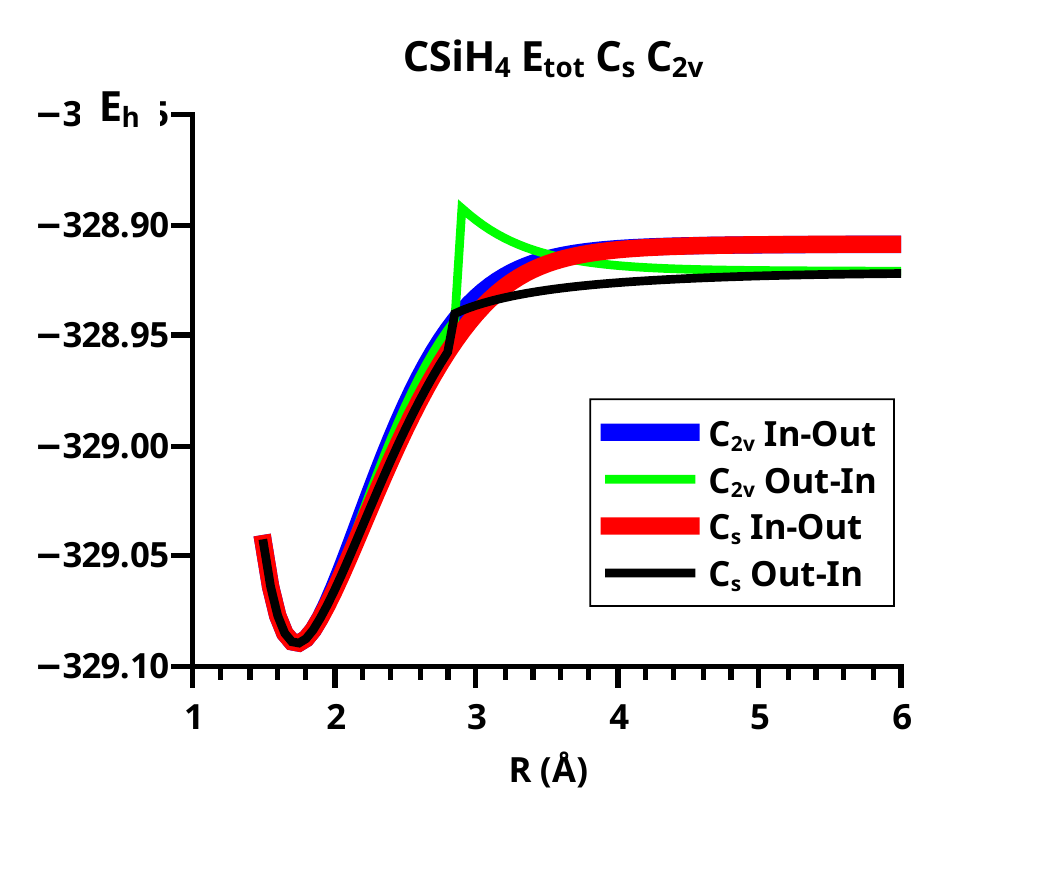}
\end{figure}

In contrast to disilene, where the difference between In-Out and Out-In disappears when the system becomes non-planar, this does not happen in case of silaethene. In-Out and Out-In occur still differently but there is no strong energy increase in case of Out-In (Figure \ref{fig:CsEtotsc}).

\section{What do the systems tell us?}
System stabilization is only possible for spatial arrangements of the interacting atoms that allow maximal interference. It is the fermionic character of electrons that is mainly responsible for such arrangements, because only when in each interacting fragment the spins of the active electrons are identical, the PEP is not operative between the fragments and they can come so close that interference of fragment states is possible. Besides these inter-fragment aspects, a high-spin arrangement has also an important intra-fragment effect: it favors a different position of the atoms surrounding the considered heavy atom, it causes changes of symmetry and strength of the perturbation potential, which eventually leads to a different hybridization and better energetic stabilization. From coordination chemistry, it is well known how strongly the electronic structure of a central atom depends on the symmetry and strength of the perturbation that is consists of the atoms positioned around the central atom. This interplay of spin rearrangement and molecular geometry
can be observed, for example, at inter-fragment distances where In-Out and Out-In reactions have similar energies and where the fragments will change from low-spin to high-spin if In-Out is energetically favored over Out-In. Or, it can be seen for decreasing inter-fragment distances when the molecular structure  changes from non-planar to planar. Good indicators that show such spin rearrangements are the geometry parameters of the fragments.

The magnitude of the projection probabilities obtained by an OVB analysis can show where a molecular state is a product of pure fragment states. All states of the dissociated systems are products of pure fragment states. If the system state is described by TT only, it is obvious that both fragments are in pure states. But also the dissociated system that consists of fragments in singlet states, is a product of pure fragment states, this show the weights of NB and DX+DX or of NB, DX(C) and DX(Si); these CSFs describe the coupled singlet states of carbene-like species.  The projection probabilities allow also to distinguish between ionic CSFs that are needed to describe delocalization and polarization and ionic CSF that are necessary to describe changes of the charge distribution from high to low molecular symmetry. Whereas the former describe covalent bonding, the latter change essentially the properties of the system, and enable, for example, the smooth dissociation of disilene by changing from planar to non-planar geometry; at large inter-fragment distances, the charge shifts described by these CSFs are obviously important to prepare the system for the spin rearrangements that occur at shorter inter-fragment distances. So, OVB CSFs are a means to analyse molecular states, they describe always coupled fragment states, therefore one cannot describe properties of fragments directly with these CSFs. If one wants  to know, for example, how the spin state of a fragment in the interacting complex changes, or the total energy of a fragment in the molecule, one has to construct the reduced density matrices for the fragments, with which one can calculate fragment properties. But this cannot be routinely done yet.

\subsection{Diabatic states}
The seemingly crossing of the In-Out and Out-In  energy curves for planar disilene, and planar and non-planar silaethene, is just an overlay of two curves that are the bottom lines of trough-like streambeds. Each reaction follows its own streambed, characterized by the molecular geometry. Another, more significant, characterization of each reaction is its different chemical characteristics. The In-Out state is dominated by  TT, the Out-In state is dominated by  NB. Geometry parameters of the fragments and the information about the local spin arrangement perfectly agree. Where TT dominates, the interacting fragments have geometries close to the geometries of non-interacting triplet species, the same holds where NB dominates. When the system follows the Out-In streambed, the geometry does not significantly  change, it is always the geometry of two fragments in their lowest singlet states. One can even follow the Out-In  streambed to inter-fragment distances where the energy of the In-Out state  is much lower, but this cannot be done to arbitrarily short distances; at a certain point the system changes to the streambed describing the dissociation.   Following the In-Out streambed, the molecular geometry resembles always two coupled fragment triplets and this geometry does not change when the fragments move apart. In contrast to Out-In, where the reduction of the inter-fragment distance increases the inter-fragment interaction and causes eventually the jump to the lower lying streambed, the inter-fragment interaction in In-Out goes to zero and therefore the geometry of the dissociated system is retained. The states describing  In-Out and Out-In are diabatic states, according to the classification by Atchity and Ruedenberg: ``..in certain regions of coordinate space, drastic changes occur in the electronic structures of the adiabatic states[...] the construction of diabatic states is guided by the goal of finding wave functions whose electronic structures maintain their essential characteristics over the entirety of such regions.''\cite{Atchity1997} The essential characteristics of the In-Out state is the dominance of coupled triplets, the essential characteristics of the Out-In state are two coupled singlets, each characteristics describes a different reaction behavior, and, therefore, I prefer to speak of the chemical characteristics of these states. If the two streambeds coalesce, one gets one adiabatic state leading from the stable molecule to the dissociation products in their lowest dissociation asymptote.  This is what one finds for non-planar disilene, but also for ethene. Here, the adiabatic state is also a diabatic state, according to the criterion of constant chemical characteristics.
If the streambeds do not coalesce there is a ridge between them, which must be surmounted, the switch of the streambeds is like changing horses in full gallop. The height of the ridge is not known, in a real reaction, the needed energy to jump over the ridge will come from internal degrees of freedom, e.g., vibrations, internal rotations or translations. Note, that this diabatic behaviour of In-Out and Out-In is found due to the constraints during calculation. If one wants to calculate diabatic states for  larger intervals of the reaction coordinate, one has make sure that the optimized state is kept orthogonal to lower lying states of the same multiplicity.

\subsection{Entanglement}
We assume that reactants, when they are far apart, are in pure states, then also the state of the molecular system is in a pure product state. With decreasing inter-fragment distance  and increasing inter-fragment interaction, the fragments become entangled; entanglement is the result of interactions between the subsystems. All molecular systems, in which reactions occur, are in entangled states, and the state describing the molecular system contains more information than a simple product of states of non-interacting subsystems. This information can be absorbed into the reduced density operator describing the subsystem, every subsystem in an entangled system is in a mixed state. The concept of entanglement was introduced by Schr\"odinger\cite{Schroedinger1935a,Schroedinger1935b} in 1935, in response to the paper by Einstein, Podolsky, and Rosen.\cite{EPR1935}In this paper, he claimed already in the second sentence "I would not call that [entanglement of states] \emph{one} but rather \emph{the} characteristic trait of quantum mechanics, the one that enforces its entire departure from classical lines of thought."\cite{Schroedinger1935a} Nevertheless, entanglement was for a long time not considered as important enough to be discussed in basic physics courses, and even today, it is dominantly mentioned in connection with quantum informatics, although any kind of interaction in chemistry and biology causes entanglement of quantum states and is therefore of utmost importance. In these branches of science, researchers are becoming increasingly, albeit gradually, aware of the importance of this aspect.\cite{Saxd}
OVB CSFs contain already information about the interaction of fragments in the LFOs, but calculation of physical quantities of a certain fragment can only be done with the reduced density matrix for this fragment. Quantities of interest could be for example the energy, the spin, or diverse electric moments; and monitoring these quantities as a function of the reaction coordinate could help to understand, ``what [really] happens to molecules, as they react''.\cite{Shaik1981}

\section{Discussion}
Understanding chemical bonding means understanding all processes that contribute to the stabilization of a molecular system. The most important quantum particles involved in these processes are electrons, which are charged fermions. Fundamental for the electron structure of atoms and molecules is the PEP, which prevents identical electrons from coming close, and which has the effect `` of a fictitious, highly effective, mutual repulsion being exerted within the system, irrespective of any other actual interactions that might be present'',\cite{Levy1990} sometimes  called ``Pauli repulsion''. It is  responsible for the shell structure of atoms, for the local spin arrangements in atoms and molecules, related to concepts like Hund's rule, dynamic spin polarization, or spin waves. And it is mainly responsible for the spatial arrangement of fragments in chemical reactions. Although Maynau \etal{} discussed in their 1983 paper spin waves in molecules with conjugated $\pi$ systems, the title of their paper is programmatic: ``Looking at Chemistry as a Spin Ordering Problem''.\cite{Maynau1983} Understanding chemical bonding means also to critically question traditional concepts and their use to explain chemistry. For example, what causes hybridization when the perturbation and its symmetry is never mentioned, which is necessary to destroy the spherical symmetry of atoms and to enable superposition of orthogonal eigenfunctions of the angular momentum operator. Is it possible to make meaningful statements about hybridization of atoms from period 3 or higher of the periodic table, unless one knows of and considers the different structures of the valence shell in period 2 and in all other periods of the periodic table? It is necessary to know that there is an interplay of PEP, hybridization as a result of perturbation, the position of an element in the periodic table, and the energetic stabilization caused by interference. The projection probabilities of the OVB CSFs are the most important source of information for changes in the spin distribution, every change of the local spin state correlates with characteristic changes of geometry data, especially bond angles. And changes in the multiplicity cause a change of the spatial extension of the electron distribution, which in turn changes characteristically the energy of the CSFs. But many more aspects must be thoroughly investigated before chemical bonding is fully understood.   For example, what triggers the change from low-spin to high-spine when fragments come close? Is it the reduction of the extension of the spatial domain lone pair electrons occupy? What indicators could be helpful to answer this question? Could the old idea of Odiot and Daudel\cite{Odiot1954} be helpful that there is an inverse relationship between the mean space  an electron can occupy in such a  domain and the (repulsive) electrostatic potential in this domain? One could assume that, if two doubly occupied lone pairs come too close, the PEP forces local spin flips which reduces the Pauli repulsion between the lone pairs. Diradicals are systems where inter-system crossing is a well known mechanism for a spin flip. For a change from spin state $S = \ket{\alpha\beta}+\ket{beta\alpha}$ to the triplet component $T_0=\ket{\alpha\beta}-\ket{beta\alpha}$ a change of the relative phase is necessary. Note, that both states are strongly entangled spin states. Using the simple model of a ``spin clock'', Turro and Kr\"autler\cite{Turro1982} discussed intersystem crossing in diradicals and radical pairs. Is simultanoeus intersystem crossing in two entangled diradical fragments a possible answer to the question how the change from coupled singlets to coupled triplets occurs? I am sure that the investigation of maximum probability domains and the calculation of fragment properties like the total energy, the total spin or electric multipole moments will be useful for a better understand of covalent bonding. Maybe the total energy of a fragment can also be a sound basis for a definition of the famous promotion energy.
All this should be considered, but the PEP ``reigns supreme''.\cite{Levy1990} Any concept that claims to be able to reveal major aspects of chemical bonding but does not acknowledge the outstanding role of the Fermi character of electrons must be considered questionable. To quote one more Lennard-Jones: ``Its [the Pauli principle's] all-pervading influence does not seem hitherto to
have been fully realized by chemists, but it is safe to say that ultimately it will be regarded as the most important property to be learned by those concerned with molecular structure.''\cite{Lennard1954} Unfortunately one has to say, here erred Lennard-Jones.

\section{Appendix}
This is a math review of sloppy rigor in the spirit of the mathematics chapter in CT.

\subsection{Vector spaces}
A vector space $\mfV$ consists of the Abelian group of vectors $\mbv$, with the vector addition as group operation, and a field \bF, in general either the field of real numbers \bR or the field of complex numbers \bC; the elements of the field are the \emph{scalars}. The conjugate complex of number $z$ is indicated as $z^*$. Depending on the field one speaks of either real or complex vector spaces. The operation connecting scalars and vectors is called multiplication of a vector by a scalar, any sum of vectors that are multiplied by a scalar, $\alpha \mba + \beta \mbb + \gamma \mbc$ with  $\mba, \mbb, \mbc \in {\mfV}$ and  $\alpha, \beta,\gamma \in \bF$,   is called a \emph{linear combination} of the vectors. Every finite linear combination of vectors is element of $\mfV$. A subset $\mfU$ of $\mfV$ having all properties of a vector space is a \emph{subspace} of $\mfV$. A set of vectors, without the null vector, is called \emph{linearly independent} if the only linear combination of these vectors giving the null vector is the one where all scalars are zero. If at least two scalars are not zero the set is called \emph{linearly dependent}.
\subsubsection{Basis}
Given a set $B=\{\mbb_1,\mbb_2,\dots\}$ of linearly independent vectors of $\mfV$, if every vector $\mbv$ can be written as linear combination of the vectors of $B$, this set is called a \emph{basis} for $\mfV$. The number of vectors in $B$ is the \emph{dimension} $\di \mfV$ of the vector space, the dimension of a vector space can be \emph{finite}, \emph{countably infinite}, or  \emph{uncountable}. The set of basis vectors can be designated $B=\{\mbb_i\}_{i\in I}$ where $I$ is an index set, this is called a family of vectors, meaning to each element $i\in I$ belongs exactly one element of $B$. In case of finite dimensions the index set is mostly a subset of non-negative integers like $\{1,2,\dots,N\}, N<\infty$; in case of countably infinite bases it is the whole set of non-negative integers \bN; for uncountable bases it is an interval of the real numbers.
If every vector $\mbv\in\mfV$ can be uniquely described as linear combination of the basis vectors, $\mbv = \sum_{i\in I} v_i\mbb_i$ with $v_i \in \bF$, the basis is called (algebraically) \emph{complete}. The set of numbers $v_i$ (coefficients of the linear combination) is called the $\mbb$-representation of $\mbv$.
\subsubsection{Direct sum of subspaces}
Any proper subset $S\subset B$ is basis of a subspace ${\mfU}_S$ of $\mfV$ having dimension $\di {\mfU}_S$. The complement $C$ of $S$ in $B$, $C=B\setminus S$, is basis of the subspace ${\mfU}_C$  with dimension $\di {\mfU}_C$; both subspaces have only the null vector in common, ${\mfU}_S \cap {\mfU}_C =\{\nullv\}$. The basis of $\mfV$ is the union of the bases for the subspaces, $B=S\cup C$, and $\di {\mfV} = \di {\mfU}_S + \di {\mfU}_C$. Every vector $\mbv\in \mfV$ is the sum of two \emph{components}, one from each subspace, $\mbv = \mbv_S+\mbv_C$.
One says, the vector space $\mfV$ is the direct sum ${\mfU}_S\oplus{\mfU}_C$ of the two subspaces ${\mfU}_S$ and ${\mfU}_C$. This can be generalized; any family of pairwise disjoint subsets $C_{i\in I}$ of $B$, with a corresponding index set $I$, induces a decomposition of $\mfV$ into a family of subspaces ${\mfU}_{i\in I}$, so that $\mfV$ is the  direct sum of the subspaces, ${\mfV} = \bigoplus_{i\in I} {\mfU}_i$; one can also write ${\mfV} = \sum_{i\in I} {\mfU}_i$ and $B = \bigcup_{i\in I} C_i$. The decomposition of a vector $\mbv$ into components is written as $\sum_{i\in I} \mbv_i$ with $\mbv_i \in {\mfU}_i$. Adding vectors in $\mfV$ is done by adding the components in each subspace and then adding these components.

\subsubsection{Linear mapping}
Given two vector spaces $\mfV$ and $\mfW$. A mapping $f: \mfV \rightarrow \mfW$ is \emph{linear}, if the image of a linear combination of vectors ist the linear combination of the images of the vectors
\begin{displaymath}
f(\alpha \mba + \beta \mbb + \gamma \mbc) = \alpha f(\mba) + \beta f(\mbb) + \gamma f(\mbc)
\end{displaymath}
it is \emph{antilinear} if
\begin{displaymath}
f(\alpha \mba + \beta \mbb + \gamma \mbc) = \alpha^* f(\mba) + \beta^* f(\mbb) + \gamma^* f(\mbc)
\end{displaymath}
Linear mappings retain the vector space structure. If mapping $f$ is bijective, it is an isomorphism, then vector spaces $\mfV$ and $\mfW$ are identical except for the choice of symbols. Then the vector spaces are isomorphic and can be identified.
\subsubsection{Scalar product}
A vector space can be endowed with a \emph{scalar product} (\emph{dot product, inner product}), which maps two vectors into the scalar field \bF. In real vector spaces, the scalar product is a bilinear form $\phi(\mba,\mbb)$, which is linear in both arguments, in complex vector spaces it is a sesquilinear form, which is linear in one argument and antilinear in the other. A vector space with a scalar product is called a \emph{scalar product space}. With the help of a scalar product it is possible to introduce the concept of an angle between two vectors. If the scalar product of a pair of vectors, neither of them is the null vector, is zero, the two vectors are   \emph{orthogonal}. The scalar product of a vector with itself is always a positive real number, the square root of it can be used to define a \emph{norm} in this vector space, which allows to attribute a length to each vector. A vector having length one is a \emph{normalized} vector. From the norm one can derive a \emph{metric} and thus a distance between vectors. A metric is necessary for the treatment of sequences and series.
In every scalar product space it is possible to choose a basis consisting of normalized vectors that are orthogonal. Such a basis is called an \emph{orthonormal} basis (ONB), ONBs are convenient to use and they allow interpretations non-orthogonal bases do not allow.
Orthogonality and normalization of an ONB can be expressed by using the Kronecker delta, which is defined as follows: $\delta_{i,j}$ is 1  if $i=j$, it is 0, if $i\ne j$. For an ONB $B=\{\mbb_i\}_{i\in I}$ one can thus write $\overl{\mbb_i}{\mbb_j} = \delta_{i,j}$.  With orthogonalization procedures it is possible to create orthogonal vectors from any set of linearly independent vectors.

\subsubsection{Dual space}
A \emph{linear functional} $f$ is a linear map from a vector space into the scalar field. The set of linear functionals $f:\mfV\rightarrow \bF$ constitutes a vector space  $\mfV^*$ called the \emph{dual space} of $\mfV$, the elements of it are called \emph{dual vectors} and designated by $\mba^*$  and instead of $\mba^*(\mbb)$ one can write $\overl{\mba^*}{\mbb}$. If $B=\{\mbb_i\}_{i\in I}$ is a basis of $\mfV$, not necessarily an ONB,  a set of dual vectors $B^* = \{\mbb^*\}$ is defined by $\overl{\mbb^{'*}}{\mbb}= \delta_{\mbb',\mbb}$, meaning $\mbb^*$ maps only basis vector $\mbb$ onto one, all other basis vectors onto zero.  $B^*$ is linearly independent in $\mfV^*$ but it is only a basis in $\mfV^*$ if $\mfV$ is finite dimensional. Then $\mfV$ and $\mfV^*$ are isomorphic.
The scalar product in a vector space, $\phi(\mba,\mbb)$, can be interpreted as linear mapping of vector $\mbb$  under the linear functional $\mba$, that is $\phi(\mba,\mbb)=\mba(\mbb)$.

\subsection{Direct sum of vector spaces}
Given a family of different vector spaces  ${\mfV}_{i\in I}$, each endowed with its own basis $B_i = \{\mbb^{(i)}_j\}_{j\in I^{(i)}}$, one can create a large vector space, called the \emph{direct sum} of the family of vector spaces,  ${\mfV} = \bigoplus_{i\in I} {\mfV}_i$.  The elements $\mbv$ of this vector space are  n-tuples of vectors; the addition and multiplication with a scalar are defined component-wise:
\begin{displaymath}
\mbv+\mbw := (\mbv_i+\mbw_i)_{i\in I},\qquad \alpha\mbv := (\alpha \mbv_i)_{i\in I}
\end{displaymath}
It is  assumed that there is always only a finite number of summands not the null vector. The dimension of the direct sum is the sum of the dimensions of the individual vector spaces, $\di \bigoplus_{i\in I} {\mfV}_i = \sum_{i\in I}\di {\mfV}_i$.

If each vector space has its own scalar product, $({\mfV}, \overl{}{})_{i\in I}$, the scalar product in $\mfV$  is defined as
$\overl{\mbv}{\mbw}:= \sum_{i\in I} \overl{\mbv_i}{\mbw_i}_i$.

\subsubsection{Embedding}
The scalar product in the direct sum does not say anything about orthogonality of vectors from different vector spaces. If this is necessary, one has to find an embedding of the direct sum into another vector space. Given a vector space $\mfW$ with basis $B_W$ and the same dimension as the direct sum $\mfV$. If there is a family of subspaces ${\mfU}_{i\in I}$ of $\mfW$ and bijective linear mappings between ${\mfV}_i$ and ${\mfU}_i$ for all $i\in I$ so that ${\mfV}_i$ and ${\mfU}_i$ are isomorphic, then $\mfW$ and ${\mfV}=\bigoplus_{i\in I} {\mfV}_i$ are isomorphic and can be identified. But then every vector $\mbv = \{\mbv_i\}_{i\in I}$ of $\mfV$ can be written as a sum of components from the vector spaces ${\mfV}_i$,  $\mbv = \sum_{i\in I}\mbv_i$. The union $\tilde{B}_W = \bigcup_{i\in I} B_i$ of the pairwise disjoint bases $B_i$ of the vector spaces ${\mfV}_i$ is a basis of $\mfW$, in general, different from basis $B_W$.

One can define in $\mfW$ a scalar product $\overl{}{}_W$ so that for each $i\in I$ $\overl{\mbv_i}{\mbw_i}_W = \overl{\mbv_i}{\mbw_i}_i$, and with this scalar product one can also check orthogonality of vector that are elements of different vector spaces, because $\overl{\mbv_i}{\mbw_j}_W$ is defined for all vectors of $\mfW$. And it may turn out that basis vectors from different vector spaces are not orthogonal, $\overl{\mbb^{(i)}_k}{\mbb^{(j)}_l}_W \ne 0$ for $i\ne j$, independently of $k$ and $l$, then $\tilde{B}_W $ is no orthogonal basis in $\mfW$.

\subsection{Hilbert space}
Given a scalar product space $\mfV$ with a norm derived from the scalar product, and a metric defined with this norm. All elements of  $\mfV$ are vectors with finite norm. If with respect to this metric  the limit of every Cauchy sequence of vectors is element of the vector space, the vector space is (metrically) \emph{complete} and called a \emph{Hilbert space}, designated $\mfH$. Topological properties of Hilbert spaces will not be discussed, only some algebraic aspects. Sometimes, only infinite-dimensional metrically complete vector spaces with scalar product are called Hilbert spaces. Frequently, the Hilbert space of square integrable functions ($L^2$) is called THE Hilbert space, which has the property that it is isomorphic to it dual space. But this is only true if the vector space is countably infinite. A complete ONB of a Hilbert space is a \emph{Hilbert basis}.
Every Hilbert space has an infinite number of different bases both orthonormal or non-orthonormal. In the following, $\{\ket{i}\}_{i\in I}$ will denote an arbitrary Hilbert basis.

\subsubsection{Dirac's bra-ket notation}
Every element of the vector space is denoted by a ket $\ket{}$, every element of the dual vector space by a bra $\bra{}$. The scalar product in $\mfH$, interpreted as the mapping of a ket on the scalar field by means of a bra is written as $\overl{a}{b}$ with $\overl{b}{a} = \overl{a}{b}^*$. Note the difference: $\bra{a}$$\ket{b}$ are simply products of a bra and a ket whereas $\overl{a}{b}$ means a scalar product.

Every ket $\ket{X}$ of a Hilbert space can be represented in a unique way as a linear combination of a Hilbert basis $\{\ket{b_i}\}_{i\in I}$
\begin{displaymath}
\ket{X} = \sum_{i\in I} c_i \ket{b_i} =  \sum_{i\in I} \overl{b_i}{X} \ket{b_i},
\end{displaymath}
called a  \emph{superposition} of the basis kets.
In quantum theory, the scalar product $ \overl{b_i}{X}$  is called the quantum amplitude or the probability amplitude.

From now, vectors of a vector space are written as kets and dual vectors as bras without boldface symbols.
\subsubsection{State space}
Eigenstates of measurable physical quantities can only be represented by vectors with finite norm, they are elements of the \emph{state space} $\BPhi$ of the system, which is a subspace of a complex Hilbert space $\mfH$, in general a Schwartz space, its elements are complex continuous functions that are infinitely differentiable and rapidly decreasing. However, many problems need considerably modified vector spaces, e.g., having elements that are vectors with infinite norm, such as eigenstates of the position or the momentum operator or plane waves. The dual vectors of such functions are  distributions or generalized functions. But distributions are linear functionals so they are elements of the dual space $\BPhi^*$ of the state space, therefore $\BPhi\subset \BPhi^*$. To avoid the mathematically correct but inconvenient asymmetry\cite{Saxa} between these two spaces, both are assumed to be equal. Vectors with infinite norm are called generalized kets, they never represent physical states.

\subsection{Operators}
Linear operators are linear mappings from a vector space onto itself. All kets are assumed to be normalized. Using the bra-ket formalism, linear operators on a Hilbert space can be written as so called \emph{outer products} of kets and bras $\proj{a}{b}$. Acting on an arbitrary ket $\ket{c}$, $\proj{a}{b}$ gives $\ket{a}$ multiplied with the scalar product of $\ket{c}$ and $\ket{b}$,   $\proj{a}{b}\ket{c} = \ket{a}\overl{b}{c}$. A projector onto  ket $\ket{a}$ reproduces $\ket{a}$ without modification, it has the form $\sfbP_a = \proj{a}{a}$. Every projector is idempotent meaning
\begin{displaymath}
\sfbP_a \sfbP_a  = \proj{a}{a} \proj{a}{a} = \ket{a}\overl{a}{a} \bra{a} = \proj{a}{a} = \sfbP_a
\end{displaymath}
\subsubsection{Hermitian operators}
To each linear operator $\sfbA$ acting on a Hilbert space $\mfH$ corresponds an adjoint operator $\sfbA^{\dagger}$, acting on the dual space. If $\sfbA\ket{a} = \ket{b}$ then also $\bra{a}\sfbA^{\dagger} = \bra{b}$. If $\sfbA=\proj{a}{b}$ then $\sfbA^{\dagger}=\proj{b}{a}$.  If $\sfbA=\sfbA^{\dagger}$ the operator is \emph{selfadjoint} or \emph{Hermitian}. Every projector is Hermitian. Linear combinations of linear operators are again linear operators, e.g., $\alpha_1 \proj{a}{b} +\alpha_2\proj{c}{d}$, the same is true for the adjoint operators, and thus for all Hermitian operators.

Linear combinations of projectors with real coefficients are Hermitian $c_1 \proj{a}{a}+c_2\proj{b}{b}$, $c_1, c_2\in \bR$ but no projectors;  linear combinations of the kind $c \proj{a}{b}+c^*\proj{b}{a}$, $c\in \bC$ are Hermitian.
\subsubsection{Unit operator}
An important operator is the sum of all projectors on the basis kets of an arbitrary Hilbert basis, $\sum_i \proj{b_i}{b_i}$, omitting the index set $I$. When this operator acts on a ket $\ket{X} = \sum_i \overl{b_i}{X}\ket{b_i}$, the ket remains unchanged
\begin{displaymath}
\left(\sum_i \proj{b_i}{b_i}\right)\ket{X} = \left(\sum_i \proj{b_i}{b_i}\right) \sum_j \overl{b_j}{X}\ket{b_j} = \sum_i\sum_j\ket{b_i}\overl{b_i}{b_j}\overl{b_j}{X} = \sum_i\ket{b_i}\overl{b_i}{X}=\ket{X}
\end{displaymath}
therefore this operator is called the \emph{identity} or \emph{unit operator} denoted by \unit. Whenever an arbitrary ONB $\{\ket{i}\}_{i\in I}$ is used, one can write $\unit = \sum_{i\in I} \proj{i}{i}$ or in short  $\unit = \sum_i \proj{i}{i}$.


\subsection{Eigenvalues, eigenkets, expectation value}
\subsubsection{Eigenvalue equation}
An equation $\sfbA\ket{a_i} = a_i\ket{a_i}$ for a Hermitian operator $\sfbA$ and a ket $\ket{a_i}$ is called  \emph{eigenvalue equation} of $\sfbA$,  the eigenvalues $a_i$ are always real and the corresponding eigenkets $\ket{a_i}$ are orthogonal; every set of normalized eigenkets of a Hermitian operator is called a \emph{complete set of orthonormal states} (CSOS), it is an ONB. If $\ket{a_i}$ is an eigenket of $\sfbA$, all multiples $\lambda\ket{a_i}$, $\lambda \in \bF$, are eigenkets to the same eigenvalue, they span a one dimensional subspace of $\mfH$ called the \emph{eigenspace} generated by $\ket{a_i}$.

If $g_k$ different eigenvectors have the same eigenvalue $a_k$, the eigenvalue is $g_k$-fold \emph{degenerate}, the number $g_k$ is the \emph{degeneracy} of the eigenvalue.

Every Hermitian operator can be written as the sum of the projectors onto the eigenkets multiplied by the corresponding eigenvalues
\begin{displaymath}
\sfbA =\sum_i a_i\proj{a_i}{a_i}.
\end{displaymath}
This is the \emph{spectral decomposition} of $\sfbA$.

The sum of the projection operators onto the eigenkets is always the unit operator in the corresponding Hilbert space.

\subsubsection{Expectation value}
The number $\brac{X}{\sfbA}{X} = \erw{\sfbA}_X$ with an arbitrary ket $\ket{X}$ is called the \emph{expectation value} of $\mcA$ in the state $\ket{X}$. Measurement of $\mcA$ in state $\ket{X}$ gives one of the possible values $a_i$ with probability $P(a_i \!\leftarrow \!\!X)$. Repeated measurement in the same initial state gives the mean value
\begin{displaymath}
\sum_i a_i P(a_i \!\leftarrow \!\!X) = \sum_i a_i |\overl{a_i}{X}|^2 = \sum_i a_i \overl{X}{a_i} \overl{a_i}{X} = \bra{X}\left( \sum_i a_i \proj{a_i}{a_i}\right) \ket{X} = \brac{X}{\sfbA}{X}
\end{displaymath}
If the ket $\ket{X}$ is an eigenket of $\sfbA$, the probability is one and the expectation value is the corresponding eigenvalue.

\subsection{Matrix representation of operators}
With the unit operator, one gets easily  \emph{matrix representations} of any operator for an arbitrary basis $\{\ket{i}\}_{i\in I}$
\begin{displaymath}
\sfbA = \unit\sfbA\unit = \sum_i\sum_j \proj{i}{i} \sfbA \proj{j}{j} = \sum_i\sum_j \brac{i}{\sfbA}{j} \proj{i}{j}=
\sum_i\sum_j A_{ij}  \proj{i}{j}
\end{displaymath}
The numbers $A_{ij}$ are called the \emph{matrix elements} of $\sfbA$ in the $\ket{i}$ basis, with $A_{ji} = A^*_{ij}$; the matrix $\mbA = (A_{ij})$ is the matrix representation of operator $\sfbA$ in basis $\ket{i}$. The diagonal matrix elements $A_{ii}$ are the expectation values of $\sfbA$ in the basis kets $\ket{i}$.

Given $\ket{X}=\sum_i \overl{i}{X} \ket{i} = \sum_i c_i \ket{i}$,  the matrix elements of the projector $\sfbP_X = \proj{X}{X}$ are
\begin{displaymath}\begin{split}
\unit\sfbP_X\unit = \sum_i \proj{i}{i} \sum_l \sum_k c_l c_k^* \proj{l}{k} \sum_j \proj{j}{j} = \\\sum_i\sum_j\sum_k\sum_l c_l c_k^*
\ket{i}\overl{i}{l}\overl{k}{j}\bra{j} = \sum_i\sum_j c_i c_j^* \proj{i}{j}
\end{split}
\end{displaymath}
with matrix elements $(P_X)_{ij} = c_i c_j^*$.

\subsection{Trace of an operator}
The trace of an operator $\sfbA$ is an invariant of the operator, meaning that it does not depend on the basis used for representation. For a given orthonormal basis $\{\ket{i}\}_{i\in I}$ it is defined as
$\Tr\sfbA = \sum_i \brac{i}{\sfbA}{i} = \sum_i A_{ii}$, that is the sum of the diagonal elements of the representation matrix.

Inserting the unit operator wrt another basis shows the invariance
\begin{displaymath}
\begin{split}
\sum_i \brac{i}{\sfbA}{i} =  \sum_i \brac{i}{\unit\sfbA}{i} =  \sum_i \brac{i}{\left(\sum_j \proj{j}{j}\right) \sfbA}{i} = \sum_i\sum_j \overl{i}{j} \brac{j} {\sfbA}{i} = \\ \sum_i\sum_j \brac{j}{\sfbA}{i}\overl{i}{j} =  \sum_j \brac{j} {\sfbA\left(\sum_i\proj{i}{i}\right)}{j} = \sum_j \brac{j}{\sfbA\hat{1}}{j} = \sum_j \brac{j}{\sfbA}{j}
\end{split}
\end{displaymath}

If  $\sfbA$ is a Hermitian operator and $\{\ket{a_i}\}_{i\in I}$ is the basis of eigenkets of $\sfbA$, the trace is the sum of the eigenvalues
$\Tr\sfbA = g_i a_i$, where $g_i$ is the degeneracy of the eigenvalue $a_i$.

Inserting the unit operator twice into the expectation value $\erw{\sfbA}_X$  gives
\begin{displaymath}
\brac{X}{\unit\sfbA \unit}{X} = \sum_i \overl{X}{i} \bra{i} \sfbA  \sum_j \ket{j} \overl{j}{X} = \sum_i \sum_j c_j c_i^* \brac{i}{\sfbA}{j} = \sum_i\sum_j (P_X)_{ji} A_{ij}
\end{displaymath}
$\sum_i (P_X)_{ji} A_{ij}$ is the trace of the matrix product $\mbP_X \mbA$, denoted by $\Tr \mbP_X \mbA$. But $\sum_i\sum_j (P_X)_{ji} A_{ij}$ is $\sum_i (A P_X )_{ii} =\Tr \mbA \mbP_X$.

\subsection{The density operator}
The density operator of a pure state $\ket{X}$ is the projector $\proj{X}{X}$. But most physical states are not pure states but  mixed states. The density operator of a mixed state is a statistical mixture of projection operators onto pure states multiplied by probabilities
\begin{displaymath}
\Brho= \sum_i w_i \proj{i}{i}\qquad \text{with}\qquad  0\le w_i \le 1, \qquad \text{and}\qquad\sum_i w_i = 1
\end{displaymath}
Since the trace of a projector onto a pure state is one, the trace of the density operator describing a mixed state is the sum of the probabilities $\Tr \Brho = \sum_i w_i \Tr  \proj{i}{i} = \sum_i w_i =1$. But $\Brho$ is not idempotent, because
$\Tr \Brho^2 =  \sum_i w_i^2 < 1$. This is a test for a quantum state of being pure or mixed.

\subsection{Position representation, wave functions}
There is no possibility to localize any quantum object in physical space exactly on a certain position with position vector $\mbr$. Therefore, the eigenvalue equation $\hat{R}\ket{\mbr} = \mbr\ket{\mbr}$  of the localization operator $\hat R$ does not have eigenkets $\ket{\mbr}$ with finite norm in the state space. The eigenvalues $\mbr$ are elements of a continuous subset $D$ of the $\bR^3$, and accordingly the eigenkets $\ket{\mbr}$ must have three continuous indices. The orthonormality condition reads for these kets  $\overl{\mbr_0}{\mbr} = \delta(\mbr_0-\mbr)=\delta(x_0-x)\delta(y_0-y)\delta(z_0-z)$, $\delta(\mbr_0-\mbr)$ is Dirac's delta function, a distribution. The probability amplitude for the projection of an arbitrary $\ket{X}$ onto $\mbr_0$, that is for the localization of $\ket{X}$ on $\mbr_0$, is the complex number $\overl{\mbr_0}{X}$, the set of the probability amplitudes for all position vectors in $D\subset \bR^3$ is a continuous mapping of $D$ into \bC, that is a complex valued function of the domain $D$, written as $X(\mbr) = \overl{\mbr}{X}$ and called the wave function of $\ket{X}$.  It is the position representation of $\ket{X}$. Aficionados of the Greek letter $\psi$ prefer to write $\psi(\mbr)$ or $\psi_X(\mbr)$ instead of $X(\mbr)$.

\subsection{Tensor products of vector spaces}
\subsubsection{Motivation}
In most physical applications, the vector spaces considered are Hilbert spaces. If the enlargement of a system is accompanied by the increase of the degrees of freedom needed to describe the system, different vector spaces must be combined to a larger vector space, called a \emph{tensor product} of vector spaces\cite{Saxb}, which is again a vector space. The individual vector spaces can be identical or completely different. For example, wave functions in one variable, e.g., $\psi(x)$, allow to describe the motion of a quantum object along the x-axis, they are elements of a Hilbert space $\mfH(x)$. To describe the motion in a plane, the new degree of freedom, motion along the y-axis, must be accounted for by elements of the Hilbert space $\mfH(y)$. The Hilbert space describing the motion in the xy-plane is the tensor product  $\mfH(xy) = \mfH(x)\otimes \mfH(y)$. For a free quantum object, both factor spaces are isomorphic. To describe motion in physical space, a third degree of freedom has to be added and the Hilbert space is  $\mfH(\mbr) = \mfH(x)\otimes \mfH(y)\otimes \mfH(z)$. If the quantum object has a spin, another (internal) degree of freedom must be considered; the corresponding spin operator acts on spin states, which are elements of a spin Hilbert space $\mfH(s)$. The Hilbert space of a free quantum object with spin is then the tensor product  $\mfH(\mbr)\otimes\mfH(s)$.
\subsubsection{Definition}
Only algebraic properties of tensor products are considered. Given three vector spaces ${\mfV}(1)$, ${\mfV}(2)$ and $\mfV$.   If every pair of vectors $(\ket{\psi(1)}, \ket{\phi(2)})$ with  $\ket{\psi(1)}\in {\mfV}(1)$ and  $\ket{\phi(2)}\in{\mfV}(2)$, is mapped onto a vector $\ket{\psi(1)}\otimes \ket{\phi(2)}$ of ${\mfV}$ using a bilinear map, this vector is called the \emph{outer product of the vectors} $\ket{\psi(1)}$ and $\ket{\phi(2)}$, and the vector space $\mfV$ is called the tensor product of  the ``factor  spaces''\cite{Castillo2005} ${\mfV}(1)$ and ${\mfV}(2)$.\cite{Saxe} As a bilinear map, the outer product of vectors has the following properties: it is linear in the first factor and it is linear in the second factor.
\begin{gather*}
\TP{\bigl(\alpha\ket{\phi(1)}+ \beta\ket{\phi(1)}\bigr)}{ \ket{\psi(2)}} = \alpha\TP{\ket{\phi(1)}}{\ket{\psi(2)}}+ \beta\TP{\ket{\phi(1)}_2}{\ket{\psi(2)}}\\
\TP{\ket{\phi(1)}}{\bigl(\alpha\ket{\psi(2)}+ \beta\ket{\psi(2)}\bigr)} = \alpha\TP{\ket{\phi(1)}}{\ket{\psi(2)}}+ \beta\TP{\ket{\phi(1)}}{\ket{\psi(2)}}
\end{gather*}
In physics, outer products of vectors $\TP{\ket{\phi(1)}}{\ket{\psi(2)}}$ are either written without the outer product symbol, that is $\ket{\phi(1)}\ket{\psi(2)}$, or as $\ket{\phi(1)\psi(2)}$ or simply as $\ket{\phi\psi}$.

\subsubsection{Bases}
If vector space ${\mfV}(1)$ has dimension $n_1=\di{\mfV}(1)$ and basis $B_1 = \{\ket{a(1)}_i\}_{i\in I_1}$ with $I_1 = [1,2,\dots,n_1]$, and vector space  ${\mfV}(2)$ has dimension $n_2=\di{\mfV}(2)$ and basis $B_2 = \{\ket{b(2)}_j\}_{j\in I_2}$ with $I_2 = [1,2,\dots,n_2]$, then is the set of outer products $B = \{\TP{\ket{a(1)}_i}{\ket{b(2)}_j}\} = \{\ket{c}_{(i,j)}\}_{(i,j)\in I_1\times I_2}$ a basis in ${\mfV}$, called the \emph{canonical basis}. The dimension of ${\mfV}$ is equal to $n_1n_2$, and $I_1\times I_2$ is the cartesian product of sets $I_1$ and $I_2$ with ordered pairs $(i,j)$ as elements.

If $B_1$ and $B_2$ are ONBs,  basis $B$ is also an ONB.
\begin{displaymath}
\overl{a_i(1) b_j(2)}{a_{i'}(1) b_{j'}(2)} = \overl{a_i(1)}{a_{i'}(1)} \overl{b_j(2)}{b_{j'}(2)} = \delta_{ii'} \delta_{jj'}
\end{displaymath}

The outer product of $\ket{\phi(1)} = \sum_{i=1}^{n_1} \alpha_i \ket{a(1)}_i$ and $\ket{\psi(2)} = \sum_{j=1}^{n_2} \beta_j \ket{b(2)}_j$ is
\begin{displaymath}
\TP{\ket{\phi(1)}}{\ket{\psi(2)}} = \sum_{i=1}^{n_1}\sum_{j=1}^{n_2}  \alpha_i \beta_j \TP{\ket{a(1)}_i}{\ket{b(2)}_j} =\sum_{ij\in I_1\times I_2}
\gamma_{(i,j)}\ket{c}_{ij}
\end{displaymath}
the coordinates of the outer product $\TP{\ket{\phi(1)}}{\ket{\psi(2)}}$ are the products of the coordinates of the two factors $\ket{\phi(1)}$ and $\ket{\psi(2)}$, $\gamma_{ij} = \alpha_i \beta_j$.

Since every vector of the tensor product ${\mfV}$ can be represented as a linear combination of outer products, it is in general not possible to write an arbitrary element of ${\mfV}$ as a single outer product.

\subsubsection{Scalar product}
If in both factor  spaces a scalar product is defined, the scalar product in the tensor product is defined as the product of the scalar products in the factor spaces
\begin{displaymath}
\overl{\phi(1)\psi(2)}{\phi'(1)\psi'(2)} = \overl{\phi(1)}{\phi'(1)} \overl{\psi(2)}{\psi'(2)}
\end{displaymath}

\subsubsection{Tensor product of operators}
If $\sfbA(1)$ is a linear operator acting on ${\mfV}(1)$ and $\sfbA(2)$ is a linear operator acting on ${\mfV}(2)$, then the tensor product of the operators $\sfbA(1)\otimes \sfbA(2)$ is acting on ${\mfV}(1)\otimes {\mfV}(2)$.
\begin{displaymath}
\bigl(\sfbA(1)\otimes \sfbA(2)\bigr) \bigl(\TP{\ket{\phi(1)}}{\ket{\psi(2)}}\bigr) = \bigl(\sfbA(1)\ket{\phi(1)}\bigr)\otimes \bigl( \sfbA(2)\ket{\psi(2)}\bigr)
\end{displaymath}

If $\sfbA(1)$ is acting on $\ket{\phi(1)}$ in ${\mfV}(1)$,  the operator $\sfbA(1)\otimes \unit(2)$, called the \emph{extension} of $\sfbA(1)$ in ${\mfV}$, is acting only on the factor space ${\mfV}(1)$  in ${\mfV}(1)\otimes {\mfV}(2)$. Similarly, $\unit(1)\otimes\sfbA(2)$ is the extension of $\sfbA(2)$ in ${\mfV}$.

In physics, tensor products of operators are written without the $\otimes$ symbol and the unit operator is omitted.

\subsubsection{Eigenvalues and eigenvectors of extended operators}
If $ \ket{a_k(1)}$ is an eigenstate of the Hermitian operator $\sfbA(1)$ acting on ${\mfV}(1)$  with a $g_k$-fold degenerate eigenvalue $a_k$
\begin{displaymath}
\sfbA(1) \ket{a_k(1)} = g_k a_k \ket{a_k(1)}, \quad k=1,2,\dots, n_1
\end{displaymath}
then has the extension of $\sfbA(1)$ in ${\mfV}$ the same eigenvalue. However, because all vectors of the form $\ket{a_i(1)}\otimes \ket{b(2)}$ with arbitrary factor $\ket{b(2)}$ are  eigenvectors to eigenvalue $a_k$, the eigenvalue $a_k$ is $(g_k n_2)$-fold degenerate in  ${\mfV}$, if $n_2$ is the dimension of ${\mfV}(2)$. The same holds for eigenvalues and eigenvectors of $\sfbA(2)$ acting on ${\mfV}(2)$.

\subsubsection{Density operators in tensor products, reduced density operators}
The concept of a partial trace  can be seen as the inverse of the construction of an extended operator: From an operator acting on the tensor product we want to derive operators that act only on the factor spaces.

If a state of a composite system is represented by a outer product $\ket{12}=\ket{\phi(1)}\otimes \ket{\psi(2)}$,  the density operator is
\begin{displaymath}
\Brho = \proj{12}{12} = \proj{1}{1}\otimes \proj{2}{2} =\Brho(1)\otimes\Brho(2)
\end{displaymath}
The matrix elements of $\Brho$ in the canonical basis $\ket{c}_{ij}$ are
\begin{displaymath}
\rho_{ij,i'j'} = \brac{c_{ij}}{\Brho}{c_{i'j'}} = \brac{a_i(1)}{\Brho(1)}{a_{i'}(1)} \; \brac{b_j(2)} {\Brho(2)}{b_{j'}(2)} = \rho_{ii'}(1) \rho_{jj'}(2)
\end{displaymath}
By calculating the trace of $\Brho(2)$
\begin{displaymath}
\sum_{j\in I_2} \brac{a_i(1)}{\Brho(1)}{a_{i'}(1)}\; \brac{b_j(2)} {\Brho(2)}{b_{j}(2)} =  \brac{a_i(1)}{\Brho(1)}{a_{i'}(1)}
\end{displaymath}
one reproduces the density operator acting in factor space 1.
\begin{displaymath}
\Tr_2\Brho = \Brho(1)
\end{displaymath}
Analogously, one gets $ \Brho(2)=\Tr_1\Brho$ and it is
\begin{displaymath}
\Tr\Brho =\Tr_1 \Brho(1) =  \Tr_2 \Brho(2) =\Tr_1\Tr_2\Brho =\Tr_2\Tr_1\Brho
\end{displaymath}

If $\Brho $ is not the tensor product of density operators acting in the respective factor spaces, we define a \emph{reduced density operator} $\tilde{\Brho}(1)$ by making the \emph{partial trace} on factor space $(2)$.
\begin{displaymath}
\tilde{\rho}_{ii'}(1) = \brac{a_i(1)}{\Brho(1)}{a_{i'}(1)}  = \sum_{j\in I_2} \brac{a_i(1) b_j(2)}{\Brho}{a_{i'}(1) b_j(2)}
\end{displaymath}
and analogously one gets $\tilde{\Brho}(2)$.

Reduced density operators have trace 1 but they are not idempotent. They represent mixed states.
\subsubsection{Expectation values in factor spaces}
The expectation value of an operator in a factor space of the tensor product is calculated as the expectation value of the extension in the tensor product. If  $\sfbA$ should be calculated in factor space 1, one needs the extension $\sfbA(1)\otimes \unit(2)$ and the unit operator defined in the tensor product, $\unit = \sum_{ij} \proj{a(1)_ib(2)_j}{a(1)_{i}b(2)_{j}}$.
\begin{gather*}
\Tr(\Brho(\sfbA(1)\otimes \unit(2))) = \Tr(\Brho\unit(\sfbA(1)\otimes \unit(2))) = \\
\sum_{i,j}\sum_{i',j'} \bigl(\brac{a_i(1)b_j(2)}{\Brho}{a_{i'}(1)b_{j'}(2)}\bigr)\times
\bigl(\brac{a_{i'}(1)b_{j'}(2)}{\sfbA(1)\otimes\unit(2)}{a_i(1)b_j(2)}\bigr)\\
= \sum_{i,j}\sum_{i',j'} \bigl(\brac{a_i(1)b_j(2)}{\Brho}{a_{i'}(1)b_{j'}(2)}\bigr)\times
\bigl(\brac{a_{i'}(1)}{\sfbA(1)}{a_i(1)}\overl{b_{j'}(2)}{ b_j(2)}\bigr)
\end{gather*}
Since the bases are ONBs, $\overl{b_{j'}(2)}{ b_j(2)} = \delta_{jj'}$ and summing over $j'$ gives
\begin{gather*}
\sum_{i}\sum_{i'}\Bigl(\sum_{j}\brac{a_i(1)b_j(2)}{\Brho}{a_{i'}(1)b_{j}(2)}\Bigr)\times
\brac{a_{i'}(1)}{\sfbA(1)}{a_i(1)} = \\
\sum_{i}\sum_{i'}\brac{a_i(1)}{\tilde{\Brho}(1)}{a_{i'}(1)}\times \brac{a_{i'}(1)}{\sfbA(1)}{a_i(1)}
=\sum_i \brac{a_i(1)}{\tilde{\Brho}(1)\sfbA(1)}{a_i(1)}
= \Tr(\tilde{\Brho}(1)\sfbA(1))
\end{gather*}
The expectation value of operator $\sfbA(1)$ is calculated with the reduced density operator $\tilde{\Brho}(1)$.

\subsubsection{Position representation and outer products}
The $m$-fold outer product of the eigenkets of the position operator $\ket{\mbr_1}\otimes \ket{\mbr_2}\otimes\dots \otimes \ket{\mbr_m}$ is often written as $\ket{\mbr_1}\ket{\mbr_2}\dots \ket{\mbr_m}$ or as $\ket{\mbr_1 \mbr_2\dots \mbr_m}$. The probability amplitude for the projection of an arbitrary ket $\ket{X}$ onto the positions of $m$ quantum objects represented by $\ket{\mbr_i}$
\begin{displaymath}
\overl{\mbr_1 \mbr_2\dots \mbr_m}{X} = X(\mbr_1, \mbr_2,\dots ,\mbr_m) = \psi_X(\mbr_1, \mbr_2,\dots ,\mbr_m)
\end{displaymath}
is the representation of $\ket{X}$ as a function of the coordinates of the $m$ quantum objects in physical space.

\subsection{Tensor spaces}
This is the special case where all factors of a tensor product are the same vector space or its dual space. Here, dual spaces will not be considered as factors.  Now again, bold face symbols will describe vectors.
\subsubsection{Definition}
Given a vector space $\mfV$ of dimension $n = \di {\mfV}$ with basis $B=\{\mbb_i\}_{i\in I}$. The  $m$-fold tensor product  ${\mfT}^{\otimes m}= \bigotimes_{i=1}^m \mfV $ is called the \emph{tensor space of order $m$}, its elements  are \emph{tensors of order} $m$.  The dimension of ${\mfT}^{\otimes m}$ is $(\di {\mfV})^m$ and its basis elements are the outer products of the basis vectors of $B$
\begin{displaymath}
\mbb_{i_1i_2\dots i_m} = \mbb_{i_1}\otimes \mbb_{i_2}\otimes \dots \otimes \mbb_{i_m}
\end{displaymath}
By convention, the tensor space of order one, ${\mfT}^{\otimes 1}$, is the vector space $\mfV$, and the tensor space of order zero, ${\mfT}^0$, is the scalar field.
Every tensor $\mbt\in{\mfT}^{\otimes m}$ is called a \emph{homogeneous tensor}, if it is a simple outer product of vectors $\mbv_i\in\mfV$, that is, $\mbt = \mbv_{1}\otimes \mbv_{2}\otimes \dots \otimes \mbv_{m}$. Every tensor $\mbt$ of ${\mfT}^{\otimes m}$ can be written as a linear combination of the homogeneous basis tensors $\mbb_{i_1i_2\dots i_m}$,
\begin{displaymath}
\mbt = \sum_{i_1} \sum_{i_2}\dots \sum_{i_m} t^{i_1i_2\dots i_m}\mbb_{i_1i_2\dots i_m}
\end{displaymath}
It is usual to write the indices of the coefficients of linear combination as upper indices lower indices are used for the vectors, then the Einstein summation convention can be applied  saying that if in a product the same index occurs as an upper and a lower index one has to sum over this index and the sum symbol can be omitted, that is $\mbt = t^{i_1i_2\dots i_m}\mbb_{i_1i_2\dots i_m}$.
\subsubsection{Tensor algebra}
A tensor of order two is the outer product of two vectors, which are tensors of order one, and in terms of tensor spaces one can write ${\mfT}^{\otimes 2} = {\mfT}^{\otimes 1}\otimes{\mfT}^{\otimes 1}$. This can be extended
${\mfT}^{\otimes (m+n)} = {\mfT}^{\otimes m}\otimes{\mfT}^{\otimes n}$, tensors of order $m+n$ are outer products of tensors of order $m$ and order $n$. Besides the two operation in a vector space, addition of vectors and multiplication of a vector with a scalar, the outer product is another important operation but the result of an outer  product of tensors is, in general, not element of either of the factor spaces. If a vector space is endowed with an additional operation, called a multiplication of vectors, one speaks of an algebra. In it, both the sum and the product of vectors are elements of the same space. If tensor multiplication should become an operation of an algebra, the algebra must be constructed appropriately to have all necessary properties. The direct sum of tensor spaces of all orders is a vector space with all  necessary properties, it is called the \emph{tensor algebra over the vector space} $\mfV$.
\begin{displaymath}
{\mfT} = \bigoplus_{m=0}^{\infty} {\mfT}^{\otimes m}
\end{displaymath}
In this vector space addition and multiplication of tensors of any order is possible.

\subsubsection{The alternating tensor algebra}
From all subalgebras of ${\mfT}$, the \emph{alternating tensor algebra} $\mfA$ is most important for the description of many-fermion systems.
Mapping an outer product $\mba\otimes \mbb$ with $\mba\ne \mbb$  onto $\mbb\otimes\mba$ is a permutation, or more precisely a transposition of the factors in the product. A second transposition reproduces the original tensor product, therefore, with each transposition the sign of the tensor product may either change or may not change. For products with more than two factors this holds for transpositions of any pair of factor. In general, $\mbb\otimes\mba$ is different from $\mba\otimes \mbb$. There are, however, tensors where only the sign may change by a transposition, these tensors are called \emph{symmetric}; if the sign changes the tensors are called \emph{antisymmetric}. Such tensors are, in general, linear combinations of homogeneous tensors. Tensors that are (anti)symmetric in every pair of factors are called \emph{totally (anti)symmetric}. For a totally antisymmetric tensor follows, that it is zero if it contains at least two identical vectors. Since transpositions are special permutations, and since every permutation $\pi$ can be written as product of say $p$ transpositions, there are $p$ changes of sign, the sign of the permutation is  $\sign\pi=(-1)^p$.  Permutations of $m$ objects constitute the symmetric group of degree $m$, symbol $S_m$,  permutations having sign $+1$ are even, permutations having sign $-1$ are odd.

If the operator
\begin{displaymath}
\tilde{\sfbA} = \sum_{{\pi}\in S_{m}}\sign({\pi}) \Bpi,
\end{displaymath}
which is the sum of the permutation operators $\Bpi$ multiplied by their sign, is applied to a homogeneous tensor $\mbt = \mbv_{1}\otimes \mbv_{2}\otimes \dots \otimes \mbv_{m}$, one gets a linear combination of permutations of the outer product of vectors multiplied by the corresponding sign of the permutations, which is called an \emph{alternating tensor}, it is the \emph{totally antisymmetric component} of $\mbt$, which is written as the \emph{wedge product} of the vectors
\begin{equation}\label{equ:wedge}
\mbv_{1}\wedge \mbv_{2}\wedge \dots \wedge \mbv_{m}.
\end{equation}
If $\tilde{\sfbA}$ is multiplied by  $1/m!$, one gets the \emph{antisymmetrization operator}
\begin{equation}
\sfbA = \frac{1}{m!}\sum_{{\pi}\in S_{m}}\sign({\pi}) \Bpi,
\end{equation}
which is a projection operator, and, if it is applied to an alternating tensor, it acts as the unit operator. Applying $\sfbA$ to $\mbt$ gives the canonical alternating tensor
\begin{equation}
\mba =  \frac{1}{m!}\mbv_{1}\wedge \mbv_{2}\wedge \dots \wedge \mbv_{m}
\end{equation}
This is the natural way to create an alternating tensor from an arbitrary tensor.  ${\mfA}^{\otimes m}$ is the tensor space of alternating tensors $\mba$ of order $m$. If the dimension of the vector space ${\mfV}$ is $\di\mfV$, the dimension of the tensor space ${\mfA}^{\otimes m}$ is  $\displaystyle \binom{\di\mfV}{m}$.

The direct sum
\begin{displaymath}
{\mfA} = \bigoplus_{m=0}^{\infty}{\mfA}^{\otimes m}
\end{displaymath}
is the alternating tensor algebra. Again, by convention is $\mfA^{\otimes 0}$ is the numeric field and $\mfA^{\otimes 1}$ is the vector space $\mfV$.

\subsubsection{Slater determinant}
If the elements of $\mfV$ are normalized orbitals $\phi_i$, the outer product $\mbt=\phi_1\otimes \phi_2\otimes\dots\otimes \phi_m$ is called the \emph{Hartree product} of the $m$ orbitals $\phi_1$, $\phi_2,\dots\phi_m$. The canonical alternating tensor $\mba=\frac{1}{m!}\phi_1\wedge \phi_2\wedge\dots\wedge \phi_m$ is not normalized, but when multiplied by $\sqrt(m!)$ it is normalized. $\sqrt{m!}\mba =
\frac{1}{\sqrt{m!}}\phi_1\wedge \phi_2\wedge\dots\wedge \phi_m$ is called a \emph{Slater determinant}.

\subsubsection{Antisymmetrization of products of antisymmetric tensors}
The outer product of  tensors of different tensor spaces $\mfA^{\otimes r}$ and $\mfA^{\otimes t}$  is not totally antisymmetric,
$\mba_{i_1,i_2\dots i_r}\otimes \mba_{j_1,j_2\dots j_t} \ne \mba_{k_1,k_2\dots k_{(r+t)}}$, the permutations of indices between the two tensor spaces are missing. To make the product totally antisymmetric one needs an antisymmetrizer $\sfbA_R$ that swaps indices corresponding to different tensor spaces. Then the antisymmetrizer in ${\mfA}^{\otimes (r+t)}$ is the  product $\sfbA_{r+t} = \sfbA_R\otimes  \sfbA_r\otimes \sfbA_t$ of the antisymmetrizers in ${\mfA}^{\otimes (r)}$ and in ${\mfA}^{\otimes (t)}$ and the antisymmetrizer $\sfbA_R$ made with the coset representatives.
\begin{equation}\label{equ:mcAR}
\sfbA_R = \frac{r!t!}{(r+t)!}\sum_{\sigma\in R} \sign(\sigma) \Bsigma
\end{equation}

Therefore is
\begin{displaymath}
\mba_{k_1,k_2\dots k_{(r+t)}} = \sfbA_R \mba_{i_1,i_2\dots i_r}\otimes \mba_{j_1,j_2\dots j_t}
\end{displaymath}

\subsubsection{Direct products of subgroups and its cosets}
A subgroup $H$ with order $\ord{H}$ of group $G$ with order $\ord{G}$ has the index $I=\ord{G}/\ord{H}$, which is the number of coset of $H$. If $H$ is multiplied from left with an element $g\in G$ one gets the left coset $gH$ of $H$, which is a subset of $G$. When $H$ is multiplied from left with all $g\in G$, one gets exactly $I$ disjoint left cosets,  $C_1, C_2, \dots, C_I$,  the union of the cosets gives $G$, $\bigcup_{i=1}^I C_i = G$. One coset is the subgroup $H$ itself. Every coset has as many elements as the subgroup $H$, and all elements of a coset $C_i$ are equivalent insofar as multiplication of $H$ by any element $t\in C_i$ gives the coset, $tH = C_i$. This is expressed by saying, that every element of the coset is a \emph{generator} of the coset. By selecting from each coset one element as representative, one gets the set $R$ of coset representatives.

Given a set of $r+t$ indices  $N= \{1,2,\dots,r, r+1, r+2,\dots, r+t\}$, the permutations acting on this set are elements of the symmetric group $S_{r+t}$ of order $(r+t)!$. The permutations acting on the disjoint subsets $N_1=\{1,2,\dots,r\}$ and $N_2=\{r+1, r+2,\dots, r+t\}$ are elements of the symmetric groups $S_r$ and $S_t$, respectively, both are subgroups of $S_{r+t}$ as is the direct product $S_r\times S_t$, the order of the direct product is $r!t!$. The permutations of the direct product act on set $N$ but never swap elements of subsets $N_1$ and $N_2$. The index of the direct product in $S_{r+t}$  is $(r+t)!/r!t!$. To get all elements of $S_{r+t}$ it is necessary to multiply all elements of the direct product from left with all elements of the coset representatives.

\bibliography{ManyElBonding}

\begin{thebibliography}{56}
\bibitem [{Sax({\natexlab{d}})}]{Saxb}%
  \BibitemOpen
  \href@noop {} {}  \bibinfo {note} {In the literature,
  ``tensor product'' means the tensor product of vector spaces. The tensor
  product of vectors is frequently called the outer product, it is a vector and
  therefore also called a tensor product vector (See Cohen-Tannoudji), whereas
  Susskind speaks of product states. In the following, tensor product will mean
  a tensor product of vector spaces and a tensor product of vectors will be
  termed as outer product of vectors.}\BibitemShut {Stop}%
\bibitem [{Sax({\natexlab{e}})}]{Saxe}%
  \BibitemOpen
  \href@noop {} {} \bibinfo {note} {In algebra, factor
  spaces are completely different objects.}\BibitemShut {Stop}%
\end{thebibliography}

\end{document}